# A Temporal Filter to Extract Doped Conducting Polymer Information Features from an Electronic Nose

Wiem Haj Ammar[a], Aicha Boujnah[a], Antoine Baron[b], Aimen Boubaker[a], Adel Kalboussi[a], Kamal Lmimouni[b] & Sébastien Pecqueur[b]*

[a] Department of Physics, University of Monastir Tunisia.

[b] Univ. Lille, CNRS, Centrale Lille, Univ. Polytechnique Hauts-de-France, UMR 8520 - IEMN, F-59000 Lille, France.

Email: sebastien.pecqueur@*iemn*.fr

## Abstract

Identifying relevant machine-learning features for multi-sensing platforms is both an applicative limitation to recognize environments and a necessity to interpret the physical relevance of transducers' complementarity in their information processing. Particularly for long acquisitions, feature extraction must be fully automatized without human intervention and resilient to perturbations without increasing significantly the computational cost of a classifier. In this study, we investigate on the relative resistance and current modulation of a 24-dimensional conductimetric electronic nose, which uses the exponential moving average as a floating reference in a low-cost information descriptor for environment recognition. In particular, we identified that depending on the structure of a linear classifier, the 'modema' descriptor is optimized for different material sensing elements' contributions to classify information patterns. The low-pass filtering optimization leads to opposite behaviors between unsupervised and supervised learning: the latter one favors longer integration of the reference, allowing to recognize five different classes over 90%, while the first one prefers using the latest events as its reference to clusterize patterns by environment nature. Its electronic implementation shall greatly diminish the computational requirements of conductimetric electronic noses for on-board environment recognition without human supervision.

Keywords: conducting polymers, electronic noses, exponential moving average, sensing, classifiers

# Introduction

Consumer electronics have experienced a revolution in the use of sensing hardware as machine-learning supported information generators, thanks to the internet-of-things. If sensors were used exclusively for metrology, sensing devices are now exploited to increase the perception field of electronics with environmentally sensitive input layers, constantly generating organic data to be classified. If smartphones and watches have now become the epidermal nerves of our modern society, it is at the cost of valuing hardware for information correlation before quantifying physics.[1, 2] Microphones, photodetectors and accelerometers are less being used to measure harmonics, photons or displacements, than to recognize spoken queries, heartbeats or gesture patterns from large populations at a given time, so the very underlying physical mechanisms ruling their functioning may appear quite irrelevant for an application. However, there is still a large part of the information we sense with our biology, which cannot be classified artificially with a hardware: this is the case of olfacto-mimetism. Smells are less something to be measured than to be classified.[3] Fragrances of volatile molecules trigger reversible changes of states on our G-coupled olfacto-receptor proteins for which the imprint encodes a specific class of stimuli in our brain, among a trillion that could be recognized.[4] These vast combinations of smells can hardly be mapped in a low dimensional space. However, as we can barely break down the chemical composition of smells, considering each molecule as a "primary scent" as we do with colors, seems quite irrelevant to constitute groups of patterns, where a pattern has multiple identities as a composition. Identifying the right sets of material sensitizers that embed enough orthogonality in their information features is a key to co-integrate electronic noses that perceive enough differences in volatile molecules imprinting. Former studies have shown that varying the dopant of a same conducting polymer allows tuning the electro-transduction of different vapors in a blow on an exposed polymer semiconductor.[5] The performance of the recognition highly depends on the software architecture of a classifier, so the same dataset can either be optimally recognized in a supervised learning scheme or far less in the case of an unsupervised classification.[6] Therefore, the structure of a classifier greatly conditions the choice of the materials that are to be co-integrated on a sensitive array. The performances of the same classifier structure on the same raw dataset also greatly depends on the information feature that is inputted.[7] By using different information features that characterize the kinetics of a phenomenon or its thermodynamics, a classifier efficiently recognize different subsets of dynamical data, depending on what part of the physics the information feature shows best. Thus, preprocessing data before classification is of great importance, in order for all complementarities of a material set to constructively express themselves within a generic information feature. In case the information feature requires some parametrization by an external operator, it limits the approach to a minimalist set of data and allows an experimentalist to manually process the set for the training and the testing or to parametrize the feature that is environment dependent. This is the case when using resistance modulations as an information feature for a conductimetric electronic nose, where each feature depends on defining a floating reference resistance $R_0$ as $R(t_0)$ for each sensing unit measured in a reference environment. This reference has to be periodically renewed and routinely defined when opening to a class-specific environment and when a sensing element reaches its steady-state, measuring $R(t_0+t)$. Both parameters are required to routinely evaluate the information feature $R((t_0+t)-R_0)/R_0$ and are highly sensitive to the $R_0$ values, that are artificially set by an experimental user upon exposures to reference environments (purges) and class-specific environments. Also, such features are inadequate when a periodic referencing before each class exposure cannot be implemented experimentally: for instance, when training an electronic nose on learning pollution peaks in an outdoor test environment.

The physical triviality of an information pattern does not condition the recognition performances of a classifier algorithm; therefore, features do not necessarily have to be interpreted *ab initio* from elementary mechanisms prior to being used as information descriptors in machine learning. This is the case in studies led by Vergara, Huerta and coworkers,[8, 9] where the exponential moving average (ema) of a resistance is used as a floating reference in the information feature of sensing elements' response in a conductimetric electronic nose classifier. In their work, the choice for such information feature was motivated by its extensive use in predictive analysis of dynamical signals where low explicability of governing physical mechanisms does not hinder classification such as in financial analysis. At any time during the measurement of a dynamical vector $X(t)$, a floating reference $ema_\alpha(X(t))$ can easily be defined as (eq. 1,2):

$$ema_\alpha\big(X(t)\big) = \alpha \cdot X(t) + (1-\alpha) \cdot ema_\alpha\big(X(t - \Delta T)\big) \qquad \text{(eq.1)}$$

$$ema_\alpha\big(X(t=0)\big) = X(t=0) \qquad \text{(eq.2)}$$

where $0 \leq \alpha \leq 1$ is the signal attenuation coefficient (or smoothing factor), here defined as a time constant and $\Delta T$ the sampling period of $X$, here set to one second. By its recursive definition, the $ema_\alpha(X(t))$ is a low-cost floating reference that is defined by attenuating a fraction of the signal X(t) with the former floating reference $ema_\alpha(X(t-\Delta t))$, which can be erased from memory once the feature is computed.

We present in this study its implementation in a conductimetric electronic nose using doped conducting polymers as chemo-sensitive electro-transducer materials, to evaluate the $modema_\alpha(X(t))$ features as information descriptors, such as (eq. 3):

$$modema_\alpha(X(t)) = \frac{X(t) - ema_\alpha\big(X(t)\big)}{ema_\alpha\big(X(t)\big)} \qquad \text{(eq.3)}$$

Also, since the ema transform is a digital low-pass filter, its physical implementation in a RC based neuron model as a circuit is discussed to optimize the design of electronic nose information generators, allowing it to behave closer to the way an actual olfactory receptor neuron preprocesses the information prior to its classification.

# Experimental Section

**Device Fabrication:** The sensing hardware microfabrication is fully described in a former study.**[5]** Concentric Au microelectrodes (28 µm in diameter, channel length $L$ = 400 nm, spiral electrodes length $W$ such that $W/L > 10^3$) are lithographically patterned by an e-beam in a cleanroom environment. Different materials are subsequently added on the different structures as clusters of 16 individual elements. The different 10 mg/mL formulations are deposited individually on the different clusters of electrodes by drop-casting, from a solution of P3HT in its pristine state (mildly doped by $O_2$ as stored in air) and metal trifluorosulfonates (salts of $Fe^{III}$, $Bi^{III}$, $Cu^{II}$, $Al^{III}$, $In^{III}$, $Dy^{III}$ and $Ce^{III}$) acting as dopants (see Fig. 1).

**Electrical Characterization:** The currents were measured versus time with an Agilent 4155 parameter analyzer. Except for the first three minutes of acquisition, the sensing elements are measured in an air blow that passes through a solvent containing vial (flow: 1 mL/s).**[5]** The control of the different exposures, each lasting three minutes, was operated manually. Acetone, ethanol and water loaded flows were exposed six times successively with permutations to avoid a systematic biasing of the data due to the analyte sequence ordering. Periods of purges (via an unloaded vial) separate each gas exposure to avoid cross-contaminations. The total acquisition consists of 6900 datapoints sampled every second, constituting five classes of environments: "no flow" during the initial three minutes (180 points), "acetone", "water", "ethanol" (each of these 6×180 points) and "purge" between each exposure (19×180+160 points).

**Data Analysis:** Raw currents are generated and used without filtering. Sensors are considered as ohmic resistors, whose resistance values are estimated from the raw current trace measured under constant bias at 10 mV. Raw current curves have already been published,**[5]** and are available on a public repository. This study used two linear software classifiers: principal component analysis (PCA) was used by means of the Clustvis online open access tool,**[10]** scaled by unit variance and computed by singular value decomposition. All PCA data are available as supplementary information. The Moore-Penrose pseudo-inverse supervised classifier was adapted from a former study,[11] to five-class recognition as follows:

The classifier output $Y(t) \in \mathbb{R}^m$ (for $m$ = 5 classes) of a given input vector $X(t) \in \mathbb{R}^n$ from the test database (for $n$ = 24 sensing units) is determined from the weight matrix $\boldsymbol{W} \in \mathbb{R}^{m \times n}$ (eq. 4):

$$Y(t) = f \circ g(\boldsymbol{W} \cdot X(t)) \qquad \text{(eq.4)}$$

The weight matrix $\boldsymbol{W}$ is computed from the pseudo-inverse matrix $\boldsymbol{X}^+ \in \mathbb{R}^{n \times p}$ for a given set of $p$ input vectors $X(t)$ from the learning database, each of them associated to an output vector $Y(t)$ defined as a vector from the standard basis $(e_i)_{i=1}^m$ of $\mathbb{R}^m$ for each of the $m$ classes. For $p$ vectors from the learning database is defined the output matrix $\boldsymbol{Y} \in \mathbb{R}^{p \times m}$ for supervised learning, so $\boldsymbol{W}$ is defined as (eq. 5):

$$\boldsymbol{W} = (\boldsymbol{X}^+ \cdot \boldsymbol{Y})^T \qquad \text{(eq.5)}$$

The Moore-Penrose pseudo-inverse matrix $\boldsymbol{X}^+$ is calculated for a given set of $p$ input vectors $X(t)$, for which is defined the input matrix $\boldsymbol{X} \in \mathbb{R}^{p \times n}$, so that (eq. 6):

$$\boldsymbol{X}^+ = (\boldsymbol{X}^T \cdot \boldsymbol{X})^{-1} \cdot \boldsymbol{X}^T \qquad \text{(eq.6)}$$

The activation function $(f \circ g)$ adjusts the output $Y(t)$ sensitivity to each class density in the learning database by the $g$ function, and input vectors are chosen randomly in the learning database. The $g$ function is defined by the diagonal matrix $\boldsymbol{D} \in \mathbb{R}^{m \times m}$ as (eq. 7, 8):

$$g(\boldsymbol{W} \cdot X(t)) = \boldsymbol{D} \cdot \boldsymbol{W} \cdot X(t) \qquad \text{(eq.7)}$$

$$\boldsymbol{D} = diag(x_i, \dots, x_m) \qquad \text{(eq.8)}$$

where $x_i$ is defined as the ratio of the number of vectors that do not belong to the i$^{th}$ class in the learning database, over $m$. The $f$ function applies a "winner takes all" normalization of the $g(\boldsymbol{W} \cdot X(t)) \in \mathbb{R}^m$ vector, such that the attributed class is defined by the coordinate of $g(\boldsymbol{W} \cdot X(t))$ which has the highest value (eq. 9-11):

$$\forall i \leq m, \qquad \exists \, i_0(t) \leq m \quad | \quad g(\boldsymbol{W} \cdot X(t))_{i=i_0(t)} = \max(g(\boldsymbol{W} \cdot X(t))_{i=1}^m) \qquad \text{(eq.9)}$$

$$if \; i = i_0(t), \quad then \; f \circ g(\boldsymbol{W} \cdot X(t))_i = 1, \quad else \; f \circ g(\boldsymbol{W} \cdot X(t))_i = 0 \qquad \text{(eq.10)}$$

$$\Leftrightarrow \; Y(t) = e_{i=i_0(t)} \qquad \text{(eq.11)}$$

# Results

**Data Collection and Feature Extraction.**

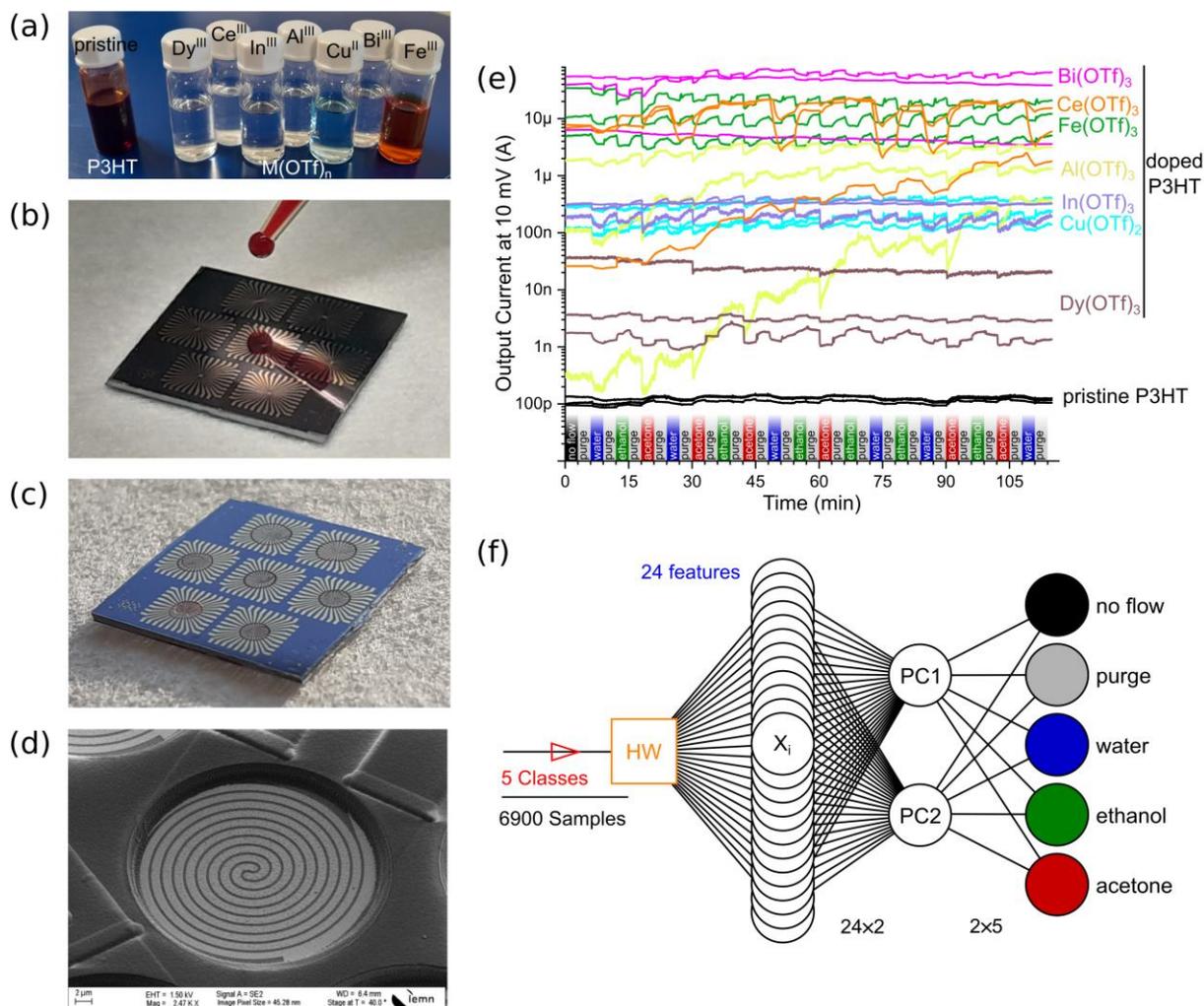

**Figure 1. Unsupervised Classifier for Solvent Vapor Recognition | a-d,** Hardware used in the classification. **a,** Photographs of the solutions used to sensitize the different clusters of conductimetric elements. **b,** Deposition of the chemosensitive materials on the individual clusters by two sequential drop castings. **c,** Seven clusters sensitized with different metal triflate doped and undoped P3HT coatings as chemospecific transducers of environmental information. **d,** Scanning electron microscope picture of the micro-electrodes of an individual sensing element used in the study. **e,** Raw current signals extracted from a population of 24 different sensing elements supplied under DC voltage bias, coated with eight different metal triflate doped and undoped P3HT coatings (three sensing elements per cluster), exposed under repeated volatile solvent vapor exposures. **f,** Software classifier structure used in the study for unsupervised classification of the five different classes of exposure, using the first two principal components of the PCA analysis. Classification of samples measured upon exposure of water (in blue), acetone (in red) or ethanol (in green), or upon purging the system (in grey) or upon starting the measurements without air flow (in black).

The hardware part of the classifier used in this study is composed of several clusters of conductimetric elements, with the metrological instrument that allows outputting the current $i(t)$ for each sensing element polarized under a steady voltage bias in a chemically varying environment.[5] Each cluster of conductimetric elements was sensitized with a conducting material and a dopant (see Fig. 1a-d). The dopants transduce to the conducting polymer specifically the chemical interactions they have with environmentally present volatile molecules. When exposed to an air blow, dopants imprint on each polymer coating the presence of volatile molecules in the carrier flow by the transience of the electric current that outputs under steady voltage addressing (see Fig. 1e). The imprint for each volatile molecule on each sensing element is specific to the doping element.[5] When acetone, water and ethanol are presented in a steady air blow on the sensing array, decreases or increases of the current arise specifically on the different elements. The response of three sensing elements coated with the same materials do not exhibit electrical properties without dispersion in current magnitude at rest and/or in current modulation during exposures. These raw data dispersions were attributed to the limitation of the deposition technique resulting in an inherent spatial heterogeneity of the materials, which conditions the reproducibility of the device performances.[5] Despite the fact that some devices even displayed significant drifts over time (see Fig. 1e for the Al(OTf)$_3$ doped P3HT traces for instance), the principal component analysis (PCA) performed on the current modulation shows that most of the information is rather conditioned by the deposited materials. To classify the five different classes of exposures that are presented to the sensing layer of the hardware ("no flow", "purge", "acetone", "ethanol" and "water"), the first two principal components were used to observe whether the information feature referenced by the ema floating point can be classified in an unsupervised way (see Fig. 1f). By clustering most of the data by the nature of different classes that are exposed, and by the nature of the materials that compose the two first principal components, the aim is to demonstrate if an ema-referenced information feature can be generically used as a descriptor to recognize dynamical environments with a conducting polymer electronic nose (see Fig. 2).

**Exponential Moving Average as a Floating Reference.**

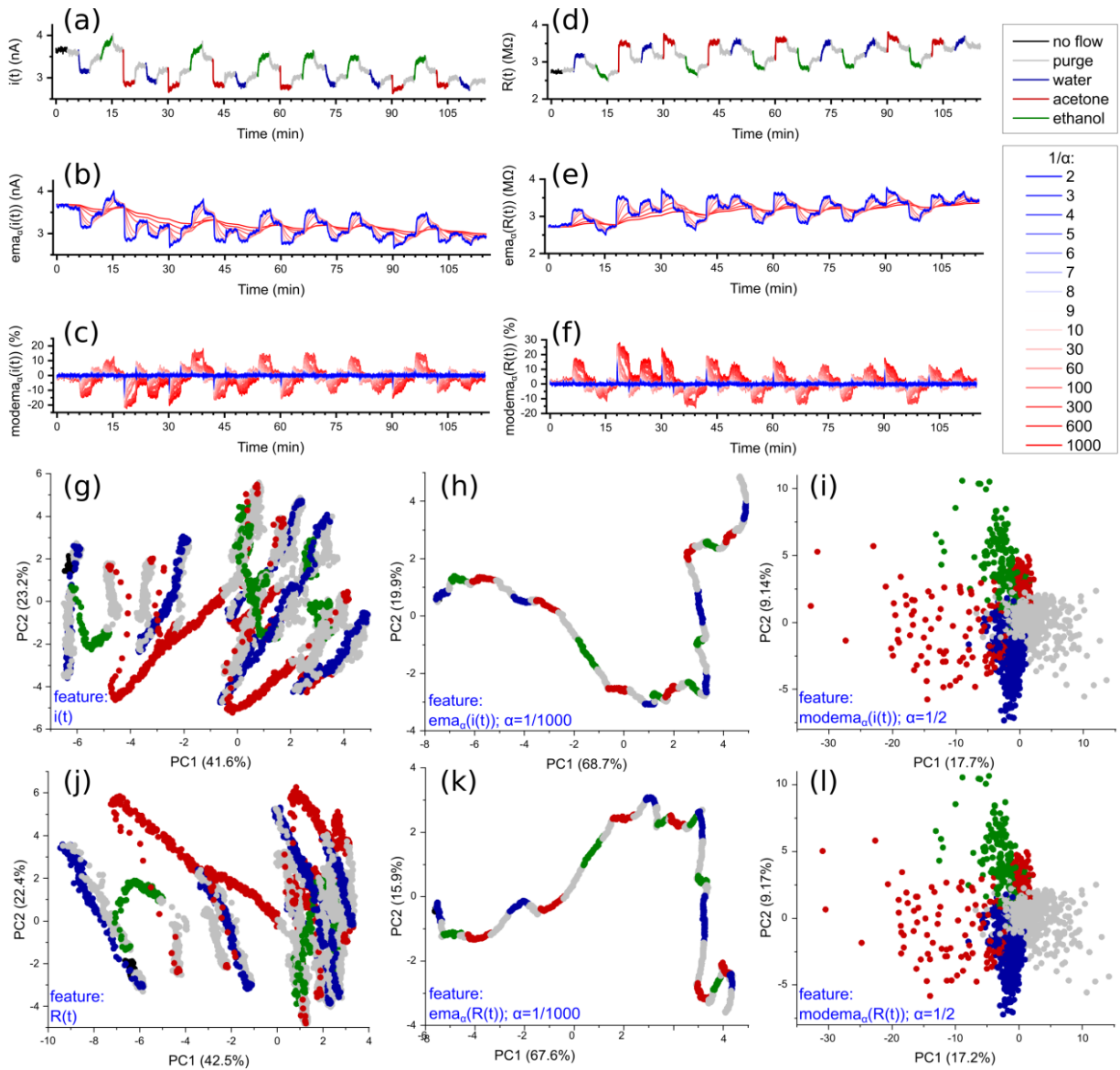

**Figure 2. The *modema*() function as a transform to condition the raw-data signal as relevant information descriptor for classification | a,d,** Raw data signal for one sensing element exposed under the different sequences, where each exposure is marked by a specific color in the single measurement trace of current (in **a**) or resistance (in **d**). **b,e,** Exponential Moving Average (ema) transform of the raw current (in **b**) and resistance (in **e**) signals displayed in **a** and **d** graphs. The effect of the $\alpha$ parameter in the $ema_\alpha()$ function is displayed in the color gradient. **c,f,** Modulation centered on the exponential moving average (modema) transform of the raw current (in **b**) and resistance (in **e**) signals displayed in **a** and **d** graphs. The effect of the $\alpha$ parameter in the $modema_\alpha()$ function is displayed in the color gradient. **g-l,** First-two-component scores of the principal component analysis for the complete 24-dimensional datasets of different information features. Scatter colors according to the legend in **a,d**. **g,j,** (PC1;PC2) planes for the raw data information *i(t)* and *R(t)*, showing poor data separation among the five environmental classes. **h,k,** (PC1;PC2) planes showing data separation of the $ema_\alpha()$ transformed *i(t)* and *R(t)* signals (for $\alpha$ = 1/1000), but poor environmental specificity. **i,l,** (PC1;PC2) planes showing data separation of the $modema_\alpha()$ transformed *i(t)* and *R(t)* signals (for $\alpha$ = 1/2) and environmental specificity for at least four different classes.

As an inverse of each other, resistances and admittances are not linearly correlated. Thus, as a linear classifier, PCA based on one will differ from a PCA based on the other. Therefore, the ema-referenced feature study was performed in parallel with the resistance $R(t)$ (in MΩ) and the current $i(t)$ (in nA) for each sensing element as raw data vectors $X(t)$ (see Fig. 2). Fig. 2a and 2d display the typical response of a sensing element without significant drift. In both Fig. 2b and 2e, the ema referencing behaves as a filter which reduces the signal noise for low $1/\alpha$ values. For values of $1/\alpha$ typically above 30, the filtering is such that the $ema_\alpha(X(t))$ presents a noticeable temporal delay compared to the signal $X(t)$, both for the resistance and the current. This delay tends to increase with a lower value of $\alpha$, where one can observe that the lower the $\alpha$, the less sensitive to upcoming exposures it becomes, leaving only a slow decay representing the sensing elements signal drift. As the smoothing quality depends highly on the signal modulation upon exposing different classes, the optimization of the filter cut-off shall be sensing element dependent. More particularly, a sensing element which displays only moderate signal changes may result in an $ema_\alpha(X(t))$ reference that is representative of the sensing element temporal drift, while a sensing element displaying large signal modulations may result in an $ema_\alpha(X(t))$ featuring a noise free image of the transient signal at a given value of $\alpha$. Therefore, it is expected that a sensing array showing large dispersion in signal modulations will have a much higher sensitivity when optimizing the $\alpha$ attenuation coefficient of an $ema_\alpha(X(t))$ floating reference than a sensing array composed of sensing elements that modulate with the same amplitudes. The sensitivity of the $ema_\alpha(X(t))$ floating reference to the $\alpha$ attenuation coefficient consequently affects the sensitivity of the $modema_\alpha(X(t))$ information feature. As show Fig. 2c and 2f, the $modema_\alpha()$ function conditions greatly the image of the sensing elements signal for both the resistance and the current as raw data information. The lower the $\alpha$ attenuation coefficient and the higher the dispersion of the $modema_\alpha(X(t))$ values over the whole acquisition. It is noticed that the information feature $modema_\alpha(X(t))$ is centered around zero. This property is a rather interesting feature if sensing elements' raw signal current or resistance displays large drifts over time, due to various physical mechanisms that may not necessarily threaten the quality of the information. It is also noticed that for any value of the $\alpha$ attenuation coefficient comprised between 1/2 and 1/1000, the noise of the $modema_\alpha(X(t))$ is relatively comparable to the one present in the raw data. As the $ema_\alpha()$ function only filters the high frequency signal of the floating reference, most of the high frequency information associated to the raw data $X(t)$ is preserved in the $modema_\alpha(X(t))$ information descriptor. As this noise seems less sensitive to $\alpha$, a large dependency of the $modema_\alpha(X(t))$ classification with $\alpha$ is expected, depending on whether the class-depending information is localized at the level of the information descriptor noise or its amplitude modulation. This has been verified by the unsupervised classification by PCA, focused strictly on the first two principal components (see supplementary material for details on all $\alpha$ cases). Fig. 2g and 2j show that the main information feature is not environment specific when using the raw data resistance or current as information descriptor, with a temporal effect on the scores displayed in (PC1;PC2). Fig. 2h and 2k show that using the floating reference $ema_\alpha(X(t))$ reduces greatly the scores dispersion in the (PC1;PC2) projection for low value of $\alpha$ (we evidenced that the lower the filtering, the closer looks the PCA with the ema to the one with the raw data). However, using the floating reference $ema_\alpha(X(t))$ still does not allow clustering the scores by the nature of the environment. The case of using the $modema_\alpha(X(t))$ as an information descriptor is substantially different to the one of the raw data $X(t)$ and the floating reference $ema_\alpha(X(t))$, for both the resistance and the current data as raw data value. One can observe in Fig.2i and 2l that at low value of $\alpha$ (effect of $\alpha$ detailed in the previous section), the information of the PCA score projection on (PC1;PC2) organizes by the nature of the environment class that is presented to the sensing hardware. PC1 discriminates specifically purge sequences from some acetone exposures, while PC2 discriminates specifically water exposures from ethanol ones, both when using the resistance and the current as raw data values. It is noticed that the dispersion of the scores is very comparable in both cases of using the resistance and the current as raw data, despite the results being different. This stresses the fact that the approach can be applied on resistance or current values without observing significant differences in the quality of the recognition, both classifications are distinct and may require a closer comparison of the explained variance for different values of the $\alpha$ attenuation coefficient in both cases (see Fig. 3).

**Relationship between Attenuation Coefficient and Environment Clustering.**

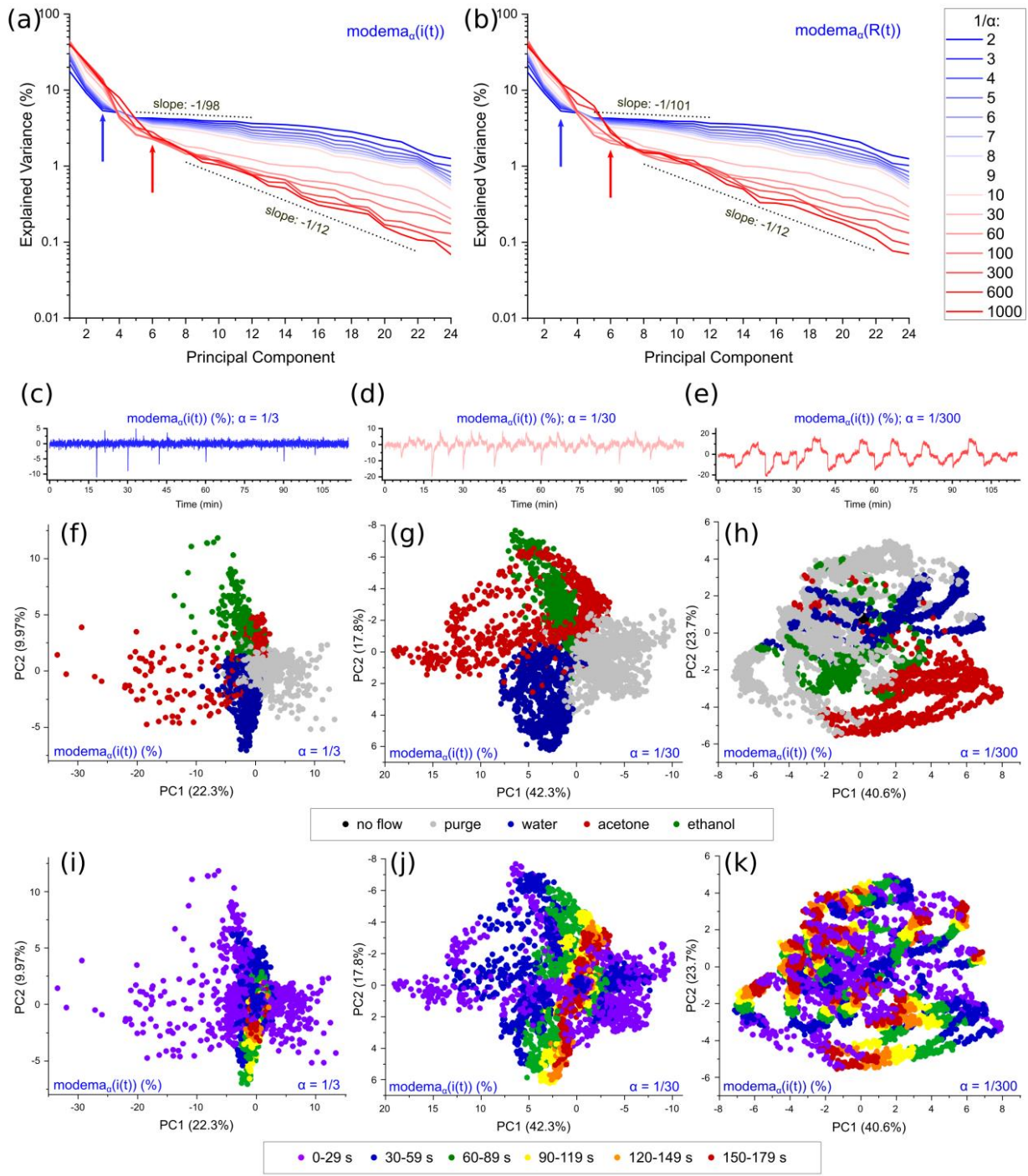

**Figure 3. Data separation of the *modema()* information descriptors depending on the attenuation coefficient α | a,b,** Scree plots for *modemaₐ(i(t))* and *modemaₐ(R(t))* with different values of α. **c-k,** First-two-component scores of the principal component analysis on the *modemaₐ(i(t))* datasets for $1/\alpha$ = 3 (**c**, **f** and **k**), 30 (**d**, **g** and **j**) and 300 (**e**, **h** and **k**). **c-e,** A single raw current data trace from Fig. 2b for different value of $1/\alpha$ = 3 (**c**), 30 (**d**) and 300 (**e**). **f-k,** (PC1;PC2) planes for the full *modemaₐ(i(t))* dataset for $1/\alpha$ = 3 (**f** and **i**), 30 (**g** and **j**) and 300 (**h** and **k**), labelled with the five environmental classes (**f**, **g** and **h**) or with six classes of acquisition times after exposure starting at $t_0$ (**i**, **j** and **k**).

The scree plots of the PCA analysis using $modema_\alpha(i(t))$ (Fig. 3a) and $modema_\alpha(R(t))$ (Fig. 3b) as distinct information descriptors quantify how much information variance is preserved in the different principal components of the scores. In both cases, very similar effects of the attenuation coefficient $\alpha$ are evidenced. The cumulative variance increases in PC1 and PC2 with the diminishing of the attenuation coefficient $\alpha$ (see Fig. 3a and 3b). As the decrease of $\alpha$ is associated to an increase of the data dispersion in the information descriptor (see Fig. 3c-3e for $\alpha$ = 1/3, 1/30 and 1/300 respectively), more information is contained in the first principal components by the choice of an attenuation coefficient that characterizes the modulations of signal upon exposures to upcoming classes. This is in line with the idea that higher values of $\alpha$ lead to lower signal-to-noise ratio (where only fast transients are detected few seconds after the change of an exposure), most of the information being yielded by stochastic processes responsible for the preponderant noise observed in $modema_\alpha(i(t))$ and $modema_\alpha(R(t))$. The variance dispersion over the different principal components shows the same dependency with the attenuation coefficient $\alpha$ in both cases of using the $modema_\alpha(i(t))$ and $modema_\alpha(R(t))$ information features (see Fig. 3a and 3b). By the elbow method (arrows displayed in Fig. 3a and 3b), one can observe that most of the explained variance is gathered within the first three principal components in case of high attenuation coefficient, while in case of applying a low attenuation coefficient, the explained variance is spread over the six principal components. This shows that when the $ema_\alpha()$ filter cuts off most of the transient signal, the information density is sparser (more noised) but the remaining explainable signal simpler (less parametric on a linear model). Instead, if the $ema_\alpha()$ only filters noise, more data can be extracted over the whole acquisition by the reading of the modulations in $modema_\alpha(X(t))$, but the information is harder to classify with only two principal components. Such property shows that various data dispersion is to be expected, depending on the attenuation coefficient $\alpha$ applied on the $ema_\alpha()$ filter conditioning the $modema_\alpha(X(t))$ information descriptors. The following results specifically focus on using currents $X(t)$ as we expect similar classification dependencies with $\alpha$ in the case of using $R(t)$ as raw data information.

Principal component scores show various organizations on the (PC1;PC2) projection depending on the attenuation coefficient $\alpha$ (see Fig. 3f-k). By looking at the gradual decrease of the attenuation coefficient $\alpha$, it can be observed that data separability by the nature of the exposed environment tends to increase (comparison between $\alpha$ = 1/3 and 1/30 in Fig. 3f and 3g), before the score clusters start collapsing in (PC1;PC2) (comparison between $\alpha$ = 1/30 and 1/300 in Fig. 3g and 3h). This indicates that the $ema_\alpha(i(t))$ shall feature a floating reference point that is filtered from the inherent signal noise, but not filtered from the whole activity of the signal. Such characteristic presents similarities of both resistance modulation ($R/R_0-1$) and drift resistance ($\dot{R}/R$) as information descriptors, which gather different aspects of the signal's physics.[7] By the structure of the feature, both $modema_\alpha(i(t))$ and $R/R_0$ normalize a value recorded at a given time to a buffered reference that changes all along the acquisition. However, both $modema_\alpha(i(t))$ and $\dot{R}/R$ are referencing to a local value all along the acquisition. $modema_\alpha(X(t))$ can represent a generic compromise to reference data without intervention of an external user (unlike setting $R/R_0-1$) with a higher representation of the steady-effect of the environment on a material (unlike $\dot{R}/R$). When looking at the temporality of the data, it appears that the most separable feature samples are generically originating from the most dynamical raw data ones (as property shared with $\dot{R}/R$).[7] Particularly for the cases of a high attenuation coefficient $\alpha$, data generated within the first 30 seconds after new class exposure at $t_0$ exhibit a higher spreading on a (PC1;PC2) projection plane when compared to points recorded much later for $\alpha$ = 1/3 and 1/30 (see Fig. 3i and 3j). In these cases, the $ema_\alpha(i(t))$ floating reference is very similar to the raw signal $i(t)$. So, the $modema_\alpha(i(t))$ is very noised except when in the transient (see Fig. 3c, 3d). For a lower attenuation coefficient $\alpha$ at 1/300, it is also noticed that PCA scores on (PC1;PC2) do not organize by acquisition times anymore, as the $ema_\alpha(i(t))$ floating reference embeds less class-specific information and $modema_\alpha(i(t))$ reflects more a drift-free relative current modulation (see Fig 3e, 3k). This is particularly interesting to observe that, despite the transduction mechanism being interpreted as a thermodynamic property,[5, 12] resistance modulation is not the main source of information that an unsupervised classifier such as the PCA exploits to sort output-current related data by the nature of the molecular environment that is exposed on the doped polymers. As the quality of the PCA clustering is sensitive to the attenuation coefficient $\alpha$, it may be important to inspect the PC loadings for different $\alpha$ to observe if the classifier uses universally a same selection of materials, independently from what physical properties they may contribute to the $ema_\alpha(i(t))$ floating reference (see Fig. 4).

**Conducting Polymer Doping Complementarity in the Principal Component Analysis.**

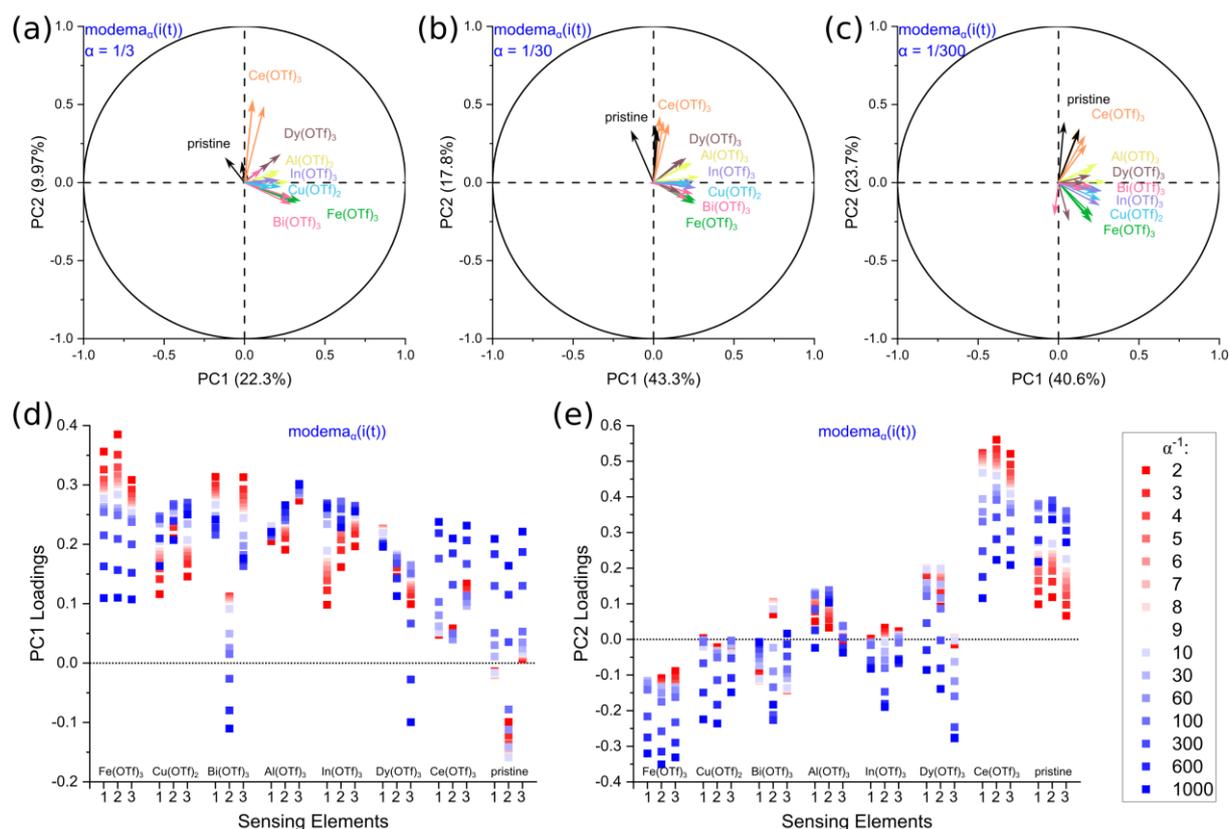

**Figure 4. Contribution of the Different Materials in the PC1 and PC2 Loadings | a-c,** Correlation circles in (PC1;PC2) for the *modema$_\alpha$(i(t))* dataset of 24 sensing elements composing eight different metal triflate doped and undoped P3HT coatings, for $1/\alpha$ = 3 (**a**), 30 (**b**) and 300 (**c**). **d,e,** Dependency of the $\alpha$ attenuation coefficient on the PC1 (**d**) and PC2 (**e**) loadings for the 24 sensing elements composed of the eight different metal triflate doped and undoped P3HT coatings, for the PCA on the *modema$_\alpha$(i(t))* feature as information descriptor.

As expected, loadings of PC1 and PC2 are very sensitive to the attenuation coefficient $\alpha$, depending on whether *ema$_\alpha$()* filters mostly the signal noise, or the whole class-specific dynamics in the floating reference point (see Fig. 4). The correlation circles show that doped polymer materials contribute differently in the principal components PC1 and PC2, with the attenuation coefficient $\alpha$ = 1/3, 1/30 or 1/300 (see Fig. 4a-c). Particularly, pristine P3HT is distinctively the only material that contributes negatively to the first principal component PC1 for lower $\alpha$ (see Fig. 4a, 4b), while for a higher attenuation coefficient $\alpha$, the material looses this property, and contribute positively to PC1 like all the doped polymers (see Fig. 4c). Surprisingly, one can observe that the direction of all loading vectors follows almost the same ordering in the correlation circles, and this, regardless of the value of the attenuation coefficient $\alpha$. Particularly, it seems that such order follows the same series of doping strength that was previously defined by the nature of the dopant.[5] So, a stronger dopant, inducing a higher value of the raw *i(t)* data (see Fig. 1e), seems to lead to a lower value for the phase coordinate on the loading vector of a sensing element in a correlation circle, while a mildlierly-doped polymer appears to induce a higher phase coordinate for its sensing element loading vector. Here, one shall not directly attribute such correlation to causality of a generic doping mechanism in a conducting polymer for sensing in a machine learning framework. The correlation made between a physical property of the material, inducing a trend of data organization in PCA, is mostly induced by the fact that doping increases the nominal value of the *i(t)* raw information, which indirectly increases the signal-to-noise ratio of the data collection. On the absolute value of the PC loadings in PC1 and PC2, all the sensing elements have different contributions depending on the value of the attenuation coefficient $\alpha$ (see Fig. 4d, 4e). Moreover, the inversion of contribution with $\alpha$ in the pristine P3HT loadings for PC1 is confirmed in Fig. 3d, such as for one Bi(OTf)$_3$ sensing element, and one Dy(OTf)$_3$ element over 15 different values of $\alpha$. Furthermore, Fig. 4e shows that an inversion of contributions with $\alpha$ in the PC2 loadings occurs for 12 different sensing elements over the 24 used in the experiment. It is observed that most contributions in the PC loadings at a given $\alpha$ are clearly material dependent (see Fig. 4d for Fe(OTf)$_3$-doped P3HT in particular, and Fig. 4e for Ce(OTf)$_3$-doped and pristine P3HT in particular). This tells that these contribution inversions induced by the $\alpha$ are no stochastic artefact and reflects a genuine property of the materials dynamics. Therefore, it seems that using an unsupervised PCA linear classifier feels inadequate for generically ranking the quality of a material population using the ema floating reference. To quantify a recognition at an optimal $\alpha$, supervised learning was performed to evaluate the potential of modema for classification (see Fig. 5).

**Supervised Training of Environment Recognition with Output Currents' Modema.**

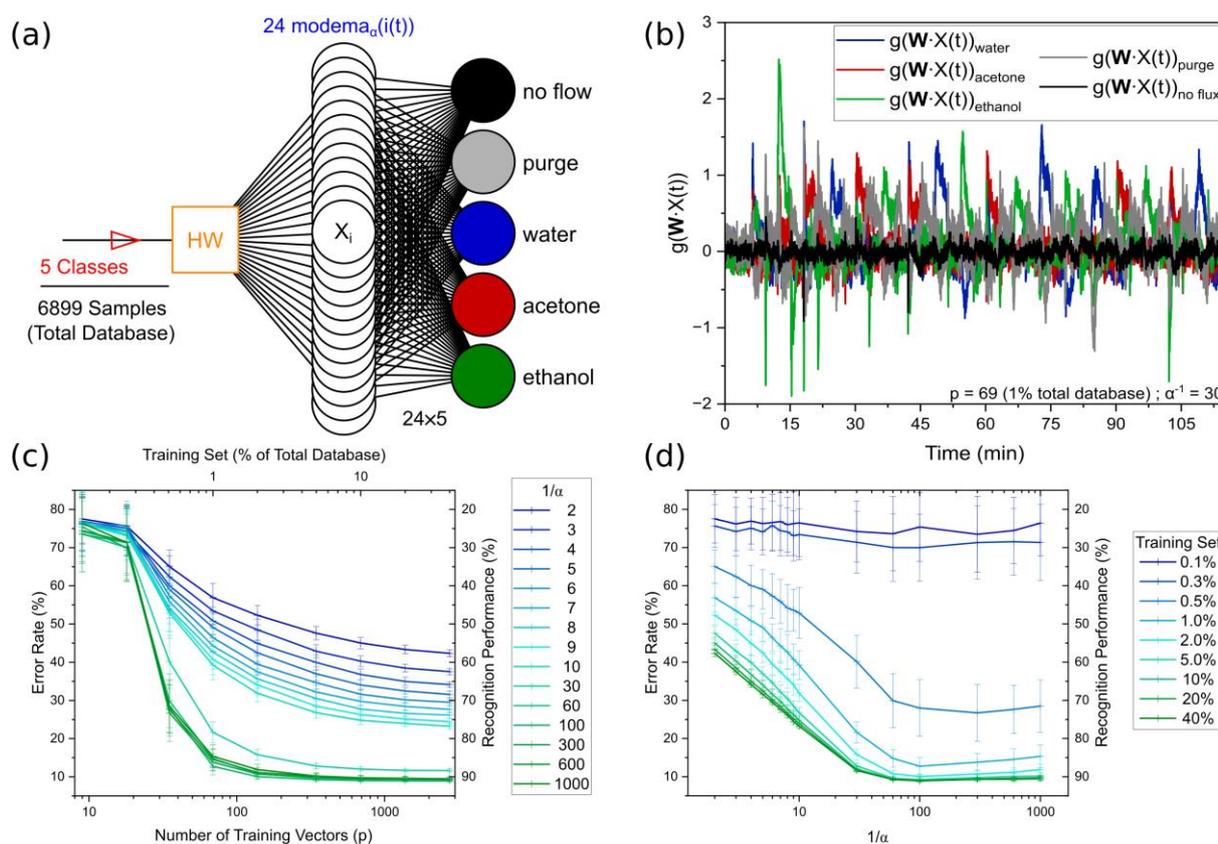

**Figure 5. Environment Recognition with the _modema$_\alpha$(i(t))_ Information Descriptor under Supervised Classification | a,** Software classifier structure used in the study for supervised classification of the five different classes of exposure, using the Moore-Penrose Pseudo-Inverse. **b,** Example of the five different coordinates of _g(**W**·X(t))_ output vector for a single training, using $1/\alpha$ = 30 and $p$ = 69 vectors. **c,** Dependency of the error rate for the five-class recognition with the size of the training set, for different values of the $\alpha$ attenuation coefficient. **d,** Dependency of the error rate for the five-class recognition with the $\alpha$ attenuation coefficient, for different sizes of the training set.

To quantify the performances of a classifier when recognizing molecular environments, the Moore-Penrose Pseudo-Inverse was used to supervise the training of the electronic nose on recognizing the five classes of environments by the use of _modema$_\alpha$(i(t))_ as the 24-dimensional information descriptor. In this case, the recognition space is not dimensionally reduced to only two principal components, so the _modema$_\alpha$(i(t))_ information is used in its entire complexity to recognize the feature vectors, whether they were recorded upon no flow, under purging or exposure of either a water, an acetone or an ethanol vapor blow (see Fig. 5a). As in former studies **[11]** and similarly as in the case of the PCA classifier, all dynamical data generated during a complete acquisition are used in a classifier. Training and testing vectors are randomly picked from the 6899 available vectors to form two complementary databases (no validation database was used to optimize the attenuation coefficient $\alpha$, as all the 15 $\alpha$ cases were generated in separated experiments as a variable of study). As a second variable of study, the size of the training dataset (related to the number of training vectors $p$) was studied to observe its impact on the classifier recognition performances at a given attenuation coefficient $\alpha$. The classification threshold was set on the weighted $g(\boldsymbol{W} \cdot X(t))$ vectors (with $modema_\alpha(i(t))$ as $X(t)$) optimized by pseudo-inversion after training, showing the different components of the five environments to express in majority at different periods of time during the acquisition (see Fig. 5b). For different training/testing cases, it was observed that the quality of the classification was highly dependent on both variables of study: the attenuation coefficient $\alpha$ of the 24-dimensional $modema_\alpha(i(t))$ input and the number of training vectors $p$ used for training/calibrating the classifier (see Fig. 5c, 5d). Naturally, the larger the dataset, the lower the error rate at a given $\alpha$. Also, the lowest attenuation coefficient $\alpha$ leads systematically to the highest recognition rate at a given size for a training dataset. Both properties were statistically verified, as each case displayed in Fig. 5c and 5d represents the average value of the error rate for 100 cases of randomly picked training vectors in each scenario of fixed $\alpha$ and $p$ parameters. The trend with $p$ is comparable to the former study on binary classification, using the same electrode structures as an OECT array to recognize projected gate voltage patterns.[11] However, it is noticed that a minimum of 0.5% of the database is required to significantly train the system recognizing the five classes, from 75±2% error (quasi independently from $\alpha$), down to 25±15% when trained with 5% of the database or higher (see Fig. 5c). The impact of the attenuation coefficient $\alpha$ is also quite significant on the error when the classifier is trained enough (see Fig. 5d). In case at least 0.5% of the database is used to train the classifier, a transition at $\alpha$ =1/30 is observed: for lower $\alpha$ values, the system does not progress further at a given $p$; and for higher $\alpha$ values, the classifier progresses at a quasi linear rate with $log(1/\alpha)$, as displays Fig. 5d. This result is quite surprising as it shows, contrary to the unsupervised classification by PCA, that the classification performs better under supervised training when the attenuation coefficient $\alpha$ is minimized. It is therefore particularly important to stress that the figure-of-merits for a sensing array material selection (see Fig. 6) are highly dependent on the machine-learning approach for a given application.

**Conducting Polymer Doping Complementarity with the Moore-Penrose Pseudo-Inverse.**

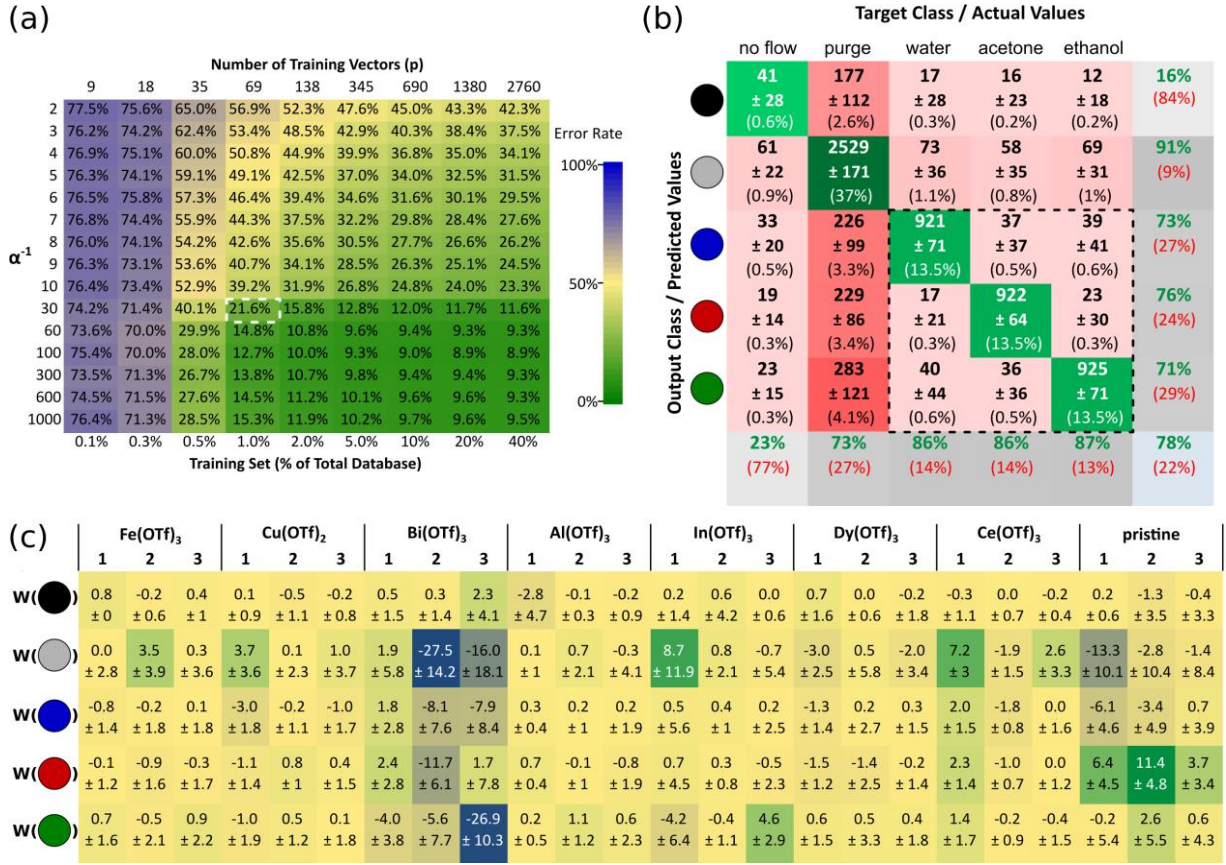

**Figure 6. Material/Environment Specificity in the Case of Supervised Classification with the *modem$\alpha_\alpha$(i(t))* Information Descriptor | a,** Correlation heatmap between the attenuation coefficient $\alpha$ and the size of the training dataset for the recognition of the five different classes of environments, using the supervised classifier. Each displayed error rate is calculated as the mean value of 100 training/testing with given $\alpha$ and $p$ values of different randomly selected sets of training vectors. **b,** Confusion matrix for the supervised classification of the five environments. The displayed values are calculated as the mean and standard deviation values of 1000 training/testing with 1/$\alpha$ = 30 and $p$ = 69 with different randomly selected sets of training vectors. **c,** Correlation heatmap between the weights connecting the different environmental classes to the 24 sensing elements composed of eight different metal triflate doped and undoped P3HT coatings, after supervised classification. The displayed values are calculated as the mean and standard deviation values of 1000 training/testing with 1/$\alpha$ = 30 and $p$ = 69 with different randomly selected sets of training vectors.

The fact that the supervised linear classifier favors the use of a highly attenuated *ema$_\alpha$(i(t))* floating reference, opposed to the unsupervised PCA classifier which favors the least attenuated *ema$_\alpha$(i(t))* motivates to further understand whether both classifiers exploit the same sensing element contributions in their weights/loadings or not (see Fig. 6). A summary of both Fig. 5c and 5d is represented (without statistics) as a correlation heatmap between $p$ and *1/$\alpha$* in Fig. 6a. Using the linear color gradients with the error rate, an optimal case between training dataset sparsity and reference signal attenuation is identified for $p$ = 69 (1% total database) and $\alpha$ = 1/30 to ensure 21.6% recognition on average. As the supervised approach requires large statistics of train-and-tests compared to PCA, the material selection optimization has been assessed only for the specific value of $\alpha$ = 1/30 for supervised classification, so no weight inversions with the attenuation coefficient is verified in the following. For the specific case of $(p;\alpha)$=(69;1/30), the confusion matrix displays the statistics of true/false positives/negatives of 1000 train-and-tests (see Fig. 6b). During the acquisition, "purge" samples are occurring the most while "no flow" ones are less common than each of the water, acetone and ethanol vapor exposures. Therefore, predictive rates are naturally higher to recognize "purge" (91%) than "no flow" (16%), with recognition of the three different vapors between 71 and 76% (see Fig. 6b). However, true positive rates are particularly in favor of classifying the three different vapors at 86-87% recognition rates, compared to "purges" (73%) and "no flow" (23%). It indicates that the classifier is particularly sensitive to exposures of vapors than either of both "no flow" and "purge" carrying no solvent vapors: upon exposure of either "acetone", "ethanol" or "water", environments are recognized in 85-86% of the cases with as few as 1% of the database for training, with $ema_\alpha(i(t))$ if $\alpha$ = 1/30. The material selection by the efficient supervised classifier at $(p;\alpha)$=(69;1/30) is a bit less difficult to interpret than the latter case of the PCA classification (see Fig. 6c). It is observed that no material is of a particularly generic relevance for identifying "no flow" (previously, this class was not separable from the rest in the (PC1;PC2) projection). For the other classes, it seems Bi(OTf)$_3$-doped P3HT (a highly doped material) was particularly useful in inducing negative weights, but not in a reproductible way between the three sensing devices. A similar statement can be done for pristine P3HT (a low doping material) which induces the highest weights on the "acetone" output class. In general, the standard deviation of each weight is quite significant compared to the actual value of the weight. As a conclusion, it seems still quite difficult to infer on the material-set quality to efficiently recognize the vapors under supervised learning with a linear classifier, despite its high sensitivity to specifically recognize "water", "acetone" and "ethanol" with the ema-floating reference in the $modema_\alpha(i(t))$ information descriptor.

# Discussion

Resistance or current modulations as classical information descriptors for conductimetric electronic nose are experiment-dependent parameters that require fixing references periodically at given times during the exposures and being synchronized with the sample exposures. As such, it can hardly be considered as generic figure-of-merits in conductimetric electronic noses. Using the $ema_\alpha()$ referencing may be better suited in this sense as it does not require external parameterizing, except setting the attenuation coefficient $\alpha$. Such value can also be data-dependent to vary dynamically and be generic in absolute. In this study, the analysis, both under unsupervised and supervised learning, shows high sensitivity of the classifier with $\alpha$. Thus, the choice for data-dependent functions that adjust $\alpha$ over time, depending on the dynamic dispersion of a high-dimensional signal at a given time, is of the highest interest to define a figure-of-merit for conductimetric electronic noses. In the case of doped conducting polymers, PCA as an unsupervised classifier uses all doped-P3HT contributions to recognize the transient points with the least attenuated floating reference. It also appeared that the doping strength of the different metal triflates has a crucial role in their different contributions to the first two principal components. The pseudo-inverse supervised classifier instead performs better with a highly attenuated floating reference. However, the involvement of the different materials in the analysis is extremely sensitive to the training database. It is very surprising to observe that for a same collection of raw data generated from a same sensing array to classify the same environment classes, two different linear classifiers may exploit very different information for recognition. It is therefore with the highest care that we need to consider $modema_\alpha(i(t))$ as a figure-of-merit in a specific classification framework, in case $\alpha$ is fixed. Although it may need to be confirmed with other materials, the quality of the classification at a fixed $\alpha$ is highly dependent on the learning scheme: supervised or not supervised.

Generating relevant information, either near-sensor or in sensor, directly from each physical device/sensing node is of a major importance to lower computational complexity of a software classifier for inferring instantly.[13, 14] Furthermore, *in materio* information preprocessing is an essential key to lower this complexity by the use of materials that specifically convolute signals thanks to inner physical mechanisms, but also to lower the fabrication costs for such classifiers.[15, 16] In this sense, using $modema_\alpha(i(t))$ represents an extraordinary opportunity to revise the way we build information generators in electronics, in a neuro-inspired way, to emulate it closer to the way an actual receptor neuron conditions classification.

On this matter, information in our olfactory cortex in a human cerebrum is generated from about $6\ 10^6$ olfactory receptor neurons gathering 339 different functional G-coupled receptors, which project the information in parallel.[17, 18] Information carriers are particularly slow for olfaction (second-scale), as metabotropic receptors are in general slower than ionotropic ones, but also because of the diffusion and retention time of volatile organic compounds in the mucus.[19, 20] Very similarly to a conductimetric nose, a biological nose may use both the chemospecificity of metabotropic receptors on each cell to recognize odors, but also the kinetics of their interactions with molecules. Also, the brain itself uses information temporality in a relevant way for classification. Many neuron models attempted to conceptualize the transience of electrical response responsible for the neurons' plasticity as a passive electric circuit. Most of them associate the output response of a neuron to the charge of a serial capacitor, such as Hodgkin & Huxley's, or the Leaky-Integrate-and-Fire model derived from Lapicque's earlier works.[21] Unconventional computing technologies mimicking such plasticity embedding RC elements in their Voigt model expression,[22] can use the temporality of the information signal to classify sensed information with doped conducting polymers.[23] Inherent electrochemical processes associated to these same doped polymers can even exhibit non-ideal capacitances,[24, 25] for which the analytical expression of their admittance in the time domain shows true biological and computational relevance to tune the synaptic temporality.[26, 27] The capability to tune such non ideality in sensing devices may be a true hardware feature to sort the information *in materio* by tuning the fading-memory time-window of the sensing element, as a physical model of a biological receptor neuron.

In this study, a strong parallel may be identified between the influence of the attenuation coefficient $\alpha$ in the ema floating reference function on the quality of an environment recognition and the influence of the non-ideality factor $\alpha$ of a fractional-leaky-integrate-and-fire (FLIF) neuron.[26, 27] In particular, this study spots the fact that controlling the ability to tune the attenuation coefficient $\alpha$ may condition the *modema$_\alpha$(i(t))* feature output to use different sensitivities from the materials depending on the learning framework. It has to be stressed at this point that the analytical expression of the *modema$_\alpha$(i(t))* information feature is not comparable to the electrical output of a FLIF neuron: the same way the attenuation coefficient of the ema function is not to be identified as the fractional order of the derivation in the FLIF neuron expression. However, the fact that doped conducting polymers are both able to classify environments depending on their attenuation coefficient, and their ability to use inherent electrochemical processes to change their capacitive properties, represent a very encouraging perspective for *in materio* information classification within the same class of material, such as doped conducting polymers in a sensing hardware.

# Conclusion

The study focuses on the 'modema' information descriptor as a potential figure-of-merit for conductimetric electronic noses. By the use of several doped conducting polymers, the ema floating point has shown to be a good reference for extracting relevant information from a whole acquisition to classify volatile molecular environments. Under supervised learning, a linear classifier is able to recognize 90% of five classes, regardless of the dynamic of the data. Under unsupervised learning, a linear classifier can clusterize data by environment identity, whether an air blow is loaded with water, ethanol, acetone vapors or not. The great advantage of such feature is that it does not need any user nor environment specific post-parameterization, which makes its application quite generic: for analyzing volatile samples intermittently for product quality monitoring, or for online analysis of time varying environment patterns such as outdoor pollution. The sensitivity of the attenuation coefficient to the learning framework (either supervised or not supervised) is both an important point to optimize prior to using such feature for environment pattern classification, and an interesting property of the descriptor to investigate on, for identifying "good" material combinations for a sensing input depending on the structure of the data. Further perspectives are to be envisioned both at software and hardware levels: on optimizing the descriptor so the attenuation coefficient can self-adapt to the data organization, and to emulate such filtering *in materio* or near the sensors at the input of an electronic nose, to mimic better a biological sense in the way environmental information is sorted.

# Acknowledgements

SP and AB thank the French Research Agency (ANR) for funding the "Sensation" project (Grant # ANR-22-CE24-0001-01). The authors thank the French Nanofabrication Network RENATECH for financial support of the IEMN cleanroom.

Very respectfully to the inspiring work of Dr. Alexander Vergara on machine olfaction.

# Competing Interests

The authors declare no competing interests.

# References

1.  Liu, Z., et al., *Multi-sensor measurement and data fusion.* IEEE Instrumentation & Measurement Magazine, 2022. **25**(1): p. 28-36. https://doi.org/ 10.1109/MIM.2022.9693406.

2.  Schiavi, A., et al. *Metrology for next generation "Phygital Sensors".* in *2023 IEEE International Workshop on Metrology for Industry 4.0 & IoT (MetroInd4. 0&IoT)*. 2023. IEEE. https://doi.org/ 10.1109/MetroInd4.0IoT57462.2023.10180196.

3.  Wise, P.M., M.J. Olsson, and W.S. Cain, *Quantification of odor quality.* Chemical senses, 2000. **25**(4): p. 429-443. https://doi.org/10.1093/chemse/25.4.429.

4.  Bushdid, C., et al., *Humans can discriminate more than 1 trillion olfactory stimuli.* Science, 2014. **343**(6177): p. 1370-1372. https://doi.org/ 10.1126/science.1249168.

5.  Boujnah, A., et al., *Mildly-doped polythiophene with triflates for molecular recognition.* Synthetic Metals, 2021. **280**: p. 116890. https://doi.org/10.1016/j.synthmet.2021.116890.

6.  Boujnah, A., et al., *An electronic nose using conductometric gas sensors based on P3HT doped with triflates for gas detection using computational techniques (PCA, LDA, and kNN).* Journal of Materials Science: Materials in Electronics, 2022. **33**(36): p. 27132-27146. https://doi.org/ 10.1007/s10854-022-09376-2.

7.  Haj Ammar, W., et al., *Steady vs. Dynamic Contributions of Different Doped Conducting Polymers in the Principal Components of an Electronic Nose's Response.* Eng, 2023. **4**(4): p. 2483-2496. https://doi.org/10.3390/eng4040141.

8.  Muezzinoglu, M.K., et al., *Acceleration of chemo-sensory information processing using transient features.* Sensors and Actuators B: Chemical, 2009. **137**(2): p. 507-512. https://doi.org/10.1016/j.snb.2008.10.065.

9.  Vergara, A., et al., *Chemical gas sensor drift compensation using classifier ensembles.* Sensors and Actuators B: Chemical, 2012. **166**: p. 320-329. https://doi.org/10.1016/j.snb.2012.01.074.

10. Metsalu, T. and J. Vilo, *ClustVis: a web tool for visualizing clustering of multivariate data using Principal Component Analysis and heatmap.* Nucleic acids research, 2015. **43**(W1): p. W566-W570. https://doi.org/10.1093/nar/gkv468.

11. Pecqueur, S., et al., *Neuromorphic time-dependent pattern classification with organic electrochemical transistor arrays.* Advanced Electronic Materials, 2018. **4**(9): p. 1800166. https://doi.org/10.1002/aelm.201800166.

12. Ferchichi, K., et al., *Concentration-control in all-solution processed semiconducting polymer doping and high conductivity performances.* Synthetic Metals, 2020. **262**: p. 116352. https://doi.org/10.1016/j.synthmet.2020.116352.

13. Zhou, F. and Y. Chai, *Near-sensor and in-sensor computing.* Nature Electronics, 2020. **3**(11): p. 664-671. https://doi.org/10.1038/s41928-020-00501-9.

14.     Chai, Y. and F. Liao, *Near-sensor and In-sensor Computing*. 2022. https://doi.org/10.1007/978-3-031-11506-6.

15.     Ghazal, M., et al., *Bio-inspired adaptive sensing through electropolymerization of organic electrochemical transistors.* Advanced Electronic Materials, 2022. **8**(3): p. 2100891. https://doi.org/10.1002/aelm.202100891.

16.     Scholaert, C., et al., *Plasticity of conducting polymer dendrites to bursts of voltage spikes in phosphate buffered saline.* Neuromorphic Computing and Engineering, 2022. **2**(4): p. 044010. https://doi.org/10.1088/2634-4386/ac9b85.

17.     Moran, D.T., et al., *The fine structure of the olfactory mucosa in man.* Journal of neurocytology, 1982. **11**: p. 721-746. https://doi.org/10.1007/BF01153516.

18.     Malnic, B., P.A. Godfrey, and L.B. Buck, *The human olfactory receptor gene family.* Proceedings of the National Academy of Sciences, 2004. **101**(8): p. 2584-2589. https://doi.org/10.1073/pnas.0307882100.

19.     Ghatpande, A.S. and J. Reisert, *Olfactory receptor neuron responses coding for rapid odour sampling.* The Journal of physiology, 2011. **589**(9): p. 2261-2273. https://doi.org/10.1113/jphysiol.2010.203687.

20.     Purves, D., Augustine, G. J., Fitzpatrick, D., Hall, W. C., LaMantia, A.-S., McNamara, J. O., & Williams, S. M. (Eds.). (2004). *Neuroscience* (3rd ed.). Sinauer Associates.

21.     Gerstner, W. and W.M. Kistler, *Spiking neuron models: Single neurons, populations, plasticity*. 2002: Cambridge university press.

22.     Pecqueur, S., et al., *The non-ideal organic electrochemical transistors impedance.* Organic Electronics, 2019. **71**: p. 14-23. https://doi.org/10.1016/j.orgel.2019.05.001.

23.     Pecqueur, S., et al., *A Neural Network to Decipher Organic Electrochemical Transistors' Multivariate Responses for Cation Recognition.* Electronic Materials, 2023. **4**(2): p. 80-94. https://doi.org/10.3390/electronicmat4020007.

24.     Janzakova, K., et al., *Dendritic organic electrochemical transistors grown by electropolymerization for 3D neuromorphic engineering.* Advanced Science, 2021. **8**(24): p. 2102973. https://doi.org/10.1038/s41467-021-27274-9.

25.     Janzakova, K., et al., *Analog programing of conducting-polymer dendritic interconnections and control of their morphology.* Nature Communications, 2021. **12**(1): p. 6898. https://doi.org/10.1002/advs.202102973.

26.     Teka, W., T.M. Marinov, and F. Santamaria, *Neuronal spike timing adaptation described with a fractional leaky integrate-and-fire model.* PLoS computational biology, 2014. **10**(3): p. e1003526. https://doi.org/10.1371/journal.pcbi.1003526.

27.     Teka, W.W., R.K. Upadhyay, and A. Mondal, *Fractional-order leaky integrate-and-fire model with long-term memory and power law dynamics.* Neural Networks, 2017. **93**: p. 110-125. https://doi.org/10.1016/j.neunet.2017.05.007.

# Supporting Information for:

# A Temporal Filter to Extract Doped Conducting Polymer Information Features from an Electronic Nose

Wiem Haj Ammar[a], Aicha Boujnah[a], Antoine Baron[b], Aimen Boubaker[a], Adel Kalboussi[a], Kamal Lmimouni[b] & Sébastien Pecqueur[b]*

[a] Department of Physics, University of Monastir Tunisia.

[b] Univ. Lille, CNRS, Centrale Lille, Univ. Polytechnique Hauts-de-France, UMR 8520 - IEMN, F-59000 Lille, France.

Email: sebastien.pecqueur@*iemn*.fr

PCA scores (a), variance for the different PC (b) and their loadings (c) are organized in the different supporting figures by acquisition time for the different information descriptors such as:

| | | | Raw Data: | |
|---|---|---|---|---|
| | | | $X(t)=i(t)$ | $X(t)=R(t)$ |
| **Descriptor :** | $modema_\alpha(X(t))$ | $\alpha = 1/2$ | Figure S1 | Figure S16 |
| | | $\alpha = 1/3$ | Figure S2 | Figure S17 |
| | | $\alpha = 1/4$ | Figure S3 | Figure S18 |
| | | $\alpha = 1/5$ | Figure S4 | Figure S19 |
| | | $\alpha = 1/6$ | Figure S5 | Figure S20 |
| | | $\alpha = 1/7$ | Figure S6 | Figure S21 |
| | | $\alpha = 1/8$ | Figure S7 | Figure S22 |
| | | $\alpha = 1/9$ | Figure S8 | Figure S23 |
| | | $\alpha = 1/10$ | Figure S9 | Figure S24 |
| | | $\alpha = 1/30$ | Figure S10 | Figure S25 |
| | | $\alpha = 1/60$ | Figure S11 | Figure S26 |
| | | $\alpha = 1/100$ | Figure S12 | Figure S27 |
| | | $\alpha = 1/300$ | Figure S13 | Figure S28 |
| | | $\alpha = 1/600$ | Figure S14 | Figure S29 |
| | | $\alpha = 1/1000$ | Figure S15 | Figure S30 |
| | $ema_\alpha(X(t))$ | $\alpha = 1/2$ | Figure S31 | Figure S33 |
| | | $\alpha = 1/1000$ | Figure S32 | Figure S34 |

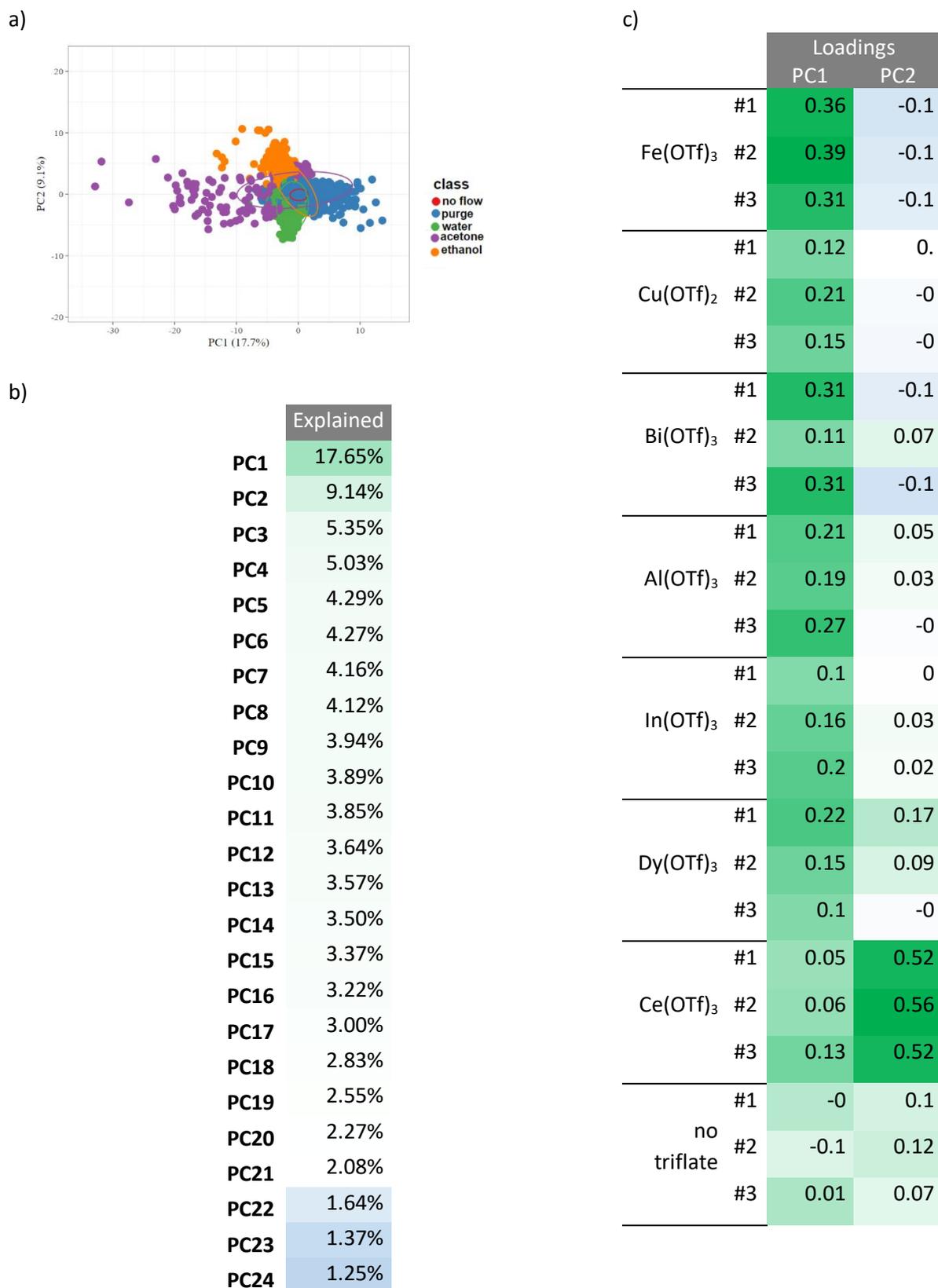

**Figure S1. PCA on "modema$_\alpha$(i(t))" for $\alpha$ = 1/2 | a,** PCA scores with 95% confidence ellipsoids. **b,** Individual variance for the different PC. **c,** PCA loadings of the different sensing elements' response for PC1 and PC2.

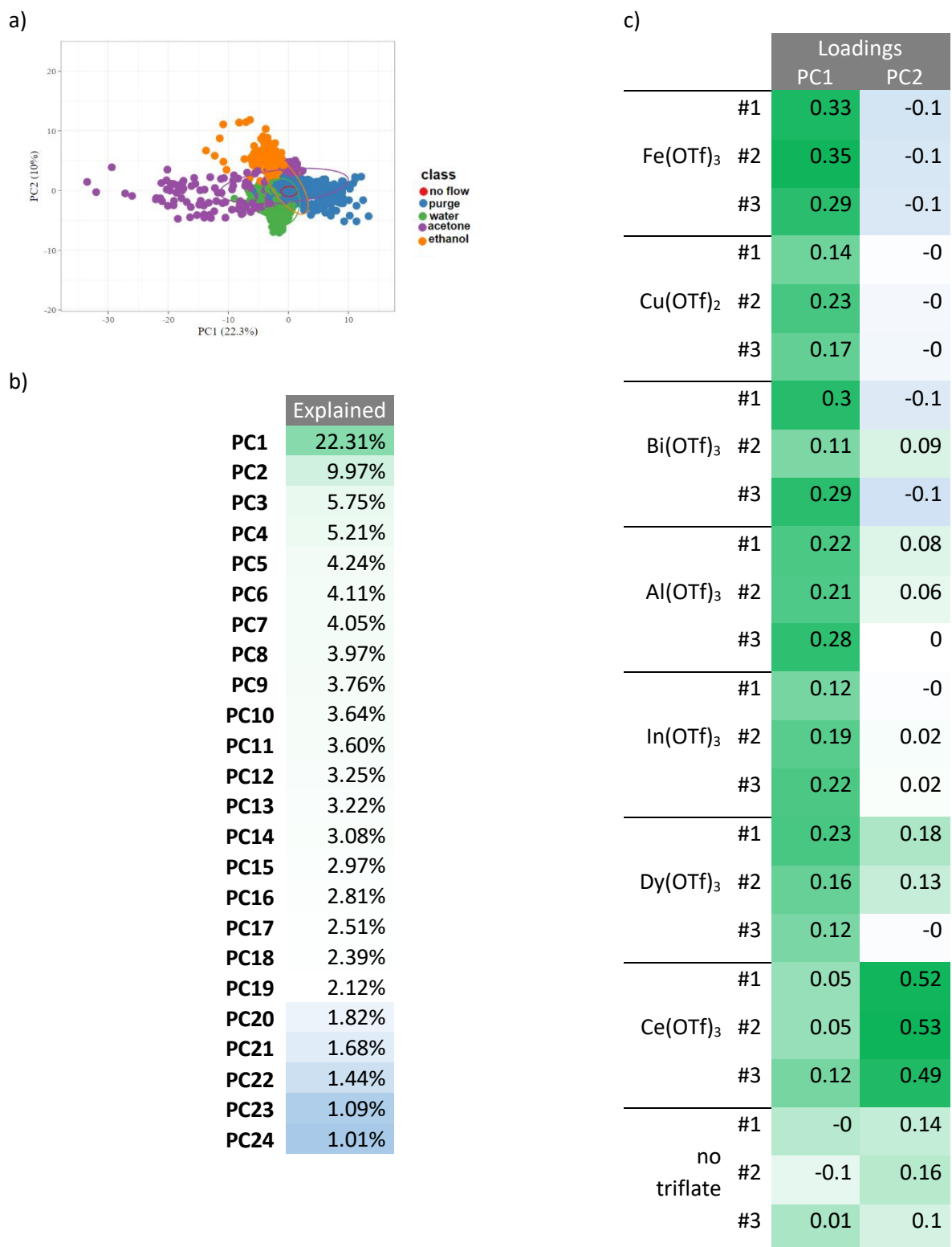

**Figure S2. PCA on "modema$_\alpha$(i(t))" for $\alpha$ = 1/3 | a**. PCA scores with 95% confidence ellipsoids. **b.** Individual variance for the different PC. **c,** PCA loadings of the different sensing elements' response for PC1 and PC2.

a)

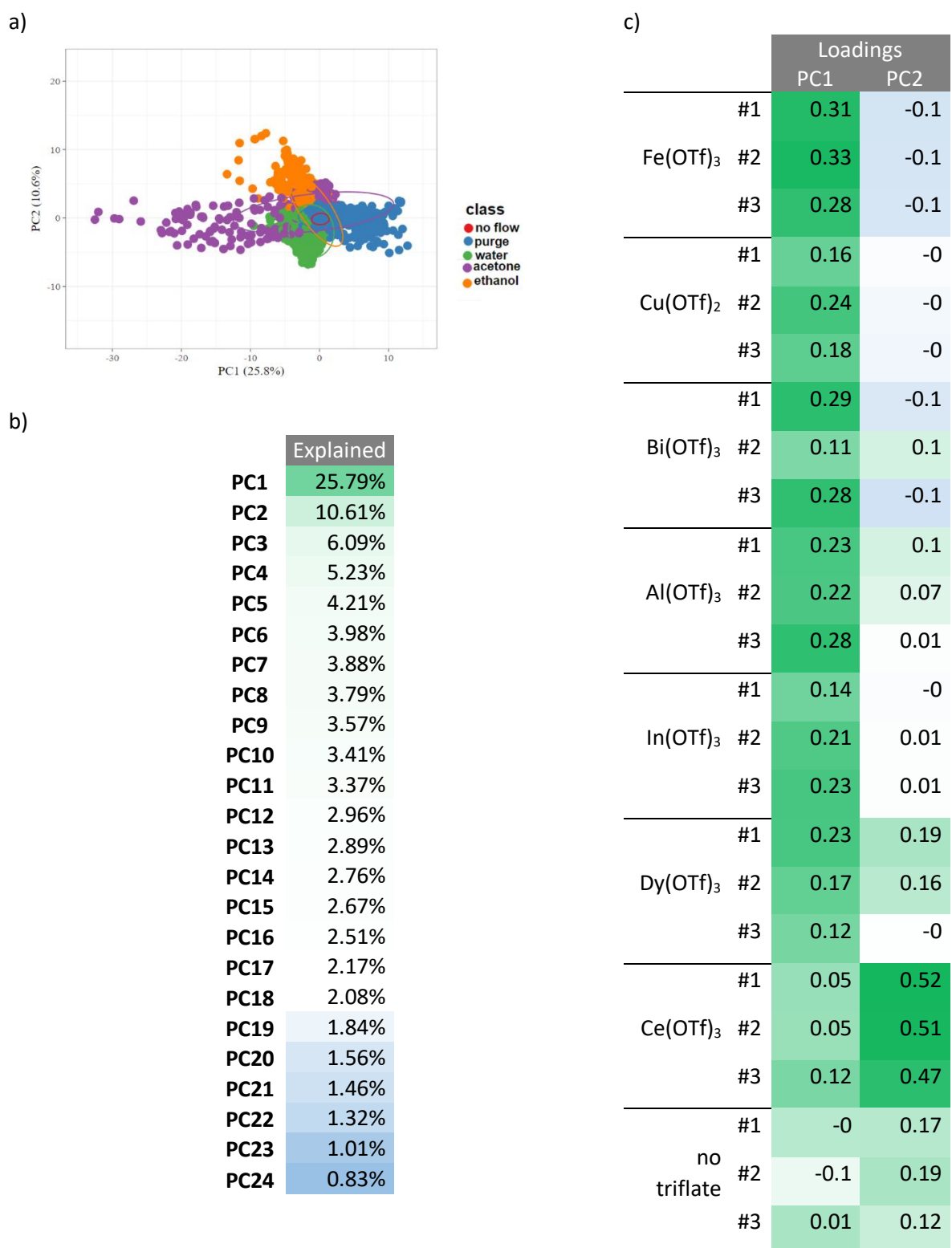

b)

| | Explained |
|---|---|
| **PC1** | 25.79% |
| **PC2** | 10.61% |
| **PC3** | 6.09% |
| **PC4** | 5.23% |
| **PC5** | 4.21% |
| **PC6** | 3.98% |
| **PC7** | 3.88% |
| **PC8** | 3.79% |
| **PC9** | 3.57% |
| **PC10** | 3.41% |
| **PC11** | 3.37% |
| **PC12** | 2.96% |
| **PC13** | 2.89% |
| **PC14** | 2.76% |
| **PC15** | 2.67% |
| **PC16** | 2.51% |
| **PC17** | 2.17% |
| **PC18** | 2.08% |
| **PC19** | 1.84% |
| **PC20** | 1.56% |
| **PC21** | 1.46% |
| **PC22** | 1.32% |
| **PC23** | 1.01% |
| **PC24** | 0.83% |

c)

| | | Loadings | |
|---|---|---|---|
| | | PC1 | PC2 |
| $Fe(OTf)_3$ | #1 | 0.31 | -0.1 |
| | #2 | 0.33 | -0.1 |
| | #3 | 0.28 | -0.1 |
| $Cu(OTf)_2$ | #1 | 0.16 | -0 |
| | #2 | 0.24 | -0 |
| | #3 | 0.18 | -0 |
| $Bi(OTf)_3$ | #1 | 0.29 | -0.1 |
| | #2 | 0.11 | 0.1 |
| | #3 | 0.28 | -0.1 |
| $Al(OTf)_3$ | #1 | 0.23 | 0.1 |
| | #2 | 0.22 | 0.07 |
| | #3 | 0.28 | 0.01 |
| $In(OTf)_3$ | #1 | 0.14 | -0 |
| | #2 | 0.21 | 0.01 |
| | #3 | 0.23 | 0.01 |
| $Dy(OTf)_3$ | #1 | 0.23 | 0.19 |
| | #2 | 0.17 | 0.16 |
| | #3 | 0.12 | -0 |
| $Ce(OTf)_3$ | #1 | 0.05 | 0.52 |
| | #2 | 0.05 | 0.51 |
| | #3 | 0.12 | 0.47 |
| no triflate | #1 | -0 | 0.17 |
| | #2 | -0.1 | 0.19 |
| | #3 | 0.01 | 0.12 |

**Figure S3. PCA on "modema$_\alpha$(i(t))" for α = 1/4 | a,** PCA scores with 95% confidence ellipsoids. **b,** Individual variance for the different PC. **c,** PCA loadings of the different sensing elements' response for PC1 and PC2.

.

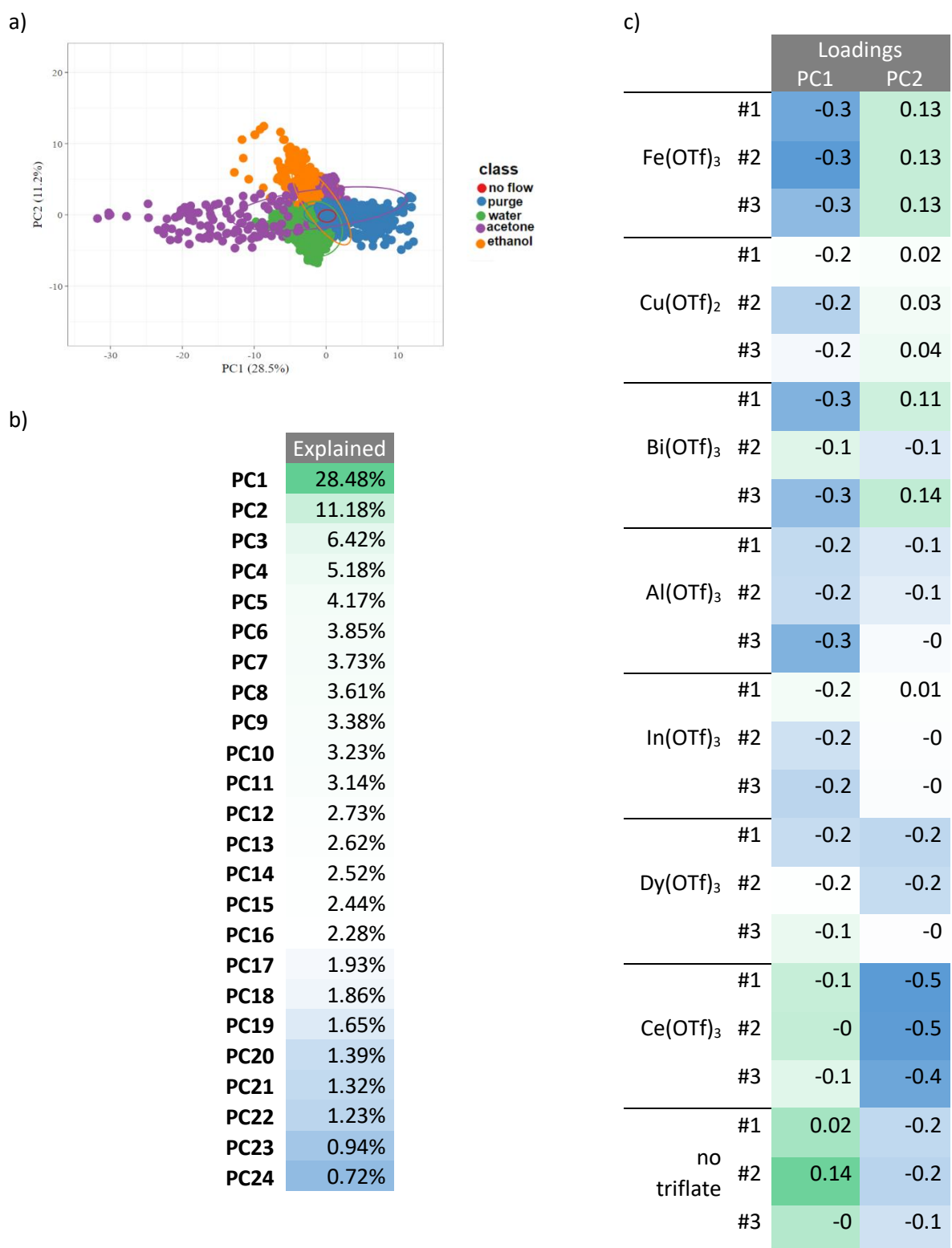

a)

b)

| | Explained |
|---|---|
| PC1 | 28.48% |
| PC2 | 11.18% |
| PC3 | 6.42% |
| PC4 | 5.18% |
| PC5 | 4.17% |
| PC6 | 3.85% |
| PC7 | 3.73% |
| PC8 | 3.61% |
| PC9 | 3.38% |
| PC10 | 3.23% |
| PC11 | 3.14% |
| PC12 | 2.73% |
| PC13 | 2.62% |
| PC14 | 2.52% |
| PC15 | 2.44% |
| PC16 | 2.28% |
| PC17 | 1.93% |
| PC18 | 1.86% |
| PC19 | 1.65% |
| PC20 | 1.39% |
| PC21 | 1.32% |
| PC22 | 1.23% |
| PC23 | 0.94% |
| PC24 | 0.72% |

c)

| | | Loadings | |
|---|---|---|---|
| | | PC1 | PC2 |
| $Fe(OTf)_3$ | #1 | -0.3 | 0.13 |
| | #2 | -0.3 | 0.13 |
| | #3 | -0.3 | 0.13 |
| $Cu(OTf)_2$ | #1 | -0.2 | 0.02 |
| | #2 | -0.2 | 0.03 |
| | #3 | -0.2 | 0.04 |
| $Bi(OTf)_3$ | #1 | -0.3 | 0.11 |
| | #2 | -0.1 | -0.1 |
| | #3 | -0.3 | 0.14 |
| $Al(OTf)_3$ | #1 | -0.2 | -0.1 |
| | #2 | -0.2 | -0.1 |
| | #3 | -0.3 | -0 |
| $In(OTf)_3$ | #1 | -0.2 | 0.01 |
| | #2 | -0.2 | -0 |
| | #3 | -0.2 | -0 |
| $Dy(OTf)_3$ | #1 | -0.2 | -0.2 |
| | #2 | -0.2 | -0.2 |
| | #3 | -0.1 | -0 |
| $Ce(OTf)_3$ | #1 | -0.1 | -0.5 |
| | #2 | -0 | -0.5 |
| | #3 | -0.1 | -0.4 |
| no triflate | #1 | 0.02 | -0.2 |
| | #2 | 0.14 | -0.2 |
| | #3 | -0 | -0.1 |

**Figure S4. PCA on "modema$_\alpha$(i(t))" for $\alpha$ = 1/5 | a,** PCA scores with 95% confidence ellipsoids. **b,** Individual variance for the different PC. **c,** PCA loadings of the different sensing elements' response for PC1 and PC2.

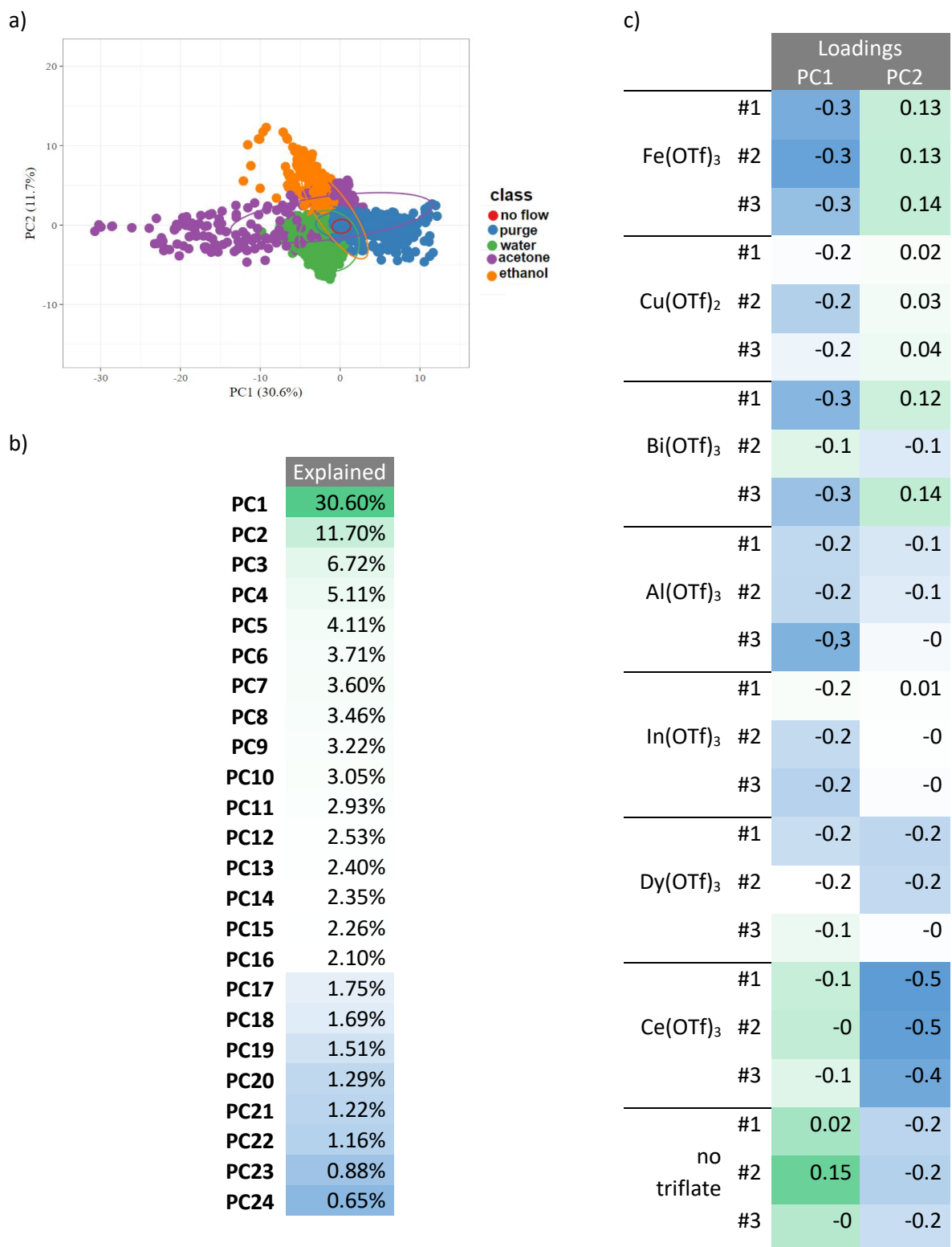

**Figure S5. PCA on "modema$_\alpha$(i(t))" for α = 1/6 | a.** PCA scores with 95% confidence ellipsoids. **b.** Individual variance for the different PC. **c.** PCA loadings of the different sensing elements' response for PC1 and PC2.

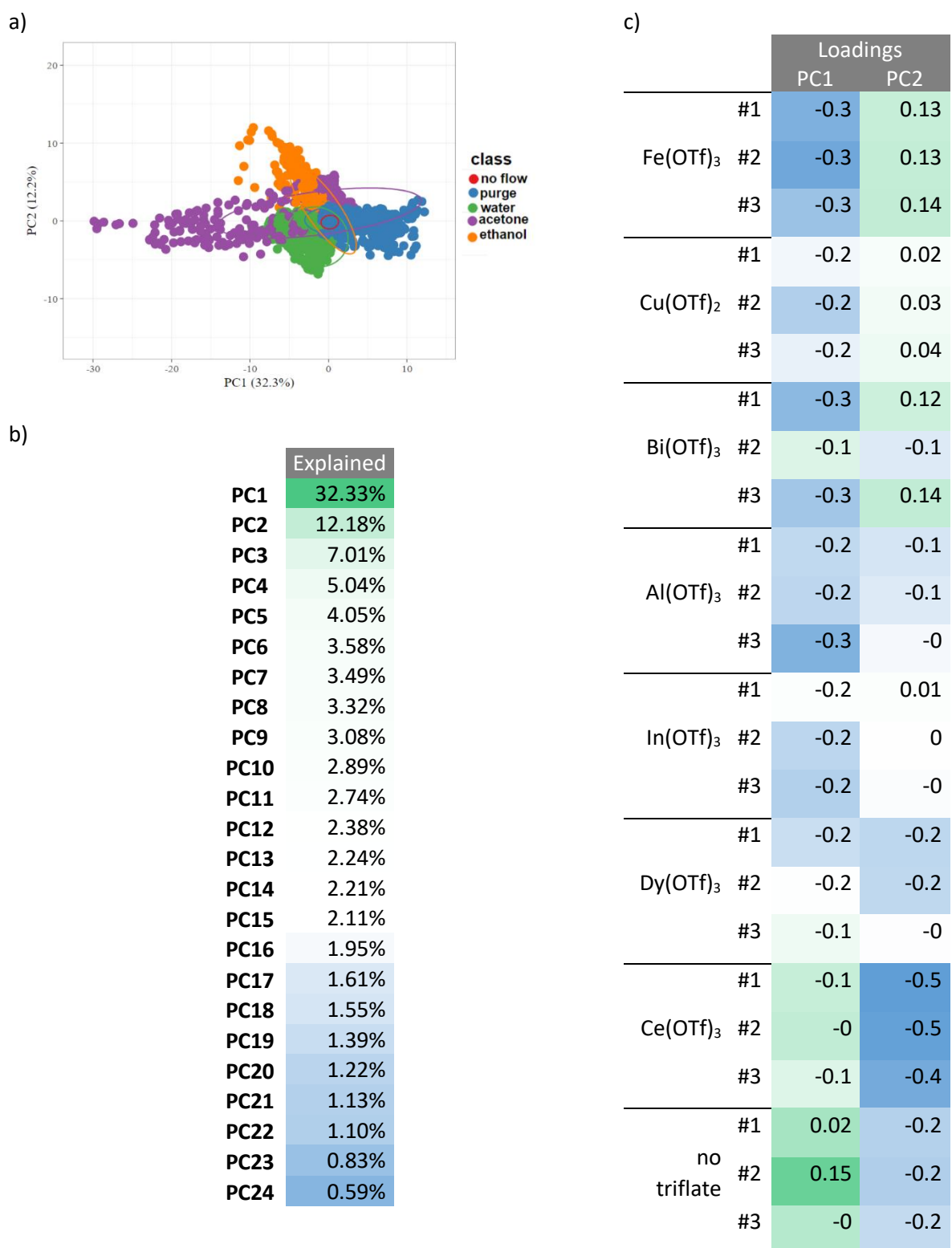

a)

b)

| | Explained |
|---|---|
| **PC1** | 32.33% |
| **PC2** | 12.18% |
| **PC3** | 7.01% |
| **PC4** | 5.04% |
| **PC5** | 4.05% |
| **PC6** | 3.58% |
| **PC7** | 3.49% |
| **PC8** | 3.32% |
| **PC9** | 3.08% |
| **PC10** | 2.89% |
| **PC11** | 2.74% |
| **PC12** | 2.38% |
| **PC13** | 2.24% |
| **PC14** | 2.21% |
| **PC15** | 2.11% |
| **PC16** | 1.95% |
| **PC17** | 1.61% |
| **PC18** | 1.55% |
| **PC19** | 1.39% |
| **PC20** | 1.22% |
| **PC21** | 1.13% |
| **PC22** | 1.10% |
| **PC23** | 0.83% |
| **PC24** | 0.59% |

c)

| | | Loadings | |
|---|---|---|---|
| | | PC1 | PC2 |
| $Fe(OTf)_3$ | #1 | -0.3 | 0.13 |
| | #2 | -0.3 | 0.13 |
| | #3 | -0.3 | 0.14 |
| $Cu(OTf)_2$ | #1 | -0.2 | 0.02 |
| | #2 | -0.2 | 0.03 |
| | #3 | -0.2 | 0.04 |
| $Bi(OTf)_3$ | #1 | -0.3 | 0.12 |
| | #2 | -0.1 | -0.1 |
| | #3 | -0.3 | 0.14 |
| $Al(OTf)_3$ | #1 | -0.2 | -0.1 |
| | #2 | -0.2 | -0.1 |
| | #3 | -0.3 | -0 |
| $In(OTf)_3$ | #1 | -0.2 | 0.01 |
| | #2 | -0.2 | 0 |
| | #3 | -0.2 | -0 |
| $Dy(OTf)_3$ | #1 | -0.2 | -0.2 |
| | #2 | -0.2 | -0.2 |
| | #3 | -0.1 | -0 |
| $Ce(OTf)_3$ | #1 | -0.1 | -0.5 |
| | #2 | -0 | -0.5 |
| | #3 | -0.1 | -0.4 |
| no triflate | #1 | 0.02 | -0.2 |
| | #2 | 0.15 | -0.2 |
| | #3 | -0 | -0.2 |

**Figure S6. PCA on "modema$_\alpha$(i(t))" for α = 1/7 | a.** PCA scores with 95% confidence ellipsoids. **b.** Individual variance for the different PC. **c.** PCA loadings of the different sensing elements' response for PC1 and PC2.

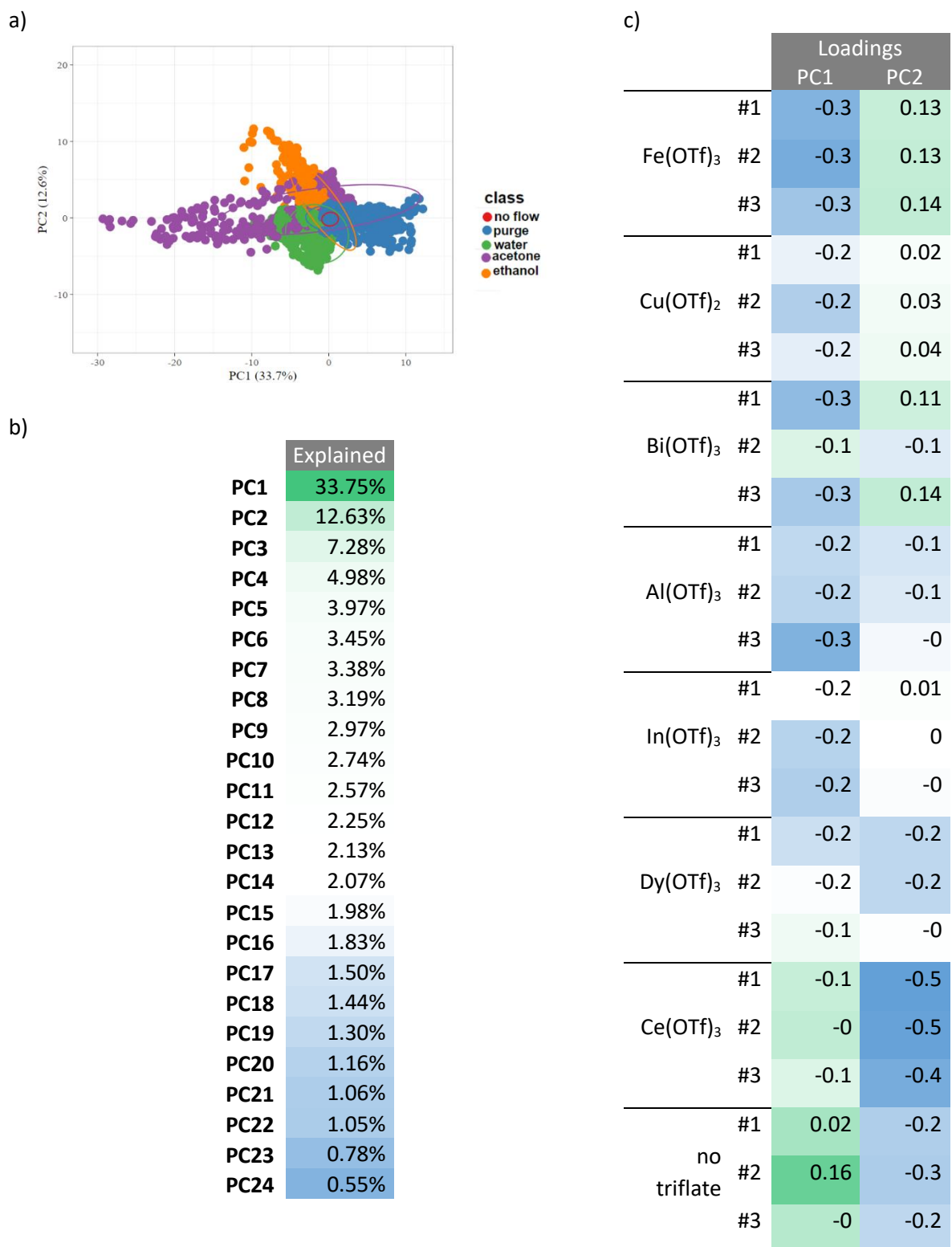

a)

b)

| | Explained |
|---|---|
| **PC1** | 33.75% |
| **PC2** | 12.63% |
| **PC3** | 7.28% |
| **PC4** | 4.98% |
| **PC5** | 3.97% |
| **PC6** | 3.45% |
| **PC7** | 3.38% |
| **PC8** | 3.19% |
| **PC9** | 2.97% |
| **PC10** | 2.74% |
| **PC11** | 2.57% |
| **PC12** | 2.25% |
| **PC13** | 2.13% |
| **PC14** | 2.07% |
| **PC15** | 1.98% |
| **PC16** | 1.83% |
| **PC17** | 1.50% |
| **PC18** | 1.44% |
| **PC19** | 1.30% |
| **PC20** | 1.16% |
| **PC21** | 1.06% |
| **PC22** | 1.05% |
| **PC23** | 0.78% |
| **PC24** | 0.55% |

c)

| | | Loadings | |
|---|---|---|---|
| | | PC1 | PC2 |
| Fe(OTf)$_3$ | #1 | -0.3 | 0.13 |
| | #2 | -0.3 | 0.13 |
| | #3 | -0.3 | 0.14 |
| Cu(OTf)$_2$ | #1 | -0.2 | 0.02 |
| | #2 | -0.2 | 0.03 |
| | #3 | -0.2 | 0.04 |
| Bi(OTf)$_3$ | #1 | -0.3 | 0.11 |
| | #2 | -0.1 | -0.1 |
| | #3 | -0.3 | 0.14 |
| Al(OTf)$_3$ | #1 | -0.2 | -0.1 |
| | #2 | -0.2 | -0.1 |
| | #3 | -0.3 | -0 |
| In(OTf)$_3$ | #1 | -0.2 | 0.01 |
| | #2 | -0.2 | 0 |
| | #3 | -0.2 | -0 |
| Dy(OTf)$_3$ | #1 | -0.2 | -0.2 |
| | #2 | -0.2 | -0.2 |
| | #3 | -0.1 | -0 |
| Ce(OTf)$_3$ | #1 | -0.1 | -0.5 |
| | #2 | -0 | -0.5 |
| | #3 | -0.1 | -0.4 |
| no triflate | #1 | 0.02 | -0.2 |
| | #2 | 0.16 | -0.3 |
| | #3 | -0 | -0.2 |

**Figure S7. PCA on "modema$_\alpha$(i(t))" for α = 1/8 | a**. PCA scores with 95% confidence ellipsoids**. b.** Individual variance for the different PC**. c**. PCA loadings of the different sensing elements' response for PC1 and PC2.

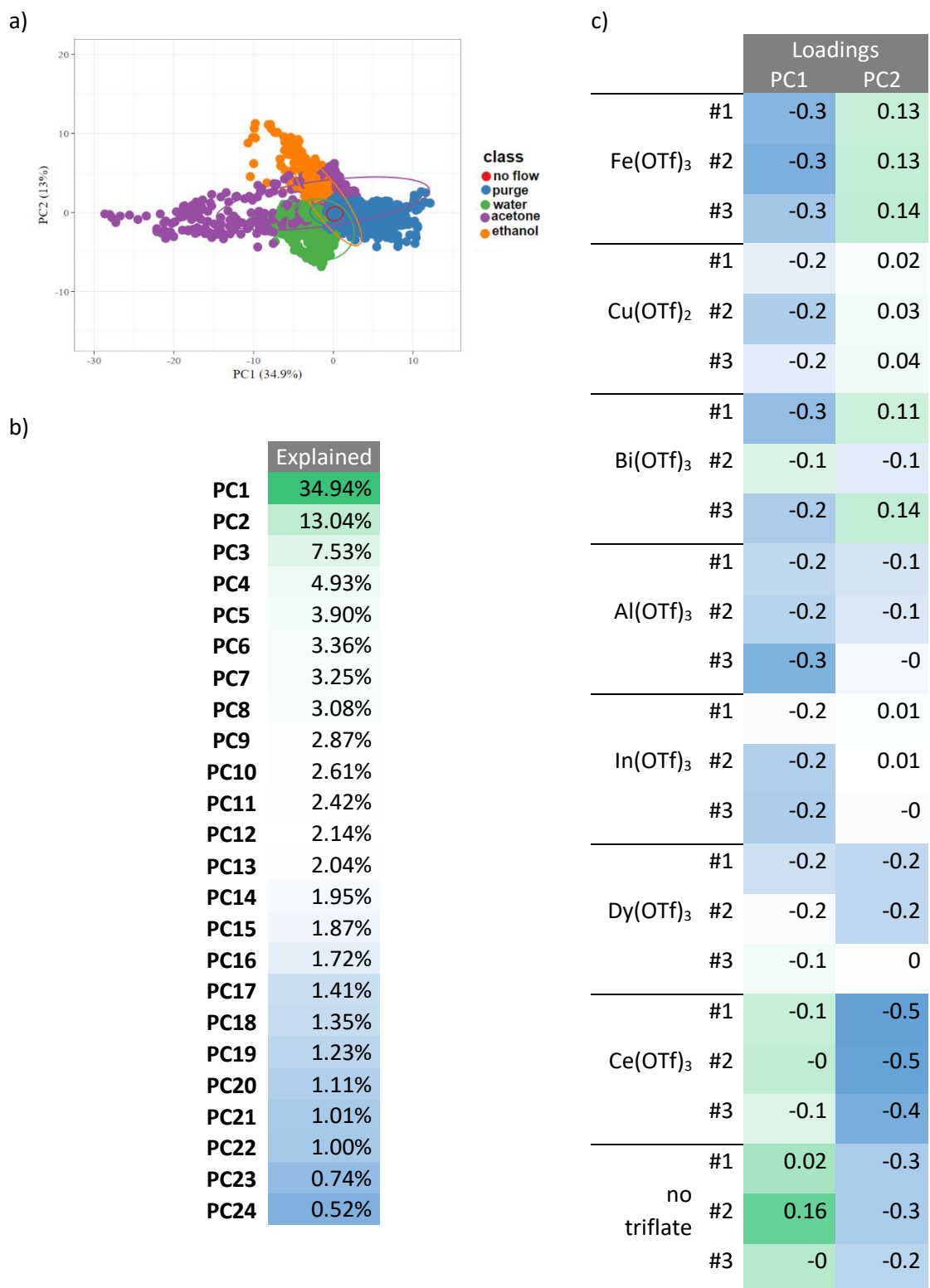

a)

b)

| | Explained |
|---|---|
| **PC1** | 34.94% |
| **PC2** | 13.04% |
| **PC3** | 7.53% |
| **PC4** | 4.93% |
| **PC5** | 3.90% |
| **PC6** | 3.36% |
| **PC7** | 3.25% |
| **PC8** | 3.08% |
| **PC9** | 2.87% |
| **PC10** | 2.61% |
| **PC11** | 2.42% |
| **PC12** | 2.14% |
| **PC13** | 2.04% |
| **PC14** | 1.95% |
| **PC15** | 1.87% |
| **PC16** | 1.72% |
| **PC17** | 1.41% |
| **PC18** | 1.35% |
| **PC19** | 1.23% |
| **PC20** | 1.11% |
| **PC21** | 1.01% |
| **PC22** | 1.00% |
| **PC23** | 0.74% |
| **PC24** | 0.52% |

c)

| | | Loadings | |
|---|---|---|---|
| | | PC1 | PC2 |
| Fe(OTf)$_3$ | #1 | -0.3 | 0.13 |
| | #2 | -0.3 | 0.13 |
| | #3 | -0.3 | 0.14 |
| Cu(OTf)$_2$ | #1 | -0.2 | 0.02 |
| | #2 | -0.2 | 0.03 |
| | #3 | -0.2 | 0.04 |
| Bi(OTf)$_3$ | #1 | -0.3 | 0.11 |
| | #2 | -0.1 | -0.1 |
| | #3 | -0.2 | 0.14 |
| Al(OTf)$_3$ | #1 | -0.2 | -0.1 |
| | #2 | -0.2 | -0.1 |
| | #3 | -0.3 | -0 |
| In(OTf)$_3$ | #1 | -0.2 | 0.01 |
| | #2 | -0.2 | 0.01 |
| | #3 | -0.2 | -0 |
| Dy(OTf)$_3$ | #1 | -0.2 | -0.2 |
| | #2 | -0.2 | -0.2 |
| | #3 | -0.1 | 0 |
| Ce(OTf)$_3$ | #1 | -0.1 | -0.5 |
| | #2 | -0 | -0.5 |
| | #3 | -0.1 | -0.4 |
| no triflate | #1 | 0.02 | -0.3 |
| | #2 | 0.16 | -0.3 |
| | #3 | -0 | -0.2 |

**Figure S8. PCA on "modema$_\alpha$(i(t))" for α = 1/9 | a.** PCA scores with 95% confidence ellipsoids. **b.** Individual variance for the different PC. **c,** PCA loadings of the different sensing elements' response for PC1 and PC2.

.

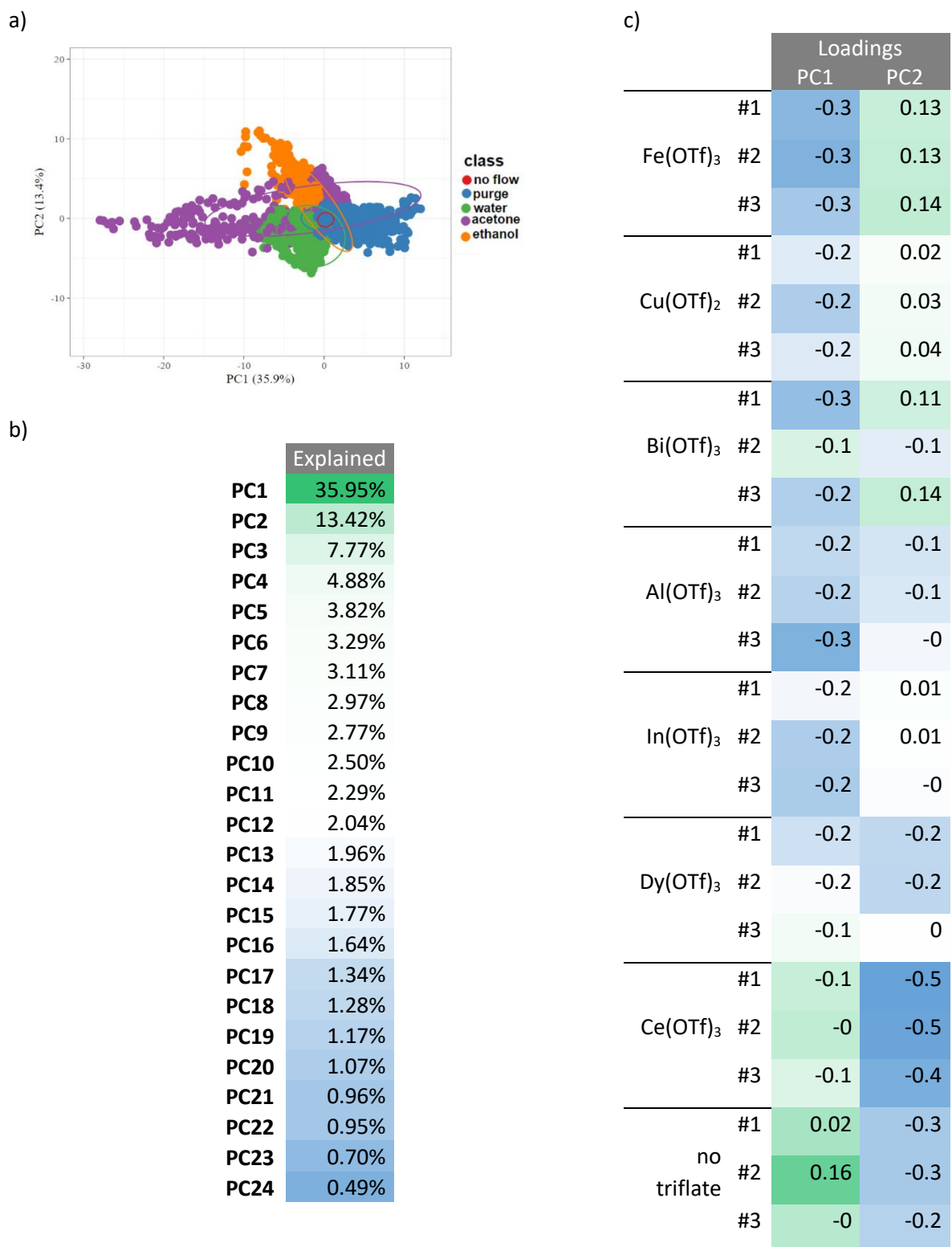

a)

b)

c)

| | | Loadings | |
|---|---|---|---|
| | | PC1 | PC2 |
| Fe(OTf)$_3$ | #1 | -0.3 | 0.13 |
| | #2 | -0.3 | 0.13 |
| | #3 | -0.3 | 0.14 |
| Cu(OTf)$_2$ | #1 | -0.2 | 0.02 |
| | #2 | -0.2 | 0.03 |
| | #3 | -0.2 | 0.04 |
| Bi(OTf)$_3$ | #1 | -0.3 | 0.11 |
| | #2 | -0.1 | -0.1 |
| | #3 | -0.2 | 0.14 |
| Al(OTf)$_3$ | #1 | -0.2 | -0.1 |
| | #2 | -0.2 | -0.1 |
| | #3 | -0.3 | -0 |
| In(OTf)$_3$ | #1 | -0.2 | 0.01 |
| | #2 | -0.2 | 0.01 |
| | #3 | -0.2 | -0 |
| Dy(OTf)$_3$ | #1 | -0.2 | -0.2 |
| | #2 | -0.2 | -0.2 |
| | #3 | -0.1 | 0 |
| Ce(OTf)$_3$ | #1 | -0.1 | -0.5 |
| | #2 | -0 | -0.5 |
| | #3 | -0.1 | -0.4 |
| no triflate | #1 | 0.02 | -0.3 |
| | #2 | 0.16 | -0.3 |
| | #3 | -0 | -0.2 |

The PC variance table (b):

| | Explained |
|---|---|
| PC1 | 35.95% |
| PC2 | 13.42% |
| PC3 | 7.77% |
| PC4 | 4.88% |
| PC5 | 3.82% |
| PC6 | 3.29% |
| PC7 | 3.11% |
| PC8 | 2.97% |
| PC9 | 2.77% |
| PC10 | 2.50% |
| PC11 | 2.29% |
| PC12 | 2.04% |
| PC13 | 1.96% |
| PC14 | 1.85% |
| PC15 | 1.77% |
| PC16 | 1.64% |
| PC17 | 1.34% |
| PC18 | 1.28% |
| PC19 | 1.17% |
| PC20 | 1.07% |
| PC21 | 0.96% |
| PC22 | 0.95% |
| PC23 | 0.70% |
| PC24 | 0.49% |

**Figure S9. PCA on "modema$_\alpha$(i(t))" for α = 1/10 | a**. PCA scores with 95% confidence ellipsoids. **b,** Individual variance for the different PC. **c,** PCA loadings of the different sensing elements' response for PC1 and PC2.

.

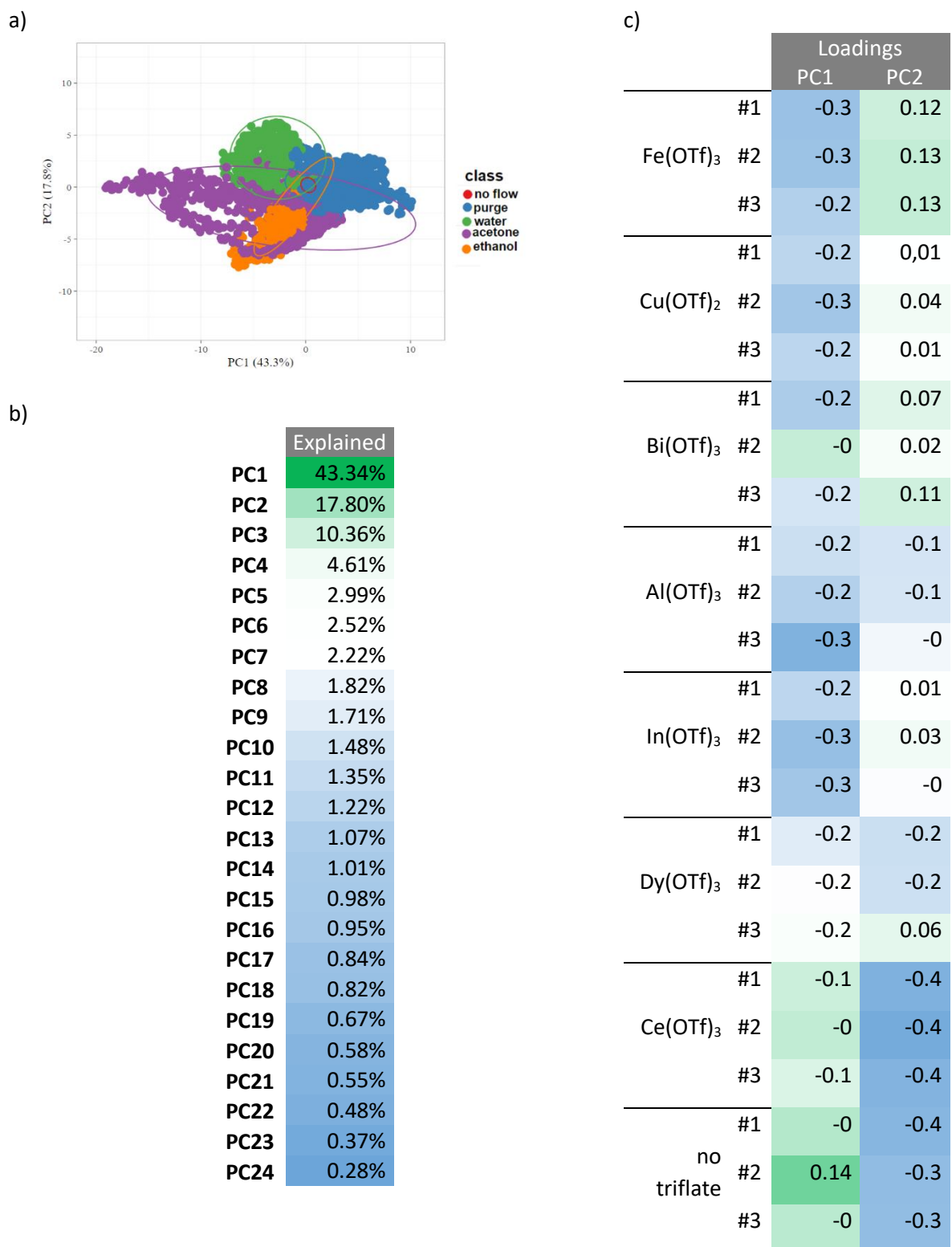

a)

b)

c)

| | | Loadings | |
|---|---|---|---|
| | | PC1 | PC2 |
| Fe(OTf)$_3$ | #1 | -0.3 | 0.12 |
| | #2 | -0.3 | 0.13 |
| | #3 | -0.2 | 0.13 |
| Cu(OTf)$_2$ | #1 | -0.2 | 0,01 |
| | #2 | -0.3 | 0.04 |
| | #3 | -0.2 | 0.01 |
| Bi(OTf)$_3$ | #1 | -0.2 | 0.07 |
| | #2 | -0 | 0.02 |
| | #3 | -0.2 | 0.11 |
| Al(OTf)$_3$ | #1 | -0.2 | -0.1 |
| | #2 | -0.2 | -0.1 |
| | #3 | -0.3 | -0 |
| In(OTf)$_3$ | #1 | -0.2 | 0.01 |
| | #2 | -0.3 | 0.03 |
| | #3 | -0.3 | -0 |
| Dy(OTf)$_3$ | #1 | -0.2 | -0.2 |
| | #2 | -0.2 | -0.2 |
| | #3 | -0.2 | 0.06 |
| Ce(OTf)$_3$ | #1 | -0.1 | -0.4 |
| | #2 | -0 | -0.4 |
| | #3 | -0.1 | -0.4 |
| no triflate | #1 | -0 | -0.4 |
| | #2 | 0.14 | -0.3 |
| | #3 | -0 | -0.3 |

| | Explained |
|---|---|
| PC1 | 43.34% |
| PC2 | 17.80% |
| PC3 | 10.36% |
| PC4 | 4.61% |
| PC5 | 2.99% |
| PC6 | 2.52% |
| PC7 | 2.22% |
| PC8 | 1.82% |
| PC9 | 1.71% |
| PC10 | 1.48% |
| PC11 | 1.35% |
| PC12 | 1.22% |
| PC13 | 1.07% |
| PC14 | 1.01% |
| PC15 | 0.98% |
| PC16 | 0.95% |
| PC17 | 0.84% |
| PC18 | 0.82% |
| PC19 | 0.67% |
| PC20 | 0.58% |
| PC21 | 0.55% |
| PC22 | 0.48% |
| PC23 | 0.37% |
| PC24 | 0.28% |

**Figure S10. PCA on "modema$_\alpha$(i(t))" for α = 1/30 | a,** PCA scores with 95% confidence ellipsoids. **b,** Individual variance for the different PC. **c,** PCA loadings of the different sensing elements' response for PC1 and PC2.

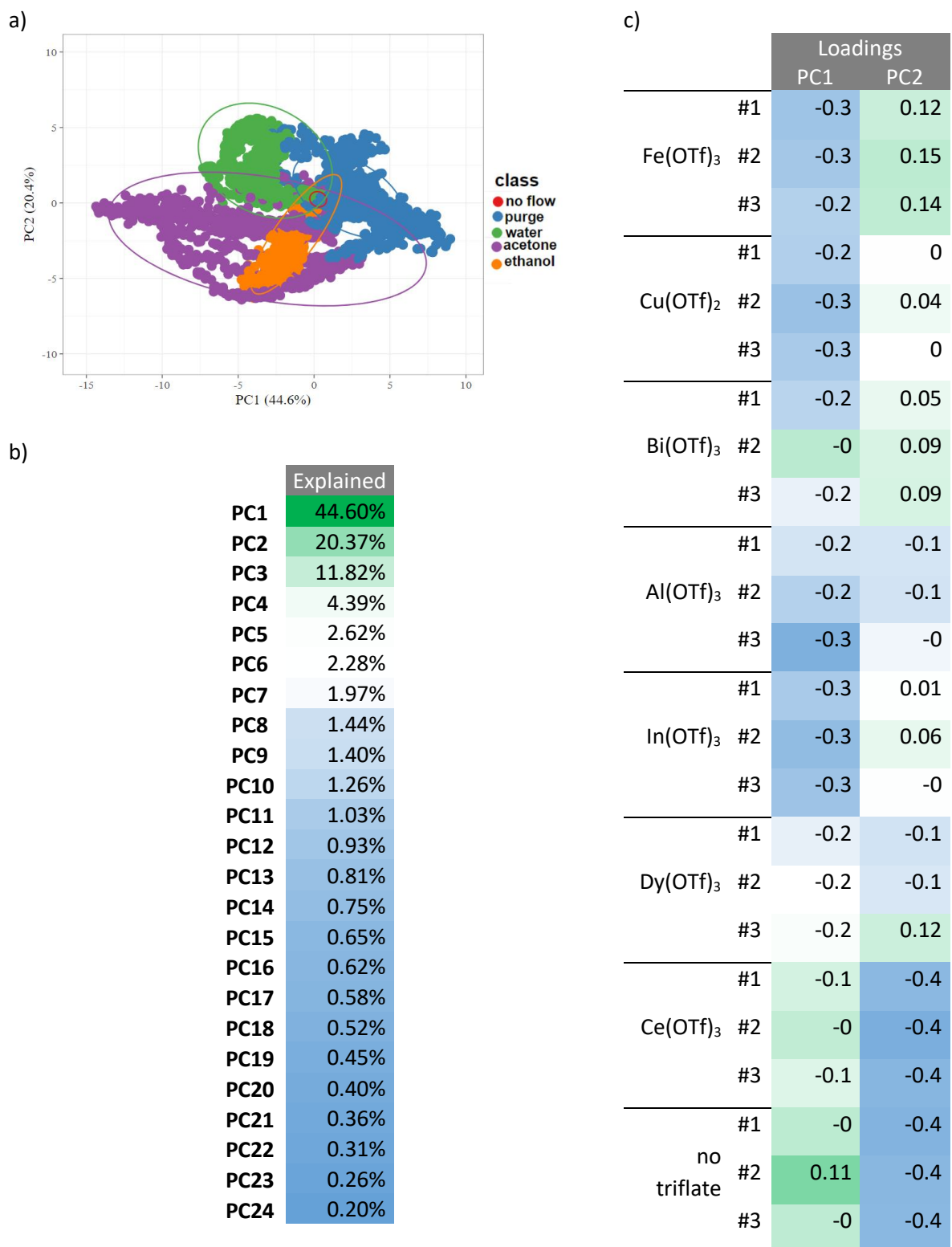

**Figure S11. PCA on "modema$_\alpha$(i(t))" for $\alpha$ = 1/60 | a,** PCA scores with 95% confidence ellipsoids. **b,** Individual variance for the different PC. **c,** PCA loadings of the different sensing elements' response for PC1 and PC2.

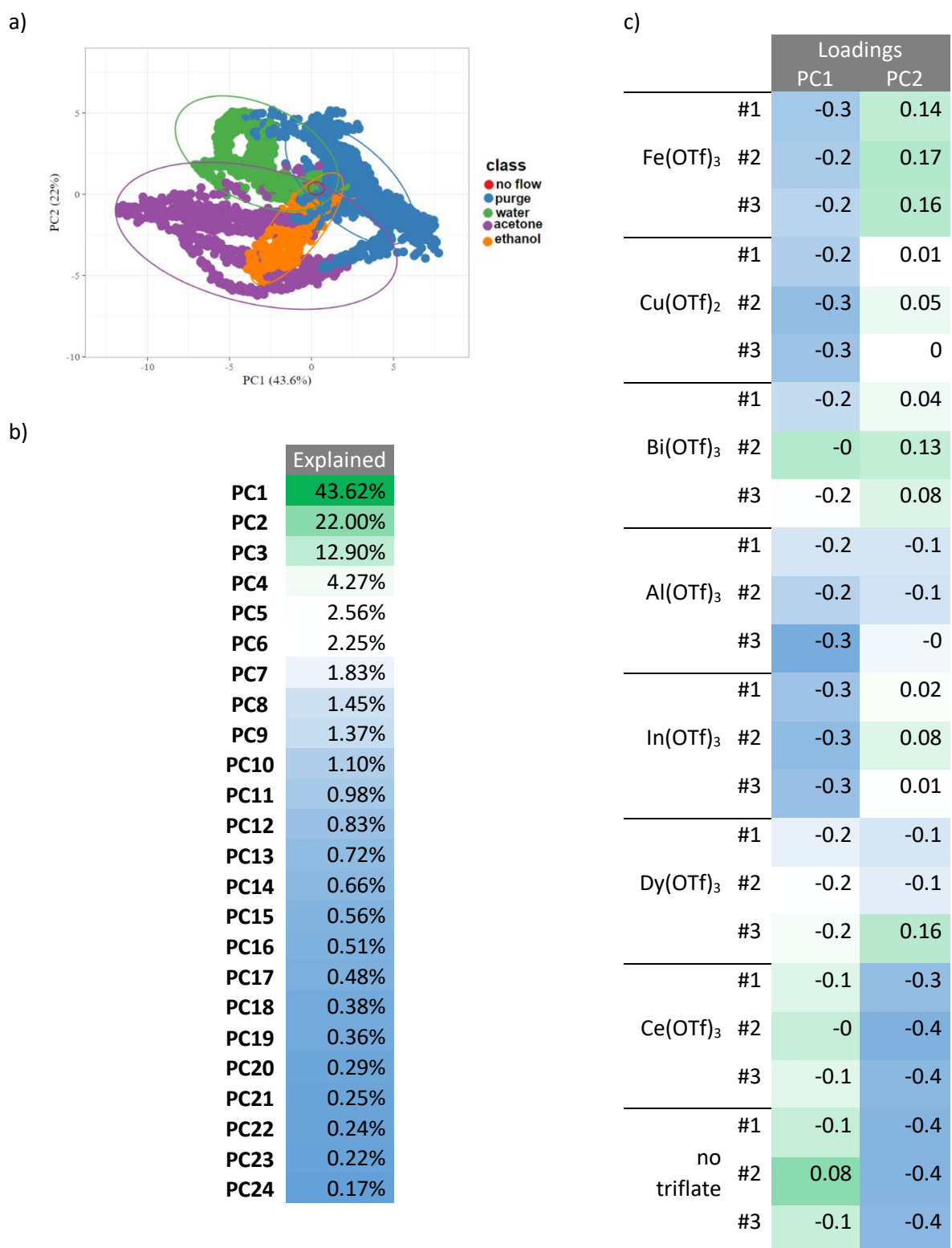

**Figure S12. PCA on "modema$_\alpha$(i(t))" for α = 1/100 | a,** PCA scores with 95% confidence ellipsoids. **b,** Individual variance for the different PC. **c,** PCA loadings of the different sensing elements' response for PC1 and PC2.

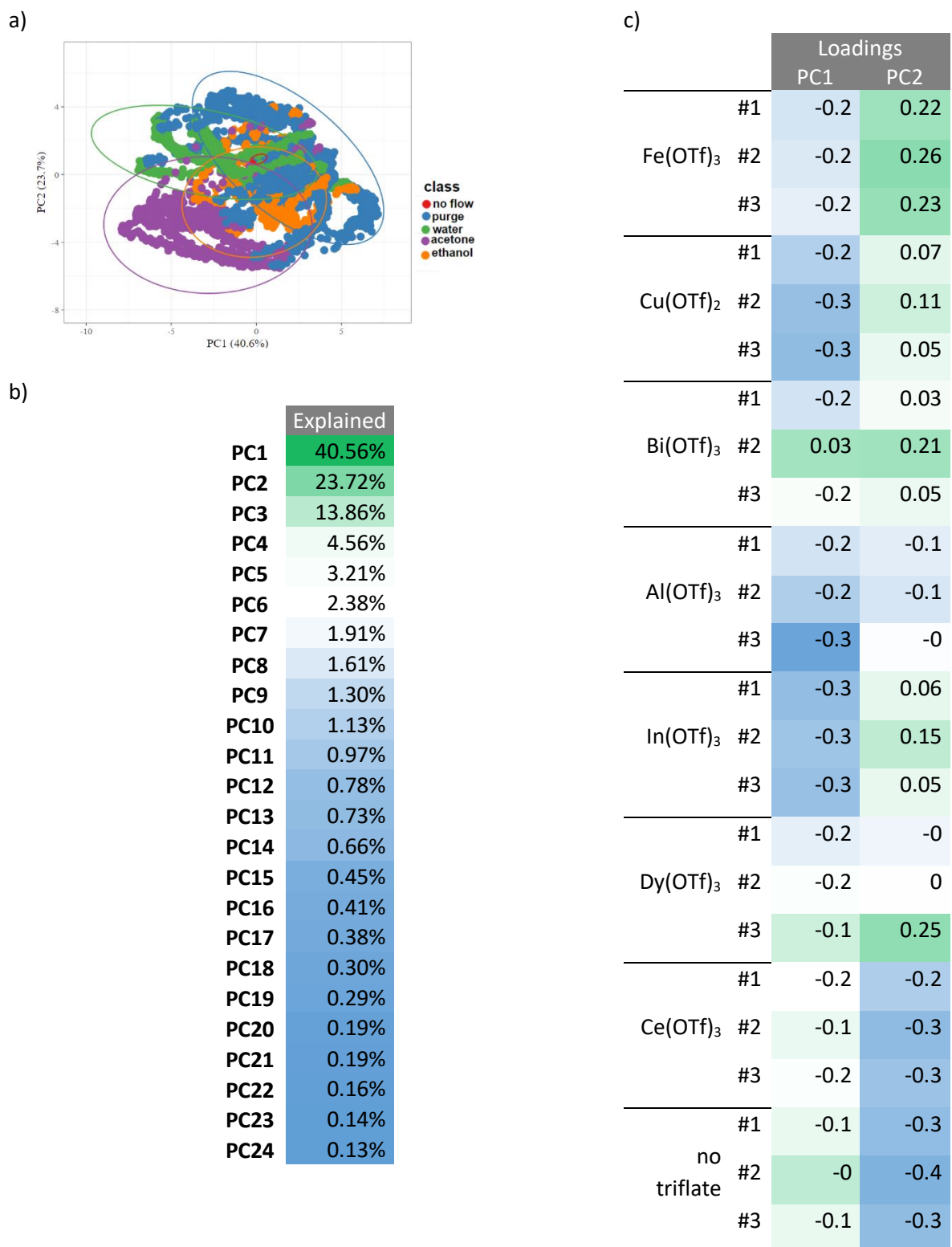

**Figure S13**. PCA on "modema$_\alpha$(i(t))" for $\alpha$ = 1/300 | **a**, PCA scores with 95% confidence ellipsoids. **b**, Individual variance for the different PC. **c**, PCA loadings of the different sensing elements' response for PC1 and PC2.

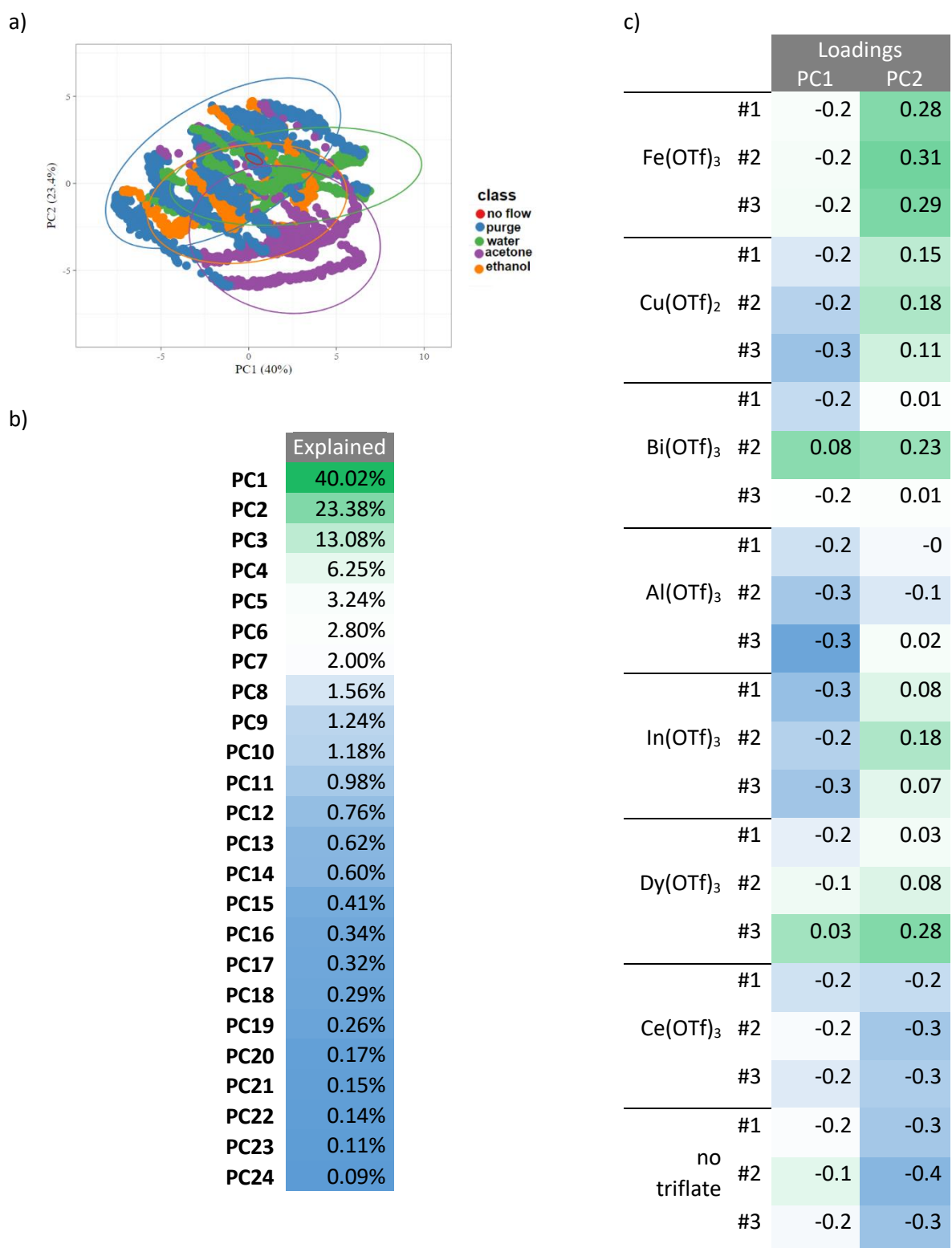

**Figure S14. PCA on "modema$_\alpha$(i(t))" for α = 1/600 | a,** PCA scores with 95% confidence ellipsoids. **b,** Individual variance for the different PC. **c,** PCA loadings of the different sensing elements' response for PC1 and PC2.

.

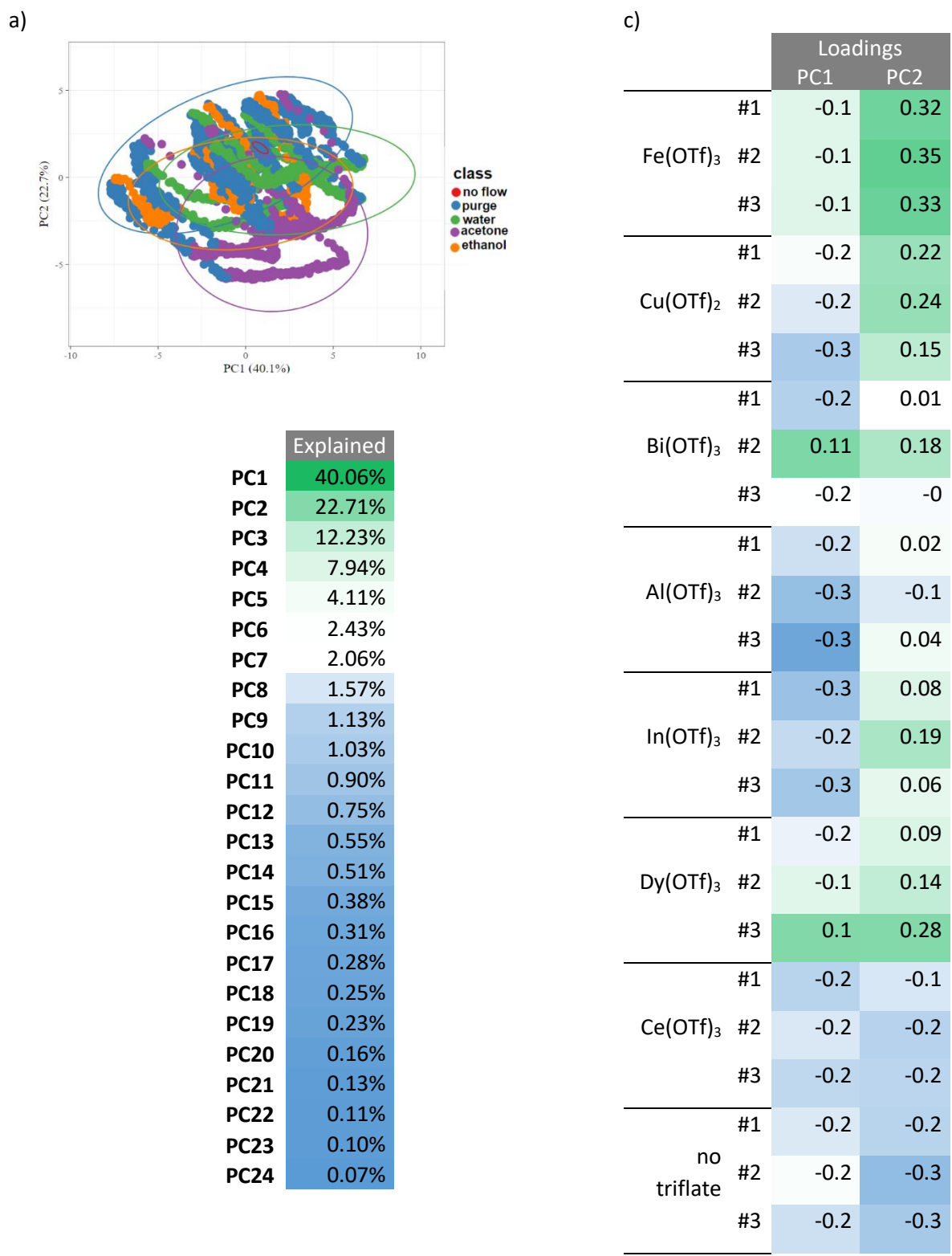

a)

c)

| | | Loadings | |
|---|---|---|---|
| | | PC1 | PC2 |
| Fe(OTf)$_3$ | #1 | -0.1 | 0.32 |
| | #2 | -0.1 | 0.35 |
| | #3 | -0.1 | 0.33 |
| Cu(OTf)$_2$ | #1 | -0.2 | 0.22 |
| | #2 | -0.2 | 0.24 |
| | #3 | -0.3 | 0.15 |
| Bi(OTf)$_3$ | #1 | -0.2 | 0.01 |
| | #2 | 0.11 | 0.18 |
| | #3 | -0.2 | -0 |
| Al(OTf)$_3$ | #1 | -0.2 | 0.02 |
| | #2 | -0.3 | -0.1 |
| | #3 | -0.3 | 0.04 |
| In(OTf)$_3$ | #1 | -0.3 | 0.08 |
| | #2 | -0.2 | 0.19 |
| | #3 | -0.3 | 0.06 |
| Dy(OTf)$_3$ | #1 | -0.2 | 0.09 |
| | #2 | -0.1 | 0.14 |
| | #3 | 0.1 | 0.28 |
| Ce(OTf)$_3$ | #1 | -0.2 | -0.1 |
| | #2 | -0.2 | -0.2 |
| | #3 | -0.2 | -0.2 |
| no triflate | #1 | -0.2 | -0.2 |
| | #2 | -0.2 | -0.3 |
| | #3 | -0.2 | -0.3 |

| | Explained |
|---|---|
| PC1 | 40.06% |
| PC2 | 22.71% |
| PC3 | 12.23% |
| PC4 | 7.94% |
| PC5 | 4.11% |
| PC6 | 2.43% |
| PC7 | 2.06% |
| PC8 | 1.57% |
| PC9 | 1.13% |
| PC10 | 1.03% |
| PC11 | 0.90% |
| PC12 | 0.75% |
| PC13 | 0.55% |
| PC14 | 0.51% |
| PC15 | 0.38% |
| PC16 | 0.31% |
| PC17 | 0.28% |
| PC18 | 0.25% |
| PC19 | 0.23% |
| PC20 | 0.16% |
| PC21 | 0.13% |
| PC22 | 0.11% |
| PC23 | 0.10% |
| PC24 | 0.07% |

**Figure S15**. **PCA on "modema$_\alpha$(i(t))" for α = 1/1000 | a,** PCA scores with 95% confidence ellipsoids. b Individual variance for the different PC. **c,** PCA loadings of the different sensing elements' response for PC1 and PC2.

.

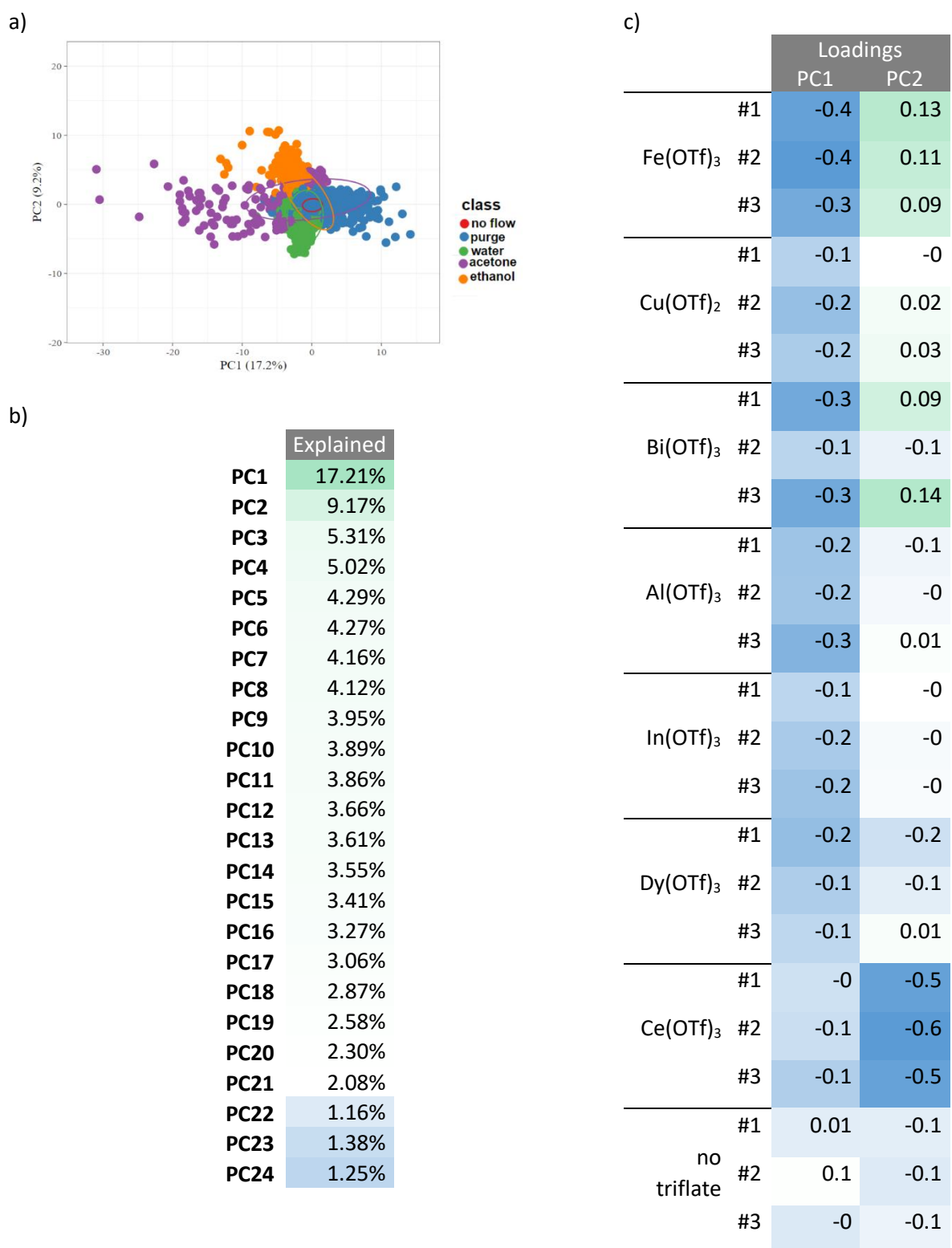

**Figure S16. PCA on "modema$_\alpha$(R(t))" for $\alpha$ = 1/2 | a,** PCA scores with 95% confidence ellipsoids. **b,** Individual variance for the different PC. **c,** PCA loadings of the different sensing elements' response for PC1 and PC2.

a)

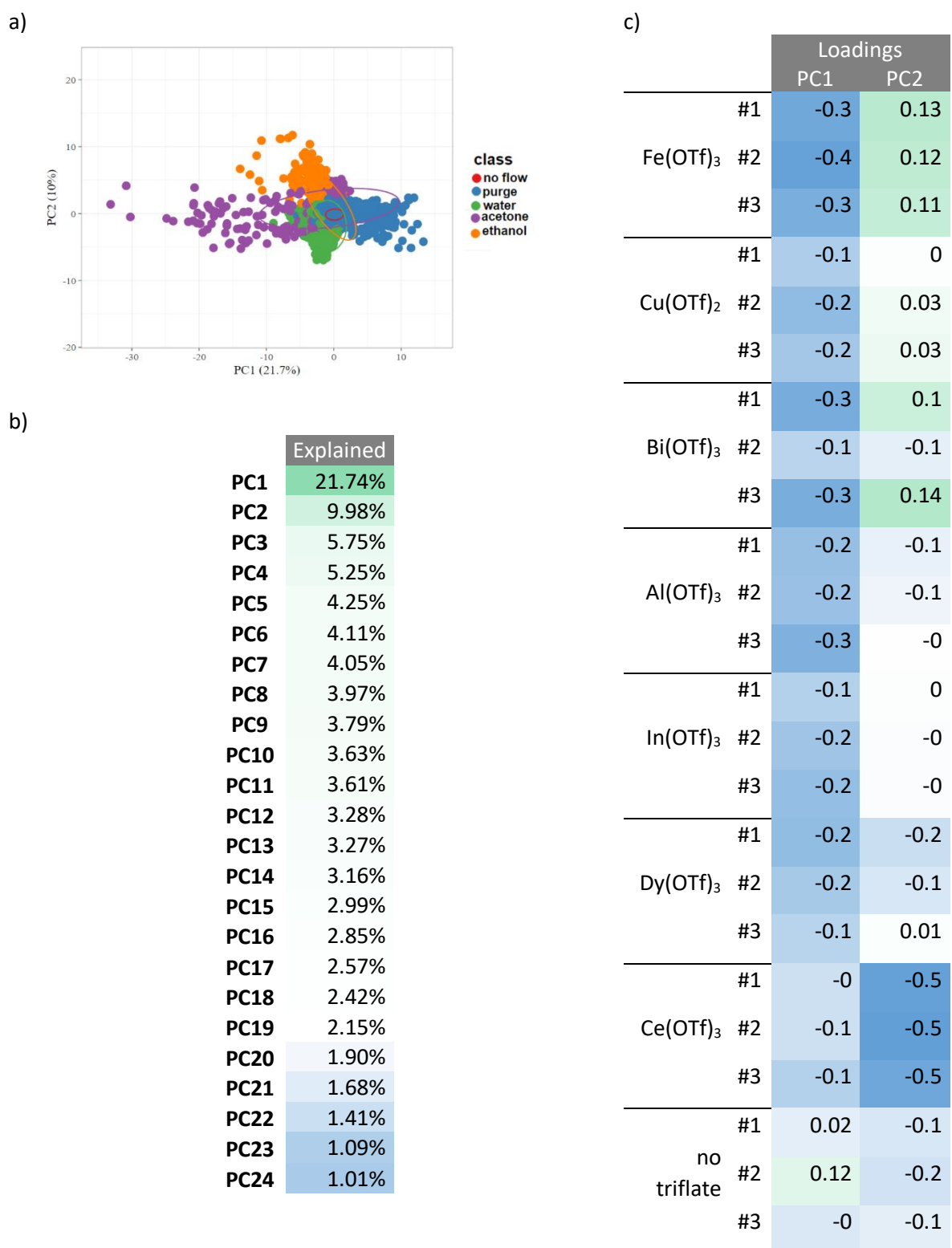

b)

| | Explained |
|---|---|
| **PC1** | 21.74% |
| **PC2** | 9.98% |
| **PC3** | 5.75% |
| **PC4** | 5.25% |
| **PC5** | 4.25% |
| **PC6** | 4.11% |
| **PC7** | 4.05% |
| **PC8** | 3.97% |
| **PC9** | 3.79% |
| **PC10** | 3.63% |
| **PC11** | 3.61% |
| **PC12** | 3.28% |
| **PC13** | 3.27% |
| **PC14** | 3.16% |
| **PC15** | 2.99% |
| **PC16** | 2.85% |
| **PC17** | 2.57% |
| **PC18** | 2.42% |
| **PC19** | 2.15% |
| **PC20** | 1.90% |
| **PC21** | 1.68% |
| **PC22** | 1.41% |
| **PC23** | 1.09% |
| **PC24** | 1.01% |

c)

| | | Loadings | |
|---|---|---|---|
| | | PC1 | PC2 |
| $Fe(OTf)_3$ | #1 | -0.3 | 0.13 |
| | #2 | -0.4 | 0.12 |
| | #3 | -0.3 | 0.11 |
| $Cu(OTf)_2$ | #1 | -0.1 | 0 |
| | #2 | -0.2 | 0.03 |
| | #3 | -0.2 | 0.03 |
| $Bi(OTf)_3$ | #1 | -0.3 | 0.1 |
| | #2 | -0.1 | -0.1 |
| | #3 | -0.3 | 0.14 |
| $Al(OTf)_3$ | #1 | -0.2 | -0.1 |
| | #2 | -0.2 | -0.1 |
| | #3 | -0.3 | -0 |
| $In(OTf)_3$ | #1 | -0.1 | 0 |
| | #2 | -0.2 | -0 |
| | #3 | -0.2 | -0 |
| $Dy(OTf)_3$ | #1 | -0.2 | -0.2 |
| | #2 | -0.2 | -0.1 |
| | #3 | -0.1 | 0.01 |
| $Ce(OTf)_3$ | #1 | -0 | -0.5 |
| | #2 | -0.1 | -0.5 |
| | #3 | -0.1 | -0.5 |
| no triflate | #1 | 0.02 | -0.1 |
| | #2 | 0.12 | -0.2 |
| | #3 | -0 | -0.1 |

**Figure S17. PCA on "modema$_\alpha$(R(t))" for α = 1/3 | a,** PCA scores with 95% confidence ellipsoids. **b,** Individual variance for the different PC. **c,** PCA loadings of the different sensing elements' response for PC1 and PC2.

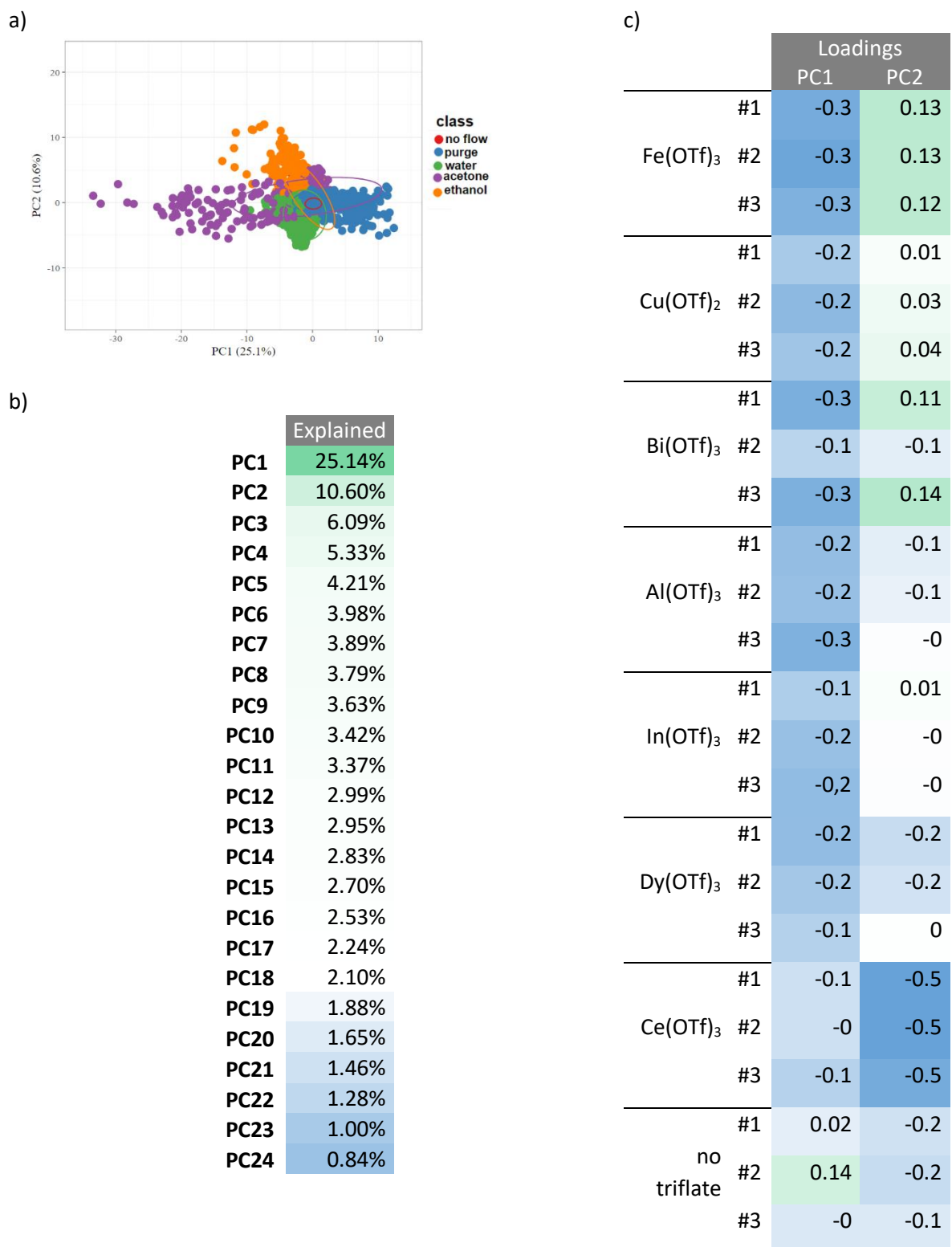

**Figure S18. PCA on "modema$_\alpha$(R(t))" for α = 1/4 | a.** PCA scores with 95% confidence ellipsoids. b, Individual variance for the different PC. **c**, PCA loadings of the different sensing elements' response for PC1 and PC2.

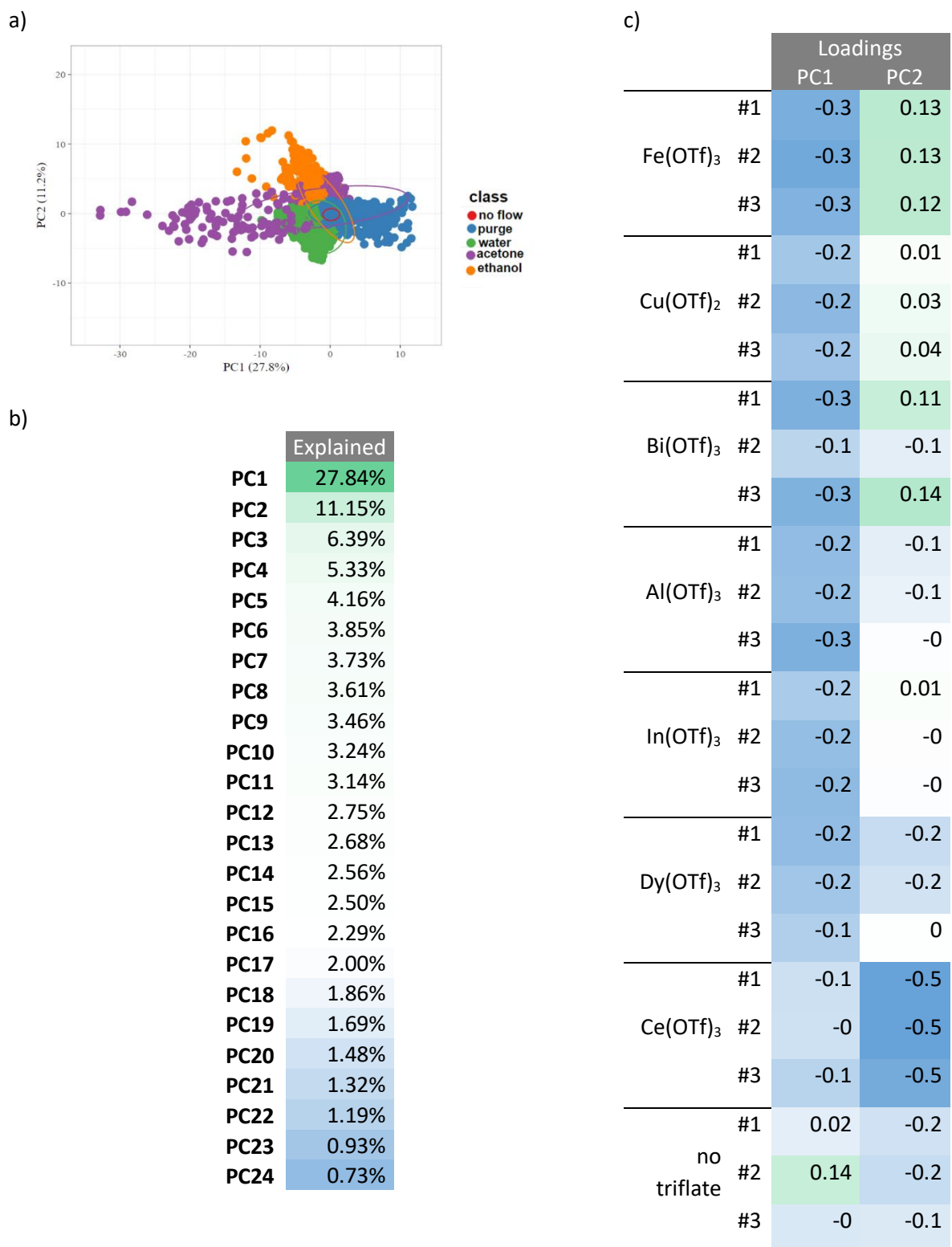

a)

b)

| | Explained |
|---|---|
| **PC1** | 27.84% |
| **PC2** | 11.15% |
| **PC3** | 6.39% |
| **PC4** | 5.33% |
| **PC5** | 4.16% |
| **PC6** | 3.85% |
| **PC7** | 3.73% |
| **PC8** | 3.61% |
| **PC9** | 3.46% |
| **PC10** | 3.24% |
| **PC11** | 3.14% |
| **PC12** | 2.75% |
| **PC13** | 2.68% |
| **PC14** | 2.56% |
| **PC15** | 2.50% |
| **PC16** | 2.29% |
| **PC17** | 2.00% |
| **PC18** | 1.86% |
| **PC19** | 1.69% |
| **PC20** | 1.48% |
| **PC21** | 1.32% |
| **PC22** | 1.19% |
| **PC23** | 0.93% |
| **PC24** | 0.73% |

c)

| | | Loadings | |
|---|---|---|---|
| | | PC1 | PC2 |
| $Fe(OTf)_3$ | #1 | -0.3 | 0.13 |
| | #2 | -0.3 | 0.13 |
| | #3 | -0.3 | 0.12 |
| $Cu(OTf)_2$ | #1 | -0.2 | 0.01 |
| | #2 | -0.2 | 0.03 |
| | #3 | -0.2 | 0.04 |
| $Bi(OTf)_3$ | #1 | -0.3 | 0.11 |
| | #2 | -0.1 | -0.1 |
| | #3 | -0.3 | 0.14 |
| $Al(OTf)_3$ | #1 | -0.2 | -0.1 |
| | #2 | -0.2 | -0.1 |
| | #3 | -0.3 | -0 |
| $In(OTf)_3$ | #1 | -0.2 | 0.01 |
| | #2 | -0.2 | -0 |
| | #3 | -0.2 | -0 |
| $Dy(OTf)_3$ | #1 | -0.2 | -0.2 |
| | #2 | -0.2 | -0.2 |
| | #3 | -0.1 | 0 |
| $Ce(OTf)_3$ | #1 | -0.1 | -0.5 |
| | #2 | -0 | -0.5 |
| | #3 | -0.1 | -0.5 |
| no triflate | #1 | 0.02 | -0.2 |
| | #2 | 0.14 | -0.2 |
| | #3 | -0 | -0.1 |

**Figure S19. PCA on "modema$_\alpha$(R(t))" for α = 1/5 | a,** PCA scores with 95% confidence ellipsoids. **b.** Individual variance for the different PC. **c,** PCA loadings of the different sensing elements' response for PC1 and PC2.

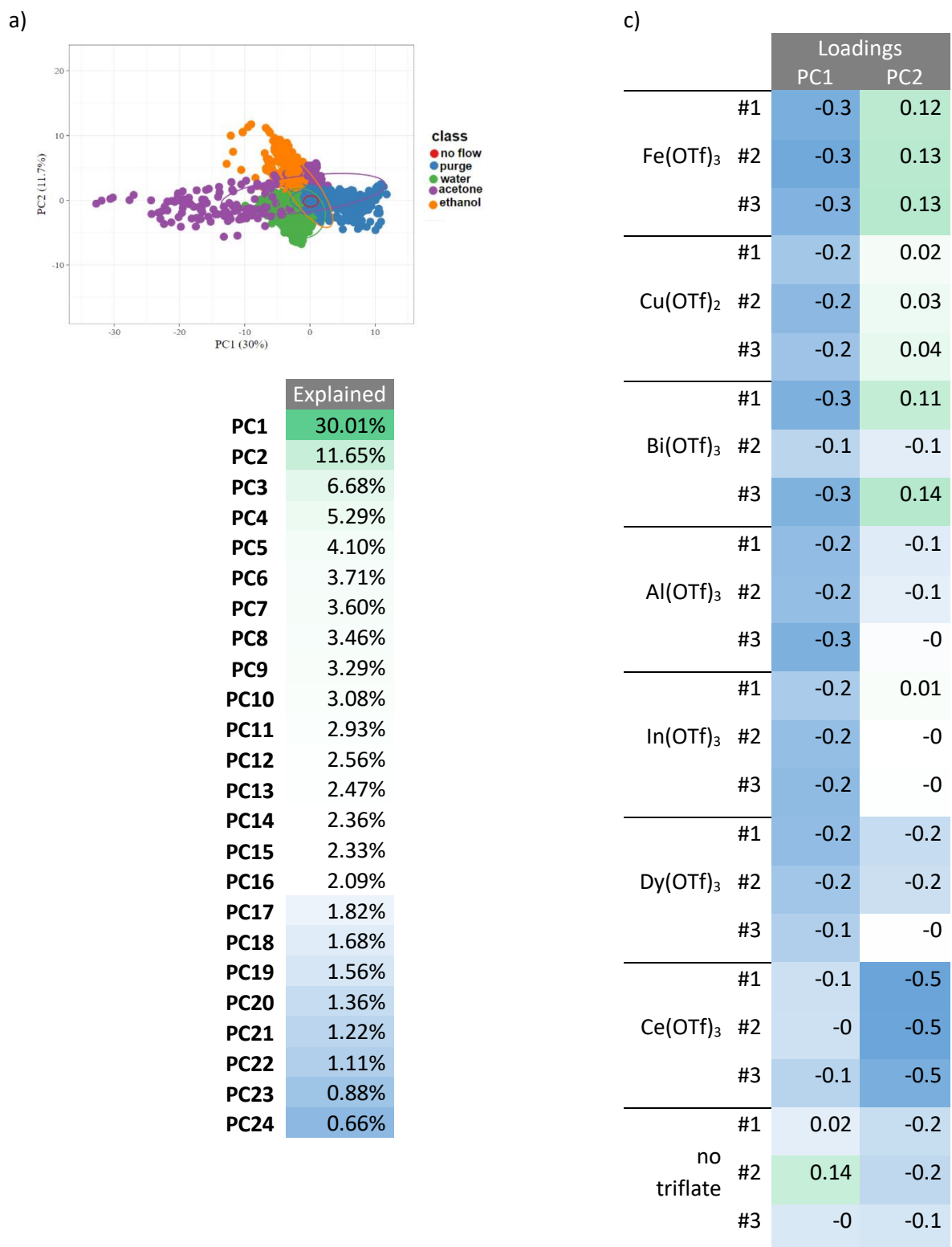

**Figure S20. PCA on "modema$_\alpha$(R(t))" for $\alpha$ = 1/6 | a**, PCA scores with 95% confidence ellipsoids. **b,** Individual variance for the different PC. **c,** PCA loadings of the different sensing elements' response for PC1 and PC2.

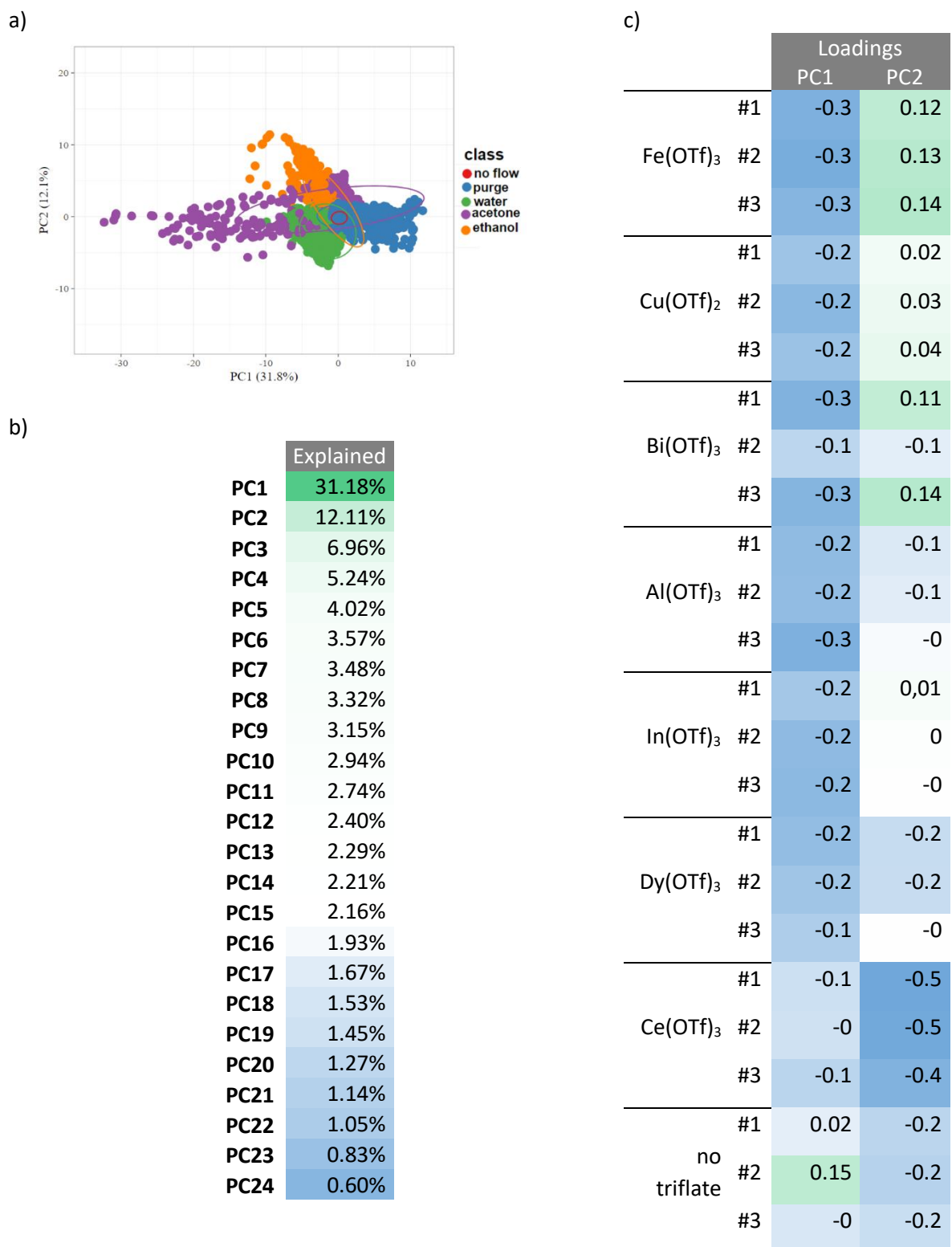

a)

b)

| | Explained |
|---|---|
| **PC1** | 31.18% |
| **PC2** | 12.11% |
| **PC3** | 6.96% |
| **PC4** | 5.24% |
| **PC5** | 4.02% |
| **PC6** | 3.57% |
| **PC7** | 3.48% |
| **PC8** | 3.32% |
| **PC9** | 3.15% |
| **PC10** | 2.94% |
| **PC11** | 2.74% |
| **PC12** | 2.40% |
| **PC13** | 2.29% |
| **PC14** | 2.21% |
| **PC15** | 2.16% |
| **PC16** | 1.93% |
| **PC17** | 1.67% |
| **PC18** | 1.53% |
| **PC19** | 1.45% |
| **PC20** | 1.27% |
| **PC21** | 1.14% |
| **PC22** | 1.05% |
| **PC23** | 0.83% |
| **PC24** | 0.60% |

c)

| | | Loadings | |
|---|---|---|---|
| | | PC1 | PC2 |
| $Fe(OTf)_3$ | #1 | -0.3 | 0.12 |
| | #2 | -0.3 | 0.13 |
| | #3 | -0.3 | 0.14 |
| $Cu(OTf)_2$ | #1 | -0.2 | 0.02 |
| | #2 | -0.2 | 0.03 |
| | #3 | -0.2 | 0.04 |
| $Bi(OTf)_3$ | #1 | -0.3 | 0.11 |
| | #2 | -0.1 | -0.1 |
| | #3 | -0.3 | 0.14 |
| $Al(OTf)_3$ | #1 | -0.2 | -0.1 |
| | #2 | -0.2 | -0.1 |
| | #3 | -0.3 | -0 |
| $In(OTf)_3$ | #1 | -0.2 | 0,01 |
| | #2 | -0.2 | 0 |
| | #3 | -0.2 | -0 |
| $Dy(OTf)_3$ | #1 | -0.2 | -0.2 |
| | #2 | -0.2 | -0.2 |
| | #3 | -0.1 | -0 |
| $Ce(OTf)_3$ | #1 | -0.1 | -0.5 |
| | #2 | -0 | -0.5 |
| | #3 | -0.1 | -0.4 |
| no triflate | #1 | 0.02 | -0.2 |
| | #2 | 0.15 | -0.2 |
| | #3 | -0 | -0.2 |

**Figure S21. PCA on "modema$_\alpha$(R(t))" for α = 1/7 | a,** PCA scores with 95% confidence ellipsoids. **b,** Individual variance for the different PC. **c,** PCA loadings of the different sensing elements' response for PC1 and PC2.

.

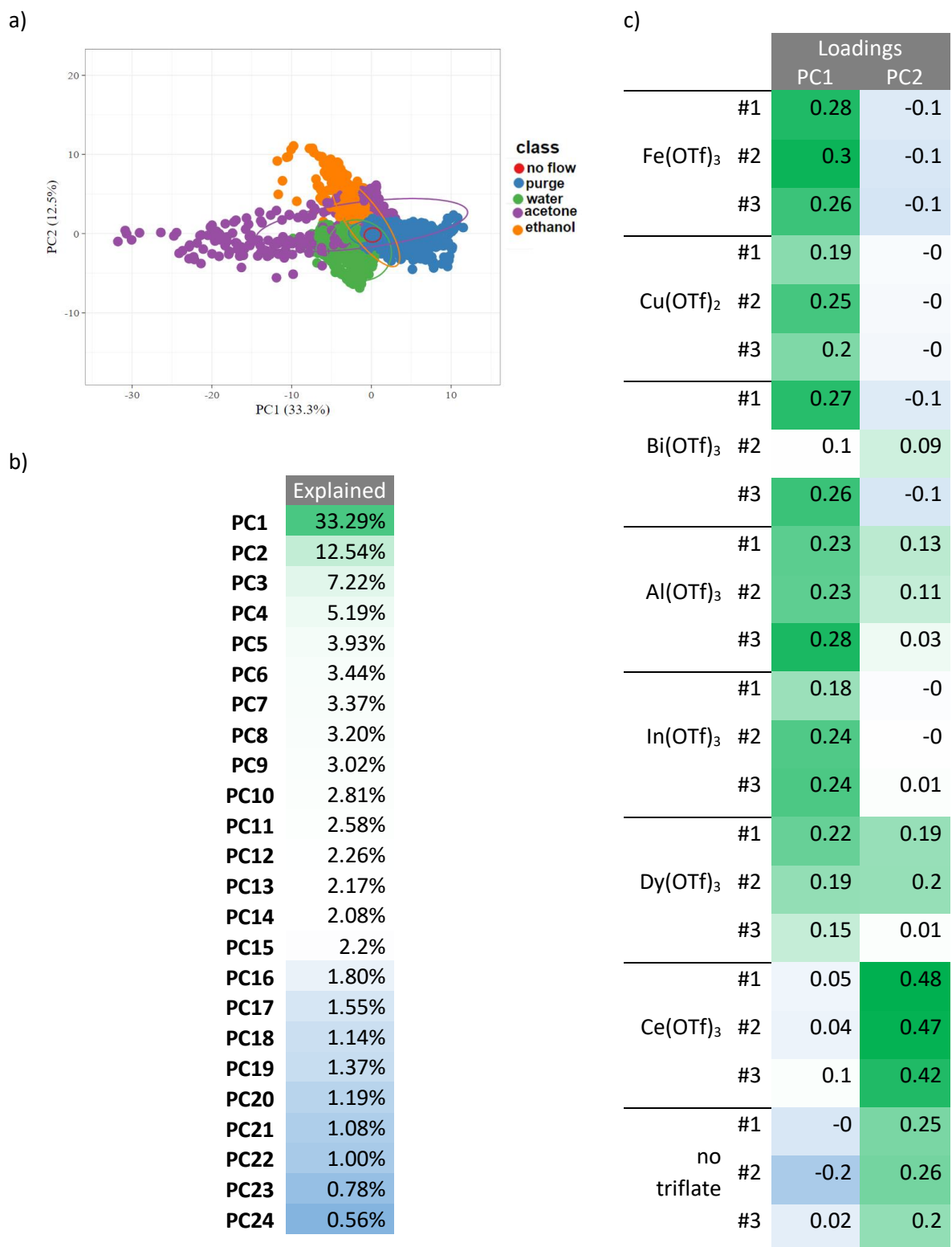

**Figure S22**. **PCA on "modema$_\alpha$(R(t))" for α = 1/8 | a,** PCA scores with 95% confidence ellipsoids. **b,** Individual variance for the different PC. **c,** PCA loadings of the different sensing elements' response for PC1 and PC2.

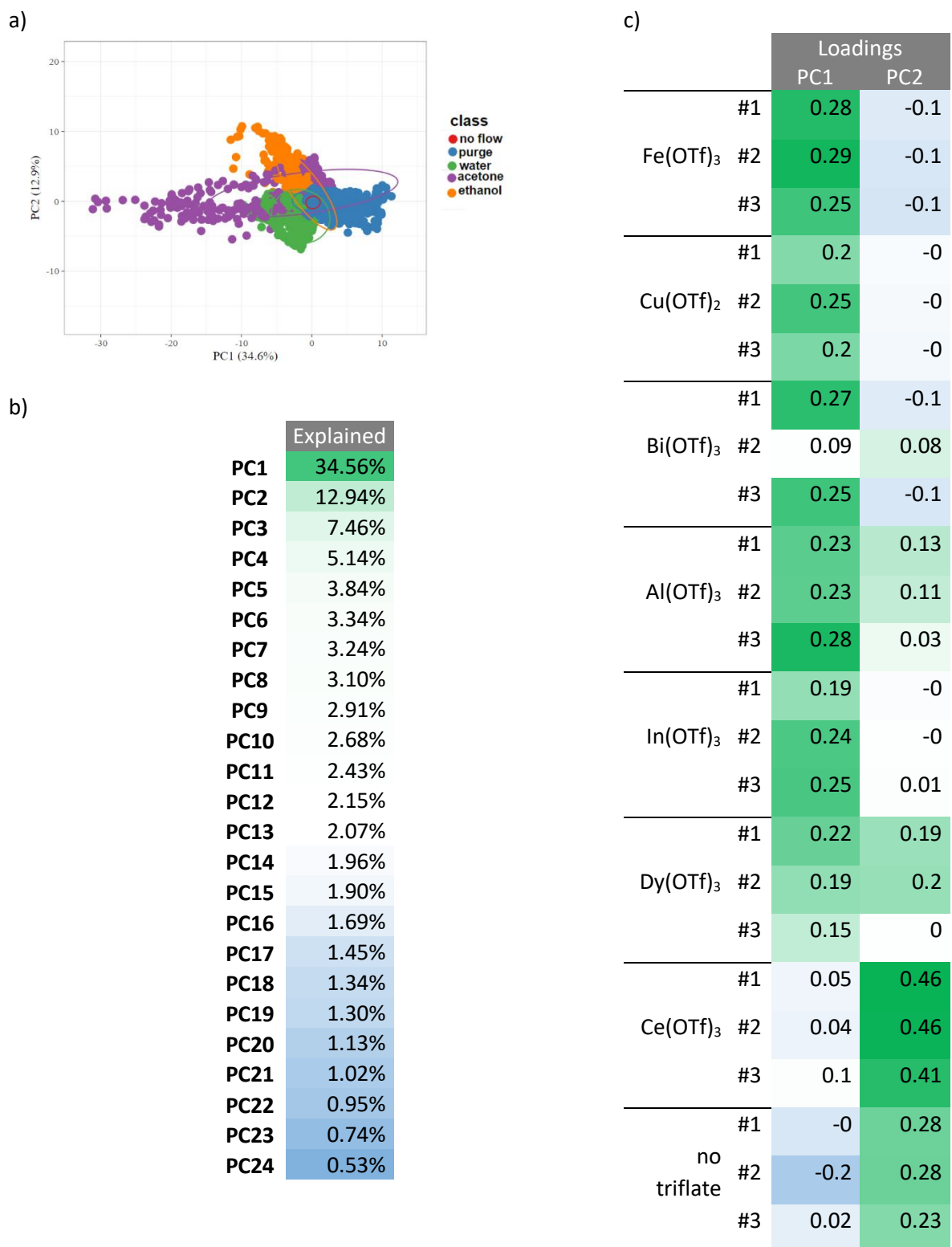

**Figure S23.** "**PCA on modema$_\alpha$(R(t))" for α = 1/9 | a,** PCA scores with 95% confidence ellipsoids. **b,** Individual variance for the different PC. **c,** PCA loadings of the different sensing elements' response for PC1 and PC2.

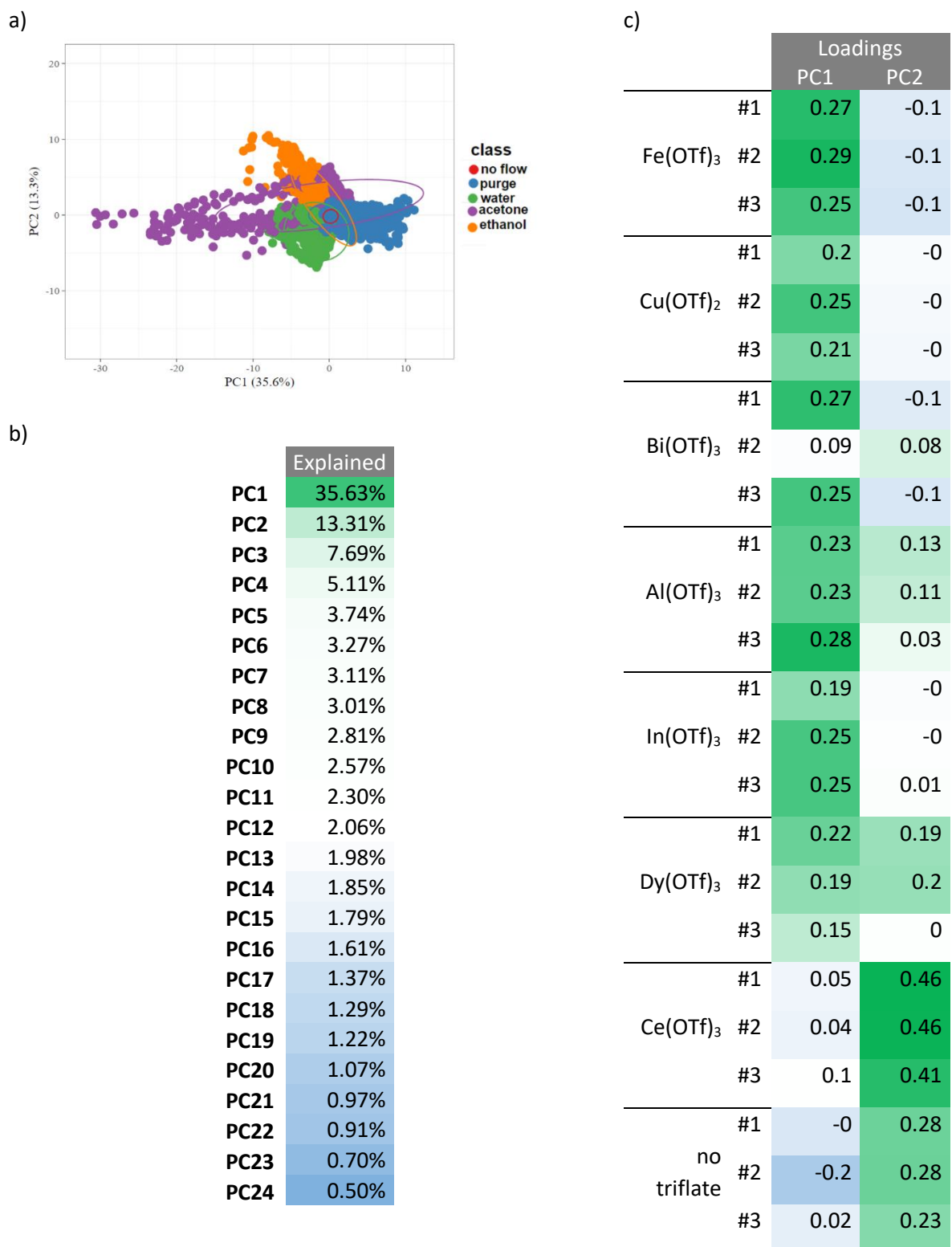

a)

b)

| | Explained |
|---|---|
| **PC1** | 35.63% |
| **PC2** | 13.31% |
| **PC3** | 7.69% |
| **PC4** | 5.11% |
| **PC5** | 3.74% |
| **PC6** | 3.27% |
| **PC7** | 3.11% |
| **PC8** | 3.01% |
| **PC9** | 2.81% |
| **PC10** | 2.57% |
| **PC11** | 2.30% |
| **PC12** | 2.06% |
| **PC13** | 1.98% |
| **PC14** | 1.85% |
| **PC15** | 1.79% |
| **PC16** | 1.61% |
| **PC17** | 1.37% |
| **PC18** | 1.29% |
| **PC19** | 1.22% |
| **PC20** | 1.07% |
| **PC21** | 0.97% |
| **PC22** | 0.91% |
| **PC23** | 0.70% |
| **PC24** | 0.50% |

c)

| | | Loadings | |
|---|---|---|---|
| | | PC1 | PC2 |
| Fe(OTf)$_3$ | #1 | 0.27 | -0.1 |
| | #2 | 0.29 | -0.1 |
| | #3 | 0.25 | -0.1 |
| Cu(OTf)$_2$ | #1 | 0.2 | -0 |
| | #2 | 0.25 | -0 |
| | #3 | 0.21 | -0 |
| Bi(OTf)$_3$ | #1 | 0.27 | -0.1 |
| | #2 | 0.09 | 0.08 |
| | #3 | 0.25 | -0.1 |
| Al(OTf)$_3$ | #1 | 0.23 | 0.13 |
| | #2 | 0.23 | 0.11 |
| | #3 | 0.28 | 0.03 |
| In(OTf)$_3$ | #1 | 0.19 | -0 |
| | #2 | 0.25 | -0 |
| | #3 | 0.25 | 0.01 |
| Dy(OTf)$_3$ | #1 | 0.22 | 0.19 |
| | #2 | 0.19 | 0.2 |
| | #3 | 0.15 | 0 |
| Ce(OTf)$_3$ | #1 | 0.05 | 0.46 |
| | #2 | 0.04 | 0.46 |
| | #3 | 0.1 | 0.41 |
| no triflate | #1 | -0 | 0.28 |
| | #2 | -0.2 | 0.28 |
| | #3 | 0.02 | 0.23 |

**Figure S24. PCA on "modema$_\alpha$(R(t))" for α = 1/10 | a,** PCA scores with 95% confidence ellipsoids. **b,** Individual variance for the different PC. **c,** PCA loadings of the different sensing elements' response for PC1 and PC2.

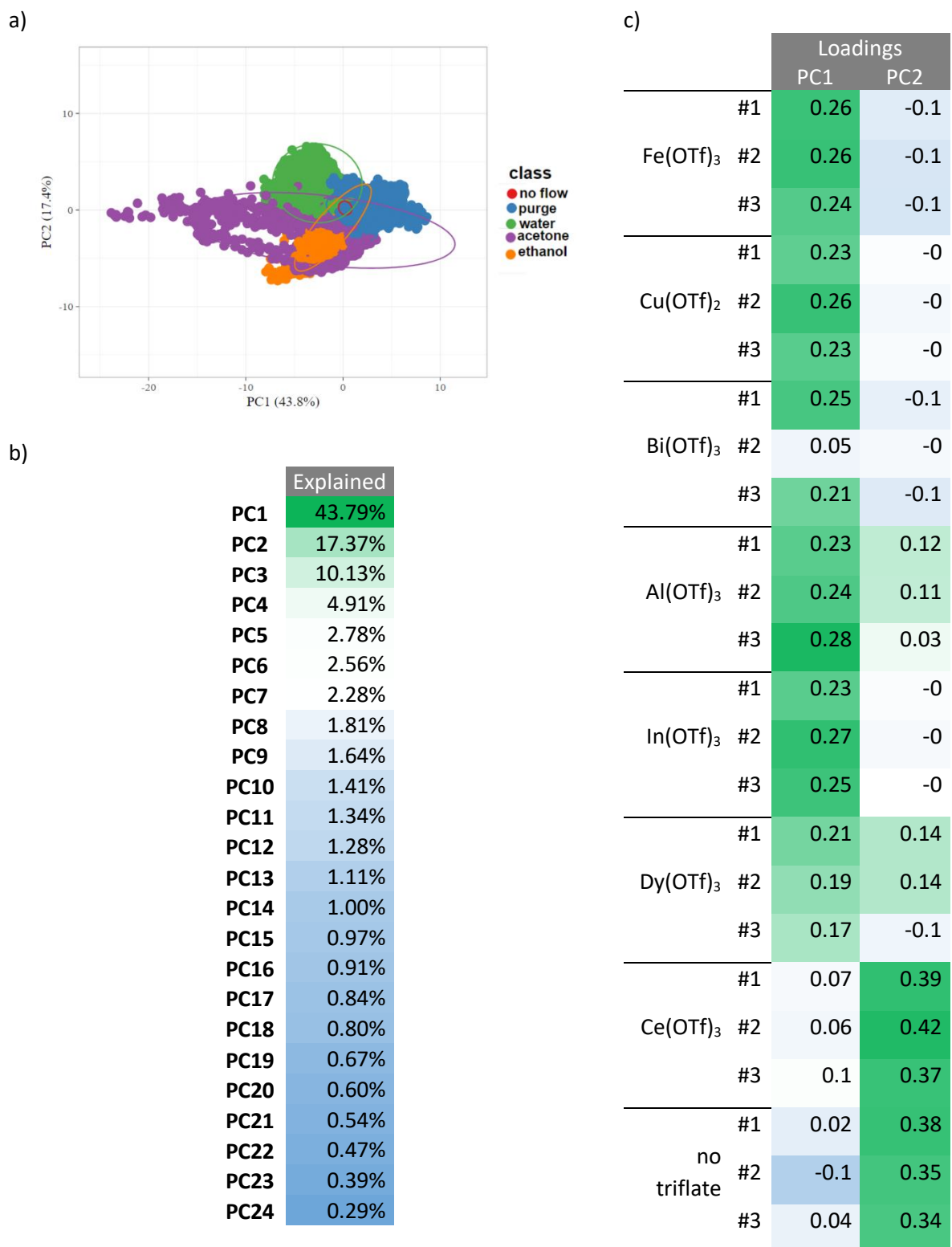

**Figure S25. PCA on "modema$_\alpha$(R(t))" for $\alpha$ = 1/30 | a,** PCA scores with 95% confidence ellipsoids. **b,** Individual variance for the different PC. **c,** PCA loadings of the different sensing elements' response for PC1 and PC2.

a)

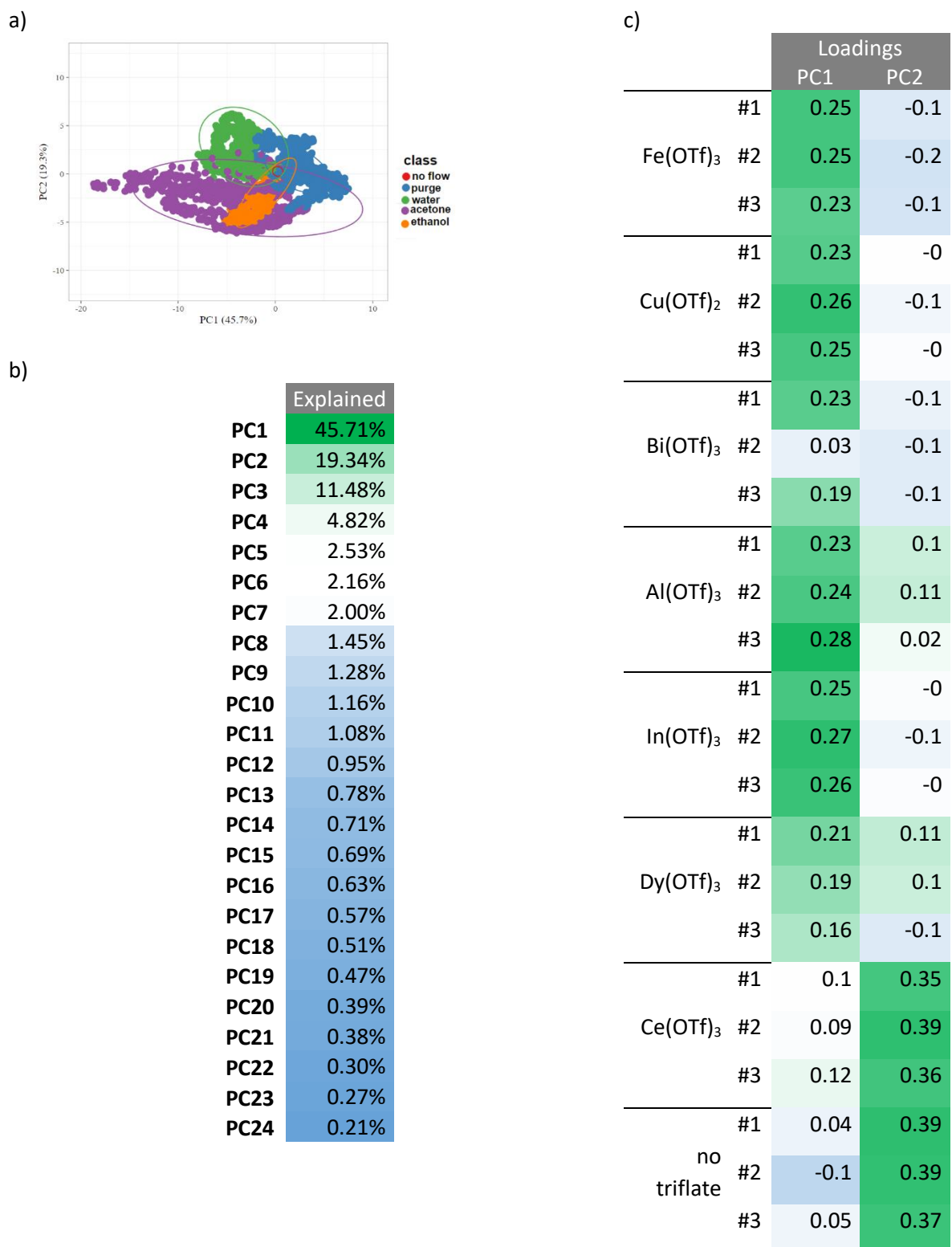

b)

| | Explained |
|---|---|
| PC1 | 45.71% |
| PC2 | 19.34% |
| PC3 | 11.48% |
| PC4 | 4.82% |
| PC5 | 2.53% |
| PC6 | 2.16% |
| PC7 | 2.00% |
| PC8 | 1.45% |
| PC9 | 1.28% |
| PC10 | 1.16% |
| PC11 | 1.08% |
| PC12 | 0.95% |
| PC13 | 0.78% |
| PC14 | 0.71% |
| PC15 | 0.69% |
| PC16 | 0.63% |
| PC17 | 0.57% |
| PC18 | 0.51% |
| PC19 | 0.47% |
| PC20 | 0.39% |
| PC21 | 0.38% |
| PC22 | 0.30% |
| PC23 | 0.27% |
| PC24 | 0.21% |

c)

| | | Loadings | |
|---|---|---|---|
| | | PC1 | PC2 |
| $Fe(OTf)_3$ | #1 | 0.25 | -0.1 |
| | #2 | 0.25 | -0.2 |
| | #3 | 0.23 | -0.1 |
| $Cu(OTf)_2$ | #1 | 0.23 | -0 |
| | #2 | 0.26 | -0.1 |
| | #3 | 0.25 | -0 |
| $Bi(OTf)_3$ | #1 | 0.23 | -0.1 |
| | #2 | 0.03 | -0.1 |
| | #3 | 0.19 | -0.1 |
| $Al(OTf)_3$ | #1 | 0.23 | 0.1 |
| | #2 | 0.24 | 0.11 |
| | #3 | 0.28 | 0.02 |
| $In(OTf)_3$ | #1 | 0.25 | -0 |
| | #2 | 0.27 | -0.1 |
| | #3 | 0.26 | -0 |
| $Dy(OTf)_3$ | #1 | 0.21 | 0.11 |
| | #2 | 0.19 | 0.1 |
| | #3 | 0.16 | -0.1 |
| $Ce(OTf)_3$ | #1 | 0.1 | 0.35 |
| | #2 | 0.09 | 0.39 |
| | #3 | 0.12 | 0.36 |
| no triflate | #1 | 0.04 | 0.39 |
| | #2 | -0.1 | 0.39 |
| | #3 | 0.05 | 0.37 |

**Figure S26. PCA on "modema$_\alpha$(R(t))" for α = 1/60 | a,** PCA scores with 95% confidence ellipsoids. **b,** Individual variance for the different PC. **c,** PCA loadings of the different sensing elements' response for PC1 and PC2.

.

a)

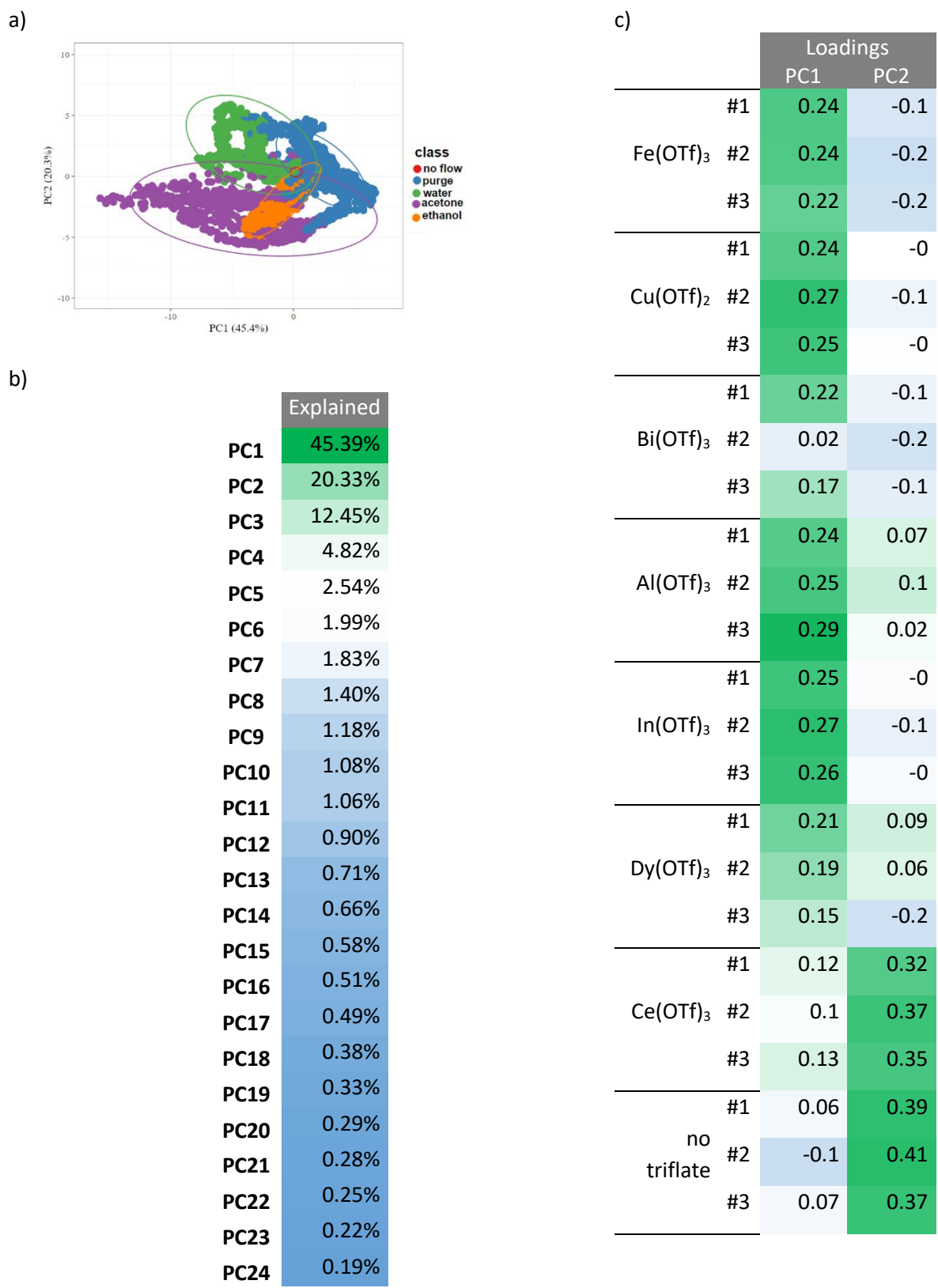

b)

| | Explained |
|---|---|
| PC1 | 45.39% |
| PC2 | 20.33% |
| PC3 | 12.45% |
| PC4 | 4.82% |
| PC5 | 2.54% |
| PC6 | 1.99% |
| PC7 | 1.83% |
| PC8 | 1.40% |
| PC9 | 1.18% |
| PC10 | 1.08% |
| PC11 | 1.06% |
| PC12 | 0.90% |
| PC13 | 0.71% |
| PC14 | 0.66% |
| PC15 | 0.58% |
| PC16 | 0.51% |
| PC17 | 0.49% |
| PC18 | 0.38% |
| PC19 | 0.33% |
| PC20 | 0.29% |
| PC21 | 0.28% |
| PC22 | 0.25% |
| PC23 | 0.22% |
| PC24 | 0.19% |

c)

| | | Loadings | |
|---|---|---|---|
| | | PC1 | PC2 |
| $Fe(OTf)_3$ | #1 | 0.24 | -0.1 |
| | #2 | 0.24 | -0.2 |
| | #3 | 0.22 | -0.2 |
| $Cu(OTf)_2$ | #1 | 0.24 | -0 |
| | #2 | 0.27 | -0.1 |
| | #3 | 0.25 | -0 |
| $Bi(OTf)_3$ | #1 | 0.22 | -0.1 |
| | #2 | 0.02 | -0.2 |
| | #3 | 0.17 | -0.1 |
| $Al(OTf)_3$ | #1 | 0.24 | 0.07 |
| | #2 | 0.25 | 0.1 |
| | #3 | 0.29 | 0.02 |
| $In(OTf)_3$ | #1 | 0.25 | -0 |
| | #2 | 0.27 | -0.1 |
| | #3 | 0.26 | -0 |
| $Dy(OTf)_3$ | #1 | 0.21 | 0.09 |
| | #2 | 0.19 | 0.06 |
| | #3 | 0.15 | -0.2 |
| $Ce(OTf)_3$ | #1 | 0.12 | 0.32 |
| | #2 | 0.1 | 0.37 |
| | #3 | 0.13 | 0.35 |
| no triflate | #1 | 0.06 | 0.39 |
| | #2 | -0.1 | 0.41 |
| | #3 | 0.07 | 0.37 |

**Figure S27. PCA on "modema$_\alpha$(R(t))" for α = 1/100 | a,** PCA scores with 95% confidence ellipsoids. **b,** Individual variance for the different PC. **c,** PCA loadings of the different sensing elements' response for PC1 and PC2.

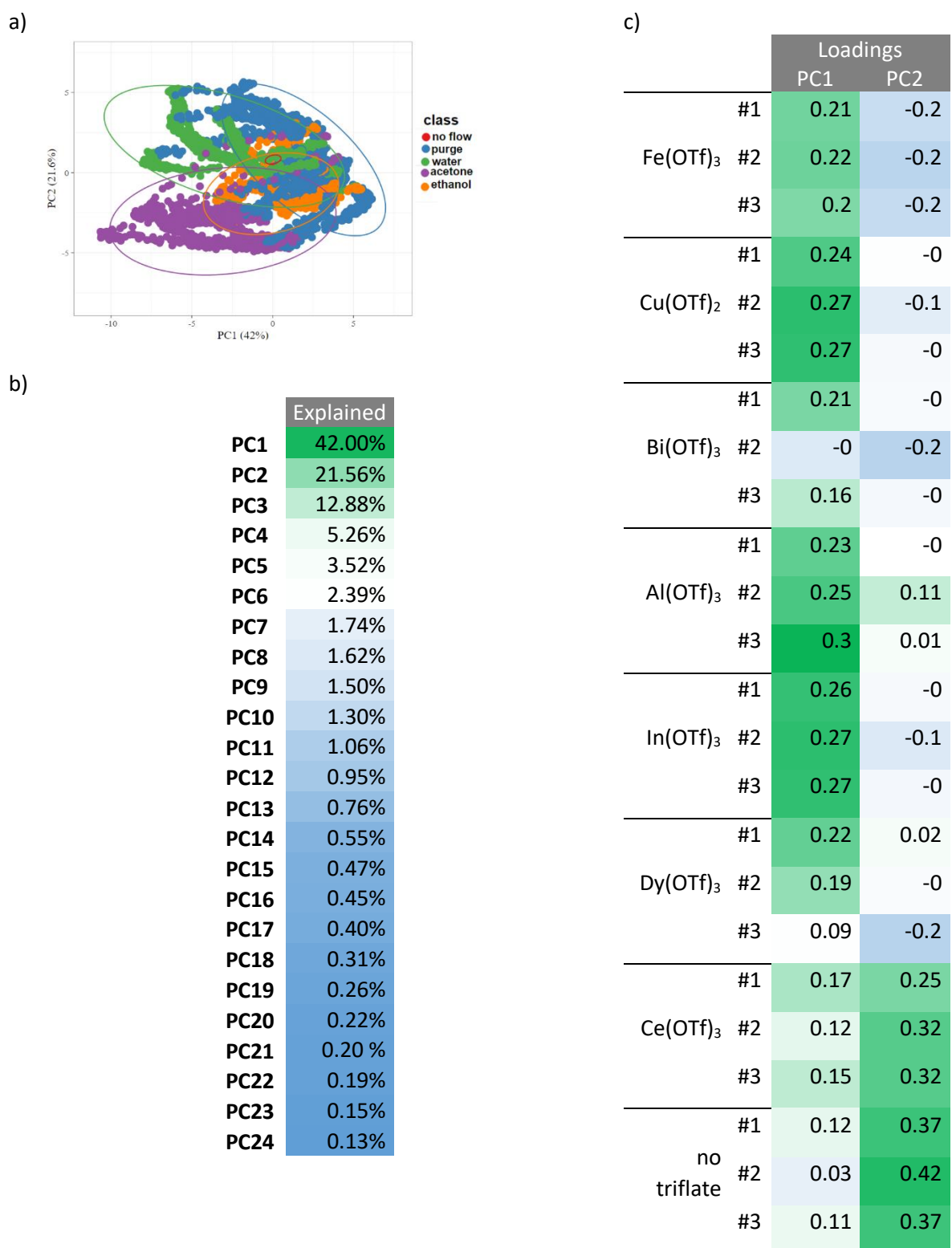

**Figure S28. PCA on "modema$_\alpha$(R(t))" for $\alpha$ = 1/300 | a,** PCA scores with 95% confidence ellipsoids. **b,** Individual variance for the different PC. **c,** PCA loadings of the different sensing elements' response for PC1 and PC2.

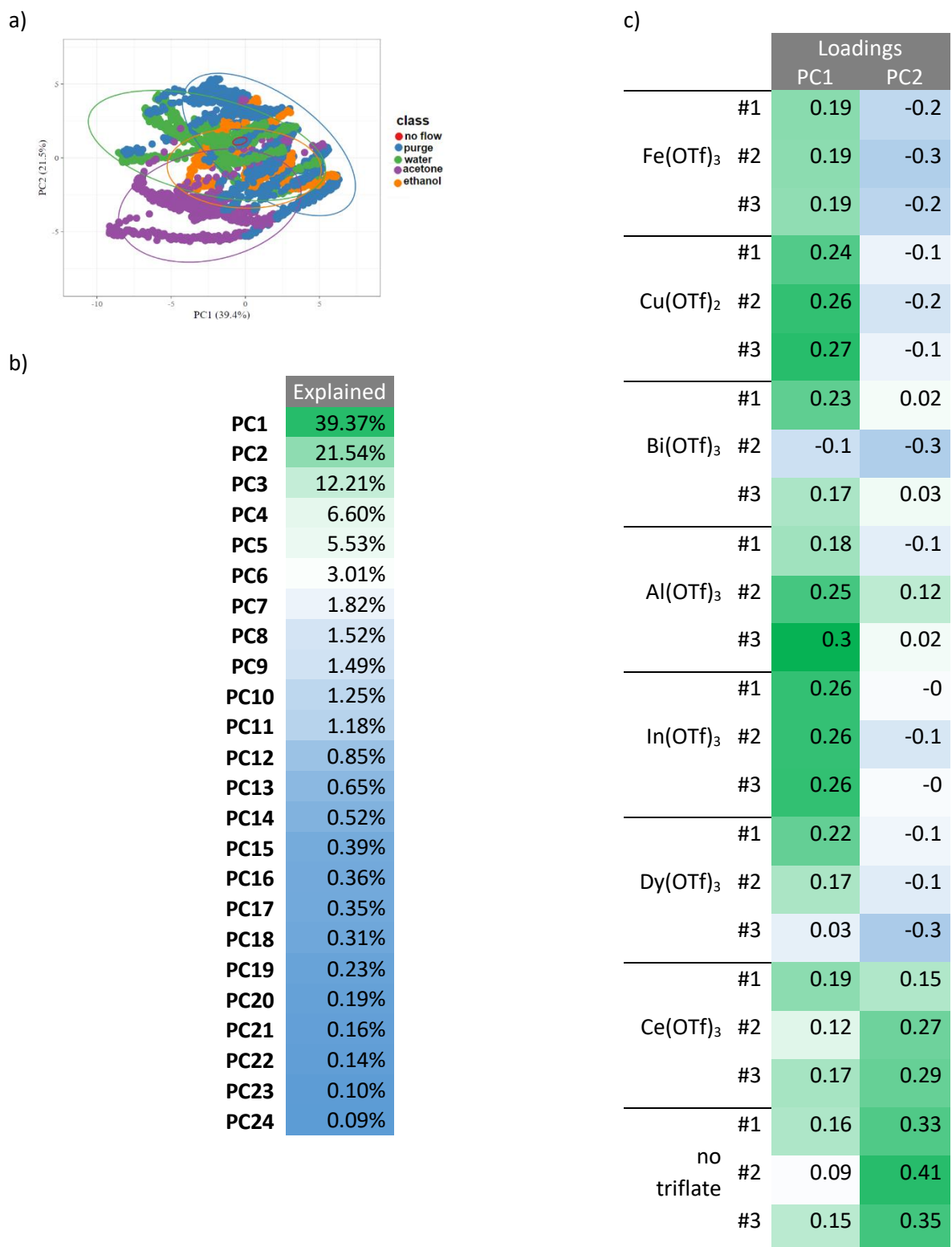

**Figure S29. PCA on "modema$_\alpha$(R(t))" for $\alpha$ = 1/600 | a,** PCA scores with 95% confidence ellipsoids. **b,** Individual variance for the different PC. **c,** PCA loadings of the different sensing elements' response for PC1 and PC2.

.

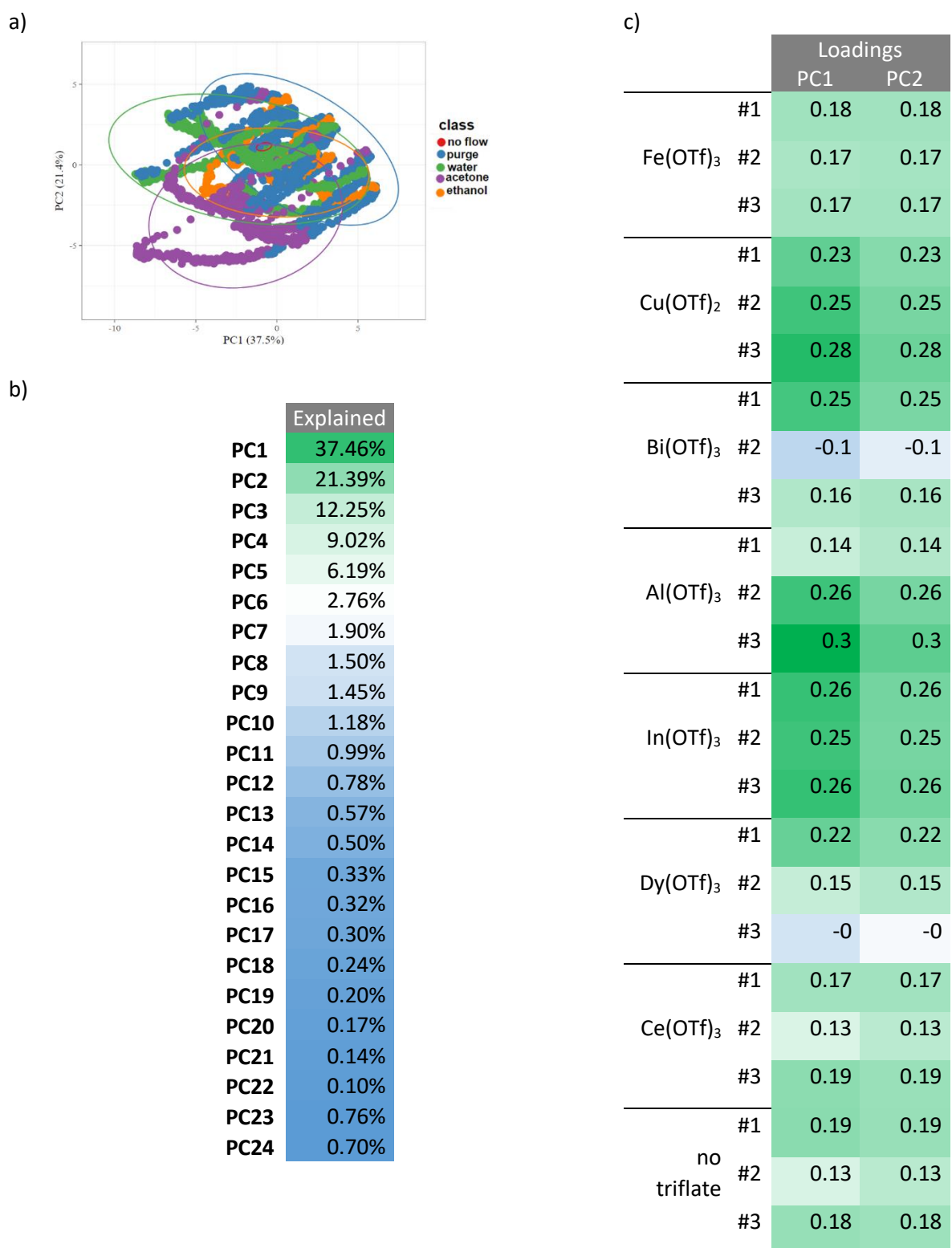

**Figure S30. PCA on "modema$_\alpha$(R(t))" for $\alpha$ = 1/1000 | a,** PCA scores with 95% confidence ellipsoids. **b,** Individual variance for the different PC. **c,** PCA loadings of the different sensing elements' response for PC1 and PC2.

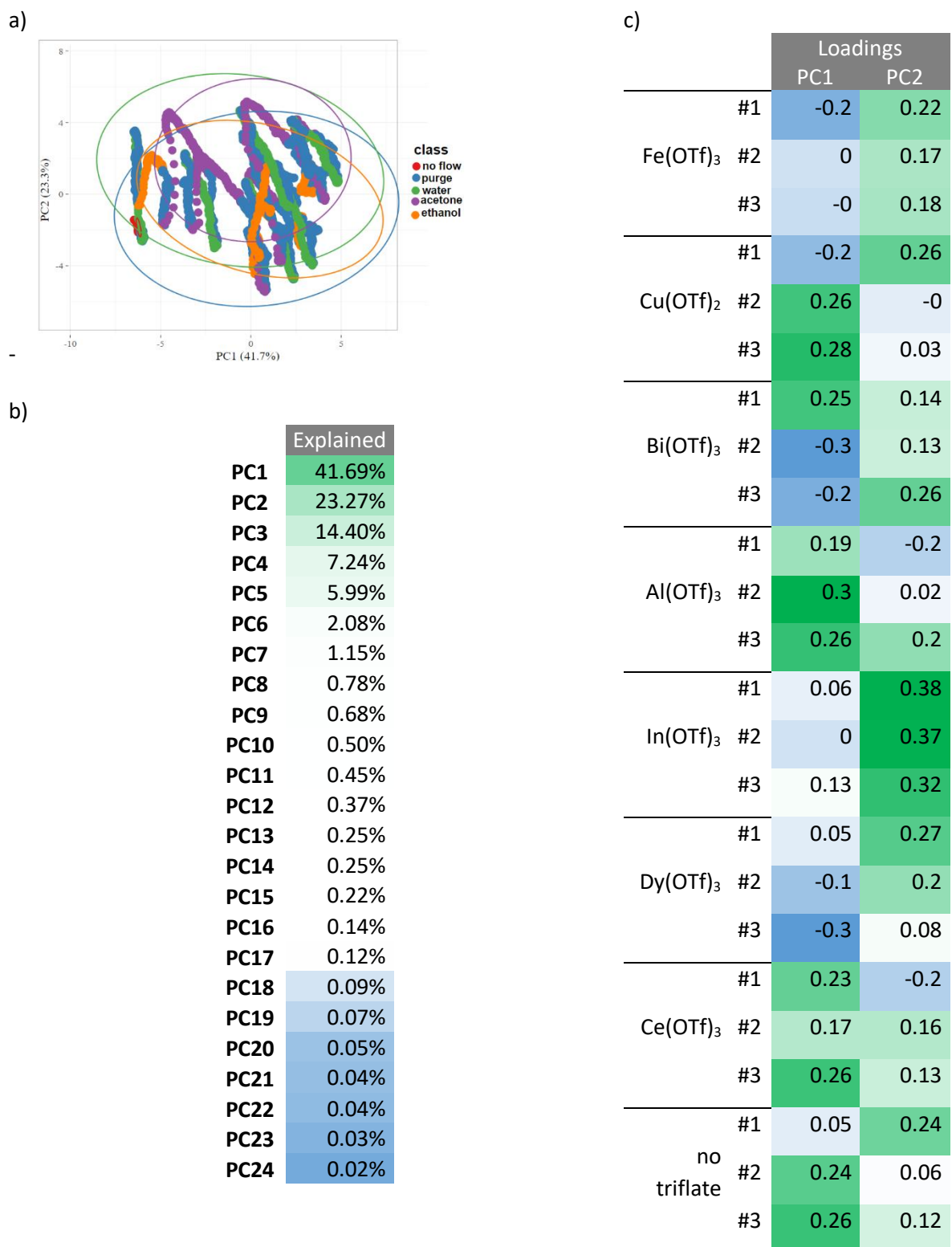

**Figure S31**. **PCA on "ema$_\alpha$(i(t))" for $\alpha$ = 1/2 | a,** PCA scores with 95% confidence ellipsoids. **b,** Individual variance for the different PC. **c,** PCA loadings of the different sensing elements' response for PC1 and PC2.

.

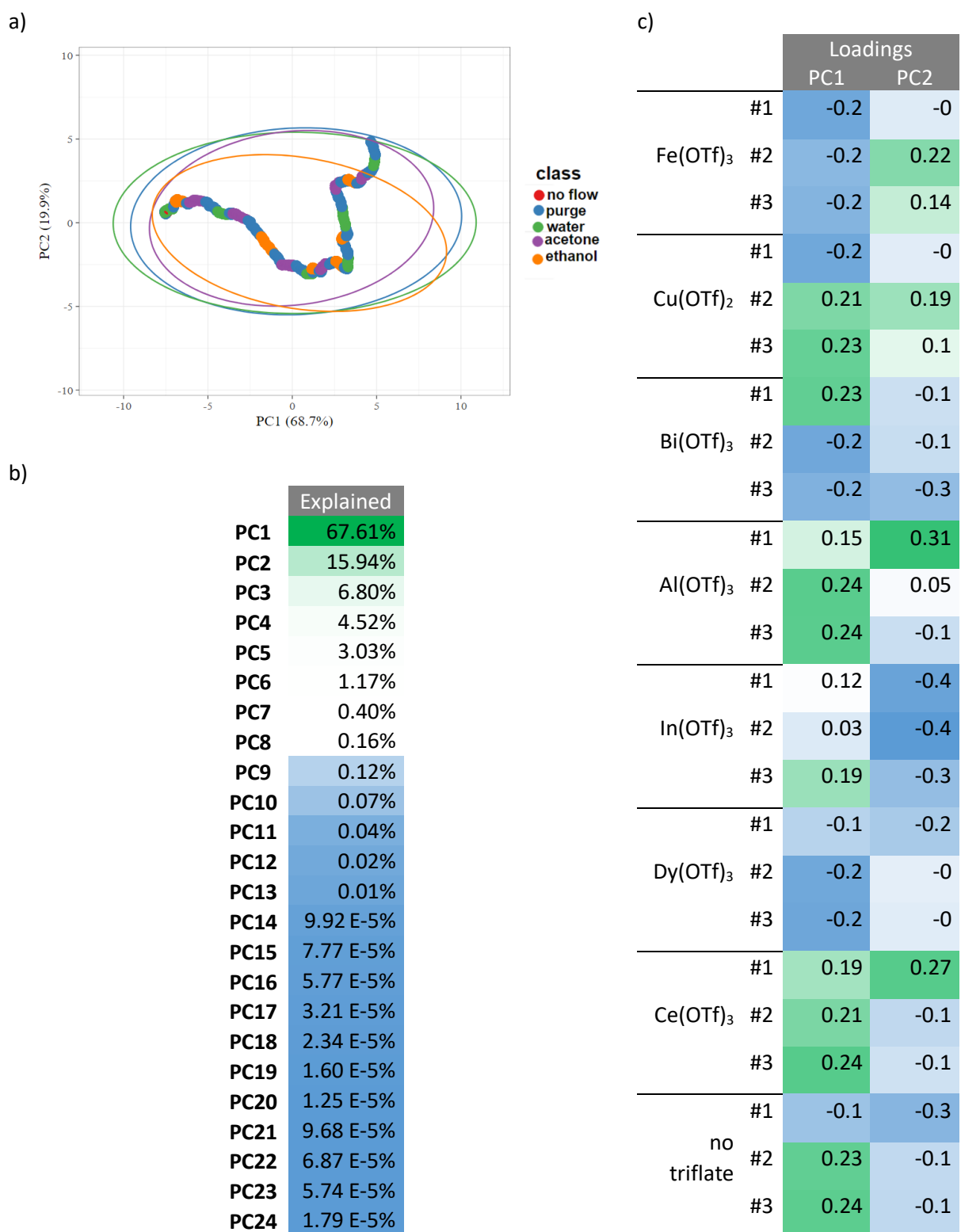

**Figure S32. PCA on "ema$_\alpha$(i(t))" for α = 1/1000. | a,** PCA scores with 95% confidence ellipsoids. **b,** Individual variance for the different PC. **c,** PCA loadings of the different sensing elements' response for PC1 and PC2.

a)

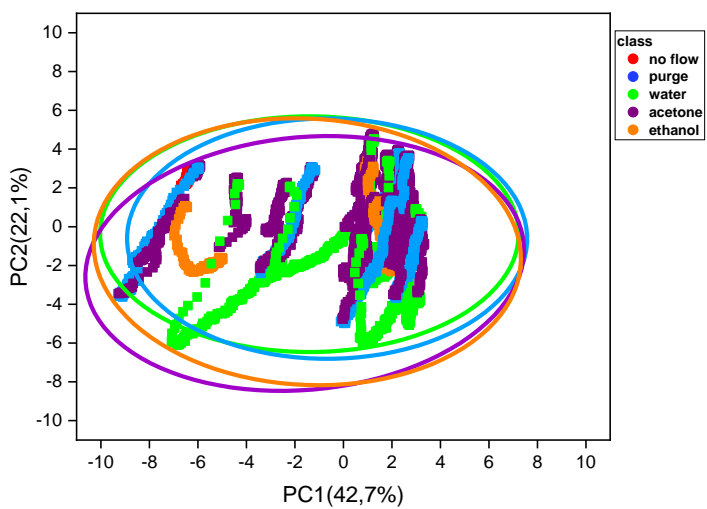

b)

| | Explained |
|---|---|
| **PC1** | 42.65% |
| **PC2** | 22.07% |
| **PC3** | 13.94% |
| **PC4** | 6.74% |
| **PC5** | 5.67% |
| **PC6** | 1.80% |
| **PC7** | 1.44% |
| **PC8** | 1.17% |
| **PC9** | 0.96% |
| **PC10** | 0.78% |
| **PC11** | 0.54% |
| **PC12** | 0.50% |
| **PC13** | 0.36% |
| **PC14** | 0.27% |
| **PC15** | 0.24% |
| **PC16** | 0.17% |
| **PC17** | 0.16% |
| **PC18** | 0.11% |
| **PC19** | 0.09% |
| **PC20** | 0.06% |
| **PC21** | 0.05% |
| **PC22** | 0.04% |
| **PC23** | 0.03% |
| **PC24** | 0.02% |

c)

| | | Loadings | |
|---|---|---|---|
| | | PC1 | PC2 |
| $Fe(OTf)_3$ | #1 | 0.16 | -0.3 |
| | #2 | 0.01 | -0.3 |
| | #3 | 0.01 | -0.2 |
| $Cu(OTf)_2$ | #1 | 0.14 | -0.3 |
| | #2 | -0.2 | -0 |
| | #3 | -0.3 | -0.1 |
| $Bi(OTf)_3$ | #1 | -0.3 | -0.1 |
| | #2 | 0.25 | -0.1 |
| | #3 | 0.18 | -0.2 |
| $Al(OTf)_3$ | #1 | -0.3 | -0 |
| | #2 | -0.3 | -0 |
| | #3 | -0.3 | -0.2 |
| $In(OTf)_3$ | #1 | -0.1 | -0.4 |
| | #2 | -0 | -0.4 |
| | #3 | -0.2 | -0.3 |
| $Dy(OTf)_3$ | #1 | -0 | -0.3 |
| | #2 | 0.11 | -0.2 |
| | #3 | 0.27 | -0.1 |
| $Ce(OTf)_3$ | #1 | -0.3 | 0.04 |
| | #2 | -0.1 | -0.1 |
| | #3 | -0.3 | -0.1 |
| no triflate | #1 | -0 | -0.2 |
| | #2 | -0.2 | 0 |
| | #3 | -0.3 | -0.1 |

**Figure S33. PCA on "ema$_\alpha$(R(t))" for α = 1/2 | a.** PCA scores with 95% confidence ellipsoids. **b.** Individual variance for the different PC. **c.** PCA loadings of the different sensing elements' response for PC1 and PC2.

.

a)

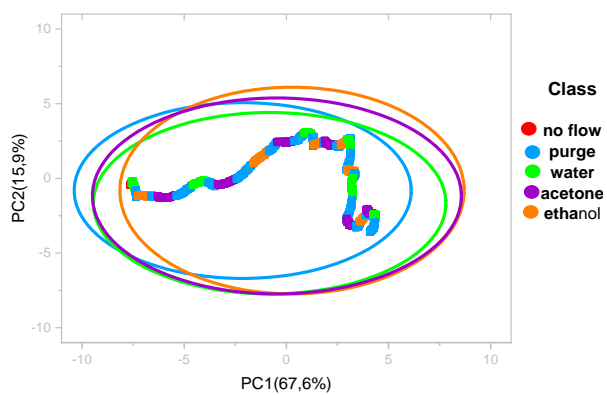

b)

| | Explained |
|---|---|
| PC1 | 67.61% |
| PC2 | 15.94% |
| PC3 | 6.80% |
| PC4 | 4.52% |
| PC5 | 3.03% |
| PC6 | 1.17% |
| PC7 | 0.40% |
| PC8 | 0.16% |
| PC9 | 0.12% |
| PC10 | 0.07% |
| PC11 | 0.04% |
| PC12 | 0.02% |
| PC13 | 0.01% |
| PC14 | 9.99 E-5% |
| PC15 | 7.77 E-5% |
| PC16 | 5.77 E-5% |
| PC17 | 3.21 E-5% |
| PC18 | 2.34 E-5% |
| PC19 | 1.60 E-5% |
| PC20 | 1.25 E-5% |
| PC21 | 9.68 E-6% |
| PC22 | 6.87 E-6% |
| PC23 | 5.74 E-6% |
| PC24 | 3.48 E-6% |

c)

| | | Loadings | |
|---|---|---|---|
| | | PC1 | PC2 |
| Fe(OTf)$_3$ | #1 | 0.24 | -0.1 |
| | #2 | 0.19 | 0.19 |
| | #3 | 0.18 | 0.15 |
| Cu(OTf)$_2$ | #1 | 0.24 | -0.1 |
| | #2 | -0.2 | 0.18 |
| | #3 | -0.2 | 0.11 |
| Bi(OTf)$_3$ | #1 | -0.2 | -0.1 |
| | #2 | 0.22 | -0.2 |
| | #3 | 0.18 | -0.3 |
| Al(OTf)$_3$ | #1 | -0.2 | 0.02 |
| | #2 | -0.2 | -0 |
| | #3 | -0.2 | -0.1 |
| In(OTf)$_3$ | #1 | -0.1 | -0.4 |
| | #2 | -0 | -0.5 |
| | #3 | -0.2 | -0.2 |
| Dy(OTf)$_3$ | #1 | 0.08 | -0.1 |
| | #2 | 0.23 | -0 |
| | #3 | 0.24 | -0,1 |
| Ce(OTf)$_3$ | #1 | -0.2 | 0.02 |
| | #2 | -0 | -0.3 |
| | #3 | -0.2 | -0.1 |
| no triflate | #1 | 0.12 | -0.4 |
| | #2 | -0.2 | -0.1 |
| | #3 | -0.2 | -0.1 |

**Figure S34. PCA on "ema$_\alpha$(R(t))" for $\alpha$ = 1/1000 | a.** PCA scores with 95% confidence ellipsoids. **b,** Individual variance for the different PC. **c,** PCA loadings of the different sensing elements' response for PC1 and PC2.